\pdfoutput=1
\documentclass[preprint,12pt,authoryear,nopreprintline]{elsarticle}
\RequirePackage{amsthm,amsmath,amsfonts,amssymb}
\RequirePackage[colorlinks,citecolor=blue,urlcolor=blue,linkcolor=blue]{hyperref}
\RequirePackage{graphicx}
\usepackage{booktabs}
\usepackage{placeins}
\usepackage{tikz}
\usetikzlibrary{arrows.meta,positioning,decorations.pathreplacing,calc}

\usepackage[table]{xcolor}
\definecolor{light-gray}{gray}{0.7}
\definecolor{light-int-gray}{gray}{0.77}
\definecolor{light-light-gray}{gray}{0.9}

\usepackage{bbm}

\DeclareMathOperator*{\esssup}{ess\,sup}

\theoremstyle{plain}

\newtheorem{proposition}{Proposition}
\newtheorem{lemma}{Lemma}
\theoremstyle{remark}

\newcommand{\codedoi}{\url{https://doi.org/10.5281/zenodo.24198622}}
\newcommand{\codelocation}{\codedoi{} (Zenodo)}

\usepackage{etoolbox}
\makeatletter
\newbox\frontsectionbox
\newenvironment{frontsection}[1]{\global\setbox\frontsectionbox=\vbox\bgroup
  \hsize=\textwidth%
  \noindent\unskip\textbf{#1}\par\medskip\noindent\unskip\ignorespaces}
 {\egroup}
\patchcmd{\pprintMaketitle}{\ifvoid\keybox}
  {\ifvoid\frontsectionbox\else\unvbox\frontsectionbox\par\vskip10pt\fi\ifvoid\keybox}
  {}{\PackageWarningNoLine{main}{Practitioner Summary not added to the title block}}
\makeatother

\begin{document}

\begin{frontmatter}

\title{Distribution-Free Prediction Bands for Multivariate Functional Time Series: an Application to the Italian Gas Market}

\author[unipd]{Jacopo Diquigiovanni}
\ead{jacopo.diquigiovanni@phd.unipd.it}
\author[UCSC,MOX]{Matteo Fontana\corref{cor1}}
\ead{matteo.fontana@unicatt.it}
\author[MOX]{Simone Vantini}
\ead{simone.vantini@polimi.it}
\cortext[cor1]{Corresponding author}

\affiliation[unipd]{organization={Department of Statistical Sciences, University of Padova},
            country={Italy}}
\affiliation[UCSC]{organization={Department of Statistical Sciences, Università Cattolica del Sacro Cuore},
            city={Milan},
            country={Italy}}
\affiliation[MOX]{organization={MOX - Department of Mathematics, Politecnico di Milano},
            country={Italy}}

\begin{abstract}
In energy auctions, offers and bids are submitted before the outcome is known, so point forecasts must be complemented by probabilistic ones. For the daily supply and demand curves of these auctions, modelled as a multivariate functional time series, we propose simultaneous prediction bands following three paradigms for nonexchangeable data. The procedures share a nonconformity score that gives closed-form bands for any point predictor. The first applies split conformal prediction with non-overlapping blocks, a special case of a known method for dependent data. Under conditions on the data and the predictor, its coverage deviation is bounded by terms vanishing as the training set and the number of blocks grow. The second applies adaptive conformal inference to split conformal bands: for any data sequence, the long-run coverage frequency approaches the nominal level, but bands can be infinite. The third, following nonexchangeable conformal prediction, weights recent calibration scores more. Its coverage loss is bounded by distances from exchangeability, and its bands are typically wider, possibly infinite. Simulations compare the procedures, also under misspecification and distribution shift. On Italian gas market data, the bands yield a prediction region for the clearing price and quantity and a what-if analysis of additional bids or offers.
\end{abstract}

\begin{frontsection}{Practitioner Summary}
In daily gas and electricity auctions, price and quantity are set where the supply curve (sale offers) meets the demand curve (purchase bids). Traders need to know how far these curves may be from forecasts.

We propose three procedures, adapted from the literature, that use past errors to turn forecasts from any model into a band of prices for every quantity, jointly for both curves.

Procedure~1 weights past errors equally. Its band contains the next day's curves with approximately a chosen probability, e.g.\ 75\%, if the data and the model satisfy certain conditions. This probability is an average over days. It suits stable markets.

Procedure~2 raises the probability after each miss (curves outside the band) and lowers it otherwise. Whatever the data, the long-run fraction of days with the curves in the band approaches the chosen one, but the band can be unbounded or empty. It suits long-term use.

Procedure~3 weights recent errors more heavily, for gradually changing markets; too strong a discount makes its band unbounded.

On Italian gas data, Procedures~1 and~2 contained the curves at approximately the chosen frequency, with similar average widths when all bands were bounded. Procedure~3 contained them at least at this frequency, with wider bands.

The intersection of both curves' bands contains the clearing point at least as often as the band contains the curves, if these cross within the quantities considered. The band allows a what-if analysis of an additional bid or offer, assuming no reaction from other participants.
\end{frontsection}


\end{frontmatter}

\section{Introduction}

In liberalised energy markets, producers, retailers, traders and storage operators submit offers and bids to daily auctions before the outcome is known. Decisions such as bidding strategies and risk management are typically formulated as optimisation problems under uncertainty, which require a description of the uncertainty rather than a single forecast \citep{wallace2003stochastic,conejo2010decision}. Forecasting is then a decision-support activity \citep{jors_fildes2008forecasting}, and point forecasts must be complemented by probabilistic forecasts, such as predictive distributions, quantiles or prediction intervals \citep{gneiting2014probabilistic,petropoulos2022forecasting}. In electricity price forecasting, \citet{weron2014electricity} identified probabilistic forecasts as an open challenge, and they have since received increasing attention \citep{hong2016probabilistic,nowotarski2018recent}. In the Operational Research literature, \citet{jors_taylor2003short} and \citet{jors_garciamartos2015electricity} forecast electricity demand and prices, respectively, and \citet{jors_wang2022uncertainty} construct prediction intervals for combined forecasts.

In auction markets for electricity and gas, the clearing price and the quantity exchanged on a given day are determined by the intersection of the aggregate supply (offer) and demand (bid) curves. A forecast of the clearing price describes the equilibrium only, whereas a forecast of the two curves also describes their shape around it, and therefore how the clearing point would move if a participant submitted an additional offer or bid. Supply and demand curves have been predicted for electricity markets \citep{ziel2016electricity,shah_forecasting_2020} and for the Italian gas market \citep{canale_constrained_2016}; the last two works treat the curves as functional data observed over time, as we do.

Functional data analysis deals with data that are functions, such as curves or surfaces \citep{ramsay_functional_2005}. Functional data observed sequentially over time form a functional time series (FTS) \citep{hormann_functional_2012}, whose literature has focused mainly on one-step-ahead point forecasting, e.g.\ through functional autoregressive models \citep[reviewed, with energy applications, by][]{chen2021review}. For multivariate functional data, i.e.\ vectors of functions possibly defined on different domains, \citet{chiou2016multivariate} propose a multivariate functional linear regression model, with prediction intervals based on asymptotic arguments. The uncertainty of FTS forecasts is usually quantified through the bootstrap: \citet{paparoditis_bootstrap_2021} construct bootstrap prediction bands for FTS, also based on asymptotic arguments, and \citet{rossini_quantifying_2019} quantify the prediction uncertainty of functional autoregressive models under shape constraints, with an application to the Italian gas market. These bands rely on assumptions on the data generating process. To the best of our knowledge, none of these approaches provides distribution-free simultaneous prediction bands for a multivariate functional response. We do not compare our bands with them, but their point predictors can be used within our procedures.

Conformal prediction (CP) \citep{gammerman_learning_1998,vovk2005algorithmic,shafer_tutorial_2008}, reviewed by \citet{fontana2023conformal} and \citet{angelopoulos2023gentle}, provides prediction sets whose marginal (also called unconditional) coverage is at least the nominal level in finite samples, for any point predictor, under the sole assumption that the data are exchangeable. In split (or inductive) conformal prediction \citep{papadopoulos2002inductive,lei2018distribution}, the point predictor is fitted once on a training set, and the prediction set is determined by the nonconformity scores of a separate calibration set. For functional data, \citet{diquigiovanni2021importance} construct finite-sample exact prediction bands, and \citet{diquigiovanni_conformal_2021} propose, for multivariate functional regression, a nonconformity score under which the split conformal prediction set is a simultaneous band in closed form. \citet{ajroldi2023conformal} construct conformal bands for functional time series defined on a two-dimensional domain, with the block-split scheme of \citet{diquigiovanni2021distribution}.

Time series are not exchangeable in general, since the observations are serially dependent and their distribution may change over time, so the finite-sample guarantee of conformal prediction does not apply directly. CP has been applied to electricity price forecasting \citep{kath2021conformal} and to the net positions of market makers \citep{wisniewski2020application} but, to the best of our knowledge, not to simultaneous prediction bands of the supply and demand curves of an energy market.

We propose simultaneous prediction bands for multivariate functional time series, such as the supply and demand curves, to support the decisions of market participants. Following three paradigms to handle nonexchangeable data, we obtain three procedures, which share the nonconformity score of \citet{diquigiovanni_conformal_2021} and differ in the way they calibrate the band. Procedure~1, block-split conformal bands (BSC), is the split (inductive) version of the non-overlapping block scheme of \citet{chernozhukov2018exact}, a special case of their framework that is only mentioned in their Section~2.4 and was written out by \citet{diquigiovanni2021distribution}. Procedure~2 (ACI) applies adaptive conformal inference \citep{gibbs2021adaptive} to the split conformal band computed on a rolling window. It controls the frequency of miscoverage over a sequence of time points, but not the coverage probability of a single band. Procedure~3 (WCP) applies the weighted split conformal prediction of \citet{barber2023beyond} to the same band, with fixed weights that decrease with the time distance from the target time point, and is intended for gradual changes in distribution. We compare the three procedures, as alternative ways of applying CP to time series, by simulation and on the offer and demand curves of the MGS (\textit{Mercato Gas in Stoccaggio}), the Italian market for stored gas. The contributions are this comparison, the statement of the procedures with their guarantees, and a prediction region for the clearing point and a what-if analysis of an additional bid or offer, which can be computed from the band of any of the three procedures.

The procedures and the nonconformity score are taken from the literature. Our guarantees concern the scalar scores, so Propositions~\ref{prop:bound}--\ref{prop:weighted} hold for any real-valued nonconformity score. The coverage bound of BSC was stated without proof by \citet{diquigiovanni2021distribution} and \citet{ajroldi2023conformal}. We give a complete proof, with smaller constants than theirs and than those of Theorem~2 of \citet{chernozhukov2018exact} (Proposition~\ref{prop:bound}), and an example in which its conditions hold with explicit rates (\ref{app:remarks}). Proposition~\ref{prop:aci} extends the bound of \citet{gibbs2021adaptive} to feedback observed with a delay. The delayed update was proposed by \citet{hallberg2024adaptive} and \citet{wang2024online} for scalar forecasts several steps ahead, without a bound of this type. Proposition~\ref{prop:weighted} is a corollary of the bounds of \citet{barber2023beyond}. We add that the mass of the test point exceeds $1-\rho$ for every size of the calibration set, which explains the overcoverage of Procedure~3 under stationarity.

\section{Methodology}
\label{sec:methods}

\subsection{Setting and notation}
\label{sec:setting}

Let $\boldsymbol{Z}_1,\dots,\boldsymbol{Z}_T$ be the observed time series, where $T \in \mathbb{N}^*$ is the number of time points, $\mathbb{N}^*$ is the set of positive integers, and $\boldsymbol{Z}_t=(\boldsymbol{X}_t,\boldsymbol{Y}_t)$ consists of covariates $\boldsymbol{X}_t$ and a multivariate functional response $\boldsymbol{Y}_t=(Y_t^{(1)},\dots,Y_t^{(p)})$ with $p \in \mathbb{N}^*$ components. For $j=1,\dots,p$, let $\mathcal{Q}_j$ be a closed and bounded subset of $\mathbb{R}^{d_j}$, $d_j \in \mathbb{N}^*$, and $L^{\infty}(\mathcal{Q}_j)$ the space of essentially bounded measurable functions $y:\mathcal{Q}_j \rightarrow \mathbb{R}$ (\ref{app:remarks}). The response $\boldsymbol{Y}_t$ takes values in $\mathcal{Y}:=L^{\infty}(\mathcal{Q}_1)\times\dots\times L^{\infty}(\mathcal{Q}_p)$; when $p>1$, the domains $\mathcal{Q}_j$, their dimensions $d_j$ and the ranges of the components may differ across $j$. The covariates are $\boldsymbol{X}_t=(X_t^{(1)},\dots,X_t^{(p)})$, where $X_t^{(j)}$ takes values in a measurable space $\mathcal{X}_j$ and collects the covariates used to predict $Y_t^{(j)}$: lagged values of the response and, possibly, exogenous scalar or functional covariates; different components may share some or all of their covariates. Realisations are denoted by lowercase letters, e.g.\ $\boldsymbol{z}_t=(\boldsymbol{x}_t,\boldsymbol{y}_t)$.

When the process $\{\boldsymbol{Z}_t\}$ is strictly stationary, the regression function of the $j$-th component, $\mu^{(j)}(x)=\mathbb{E}[Y_t^{(j)} \mid X_t^{(j)}=x]$, $x \in \mathcal{X}_j$, does not depend on $t$; its value at $q \in \mathcal{Q}_j$ is $[\mu^{(j)}(x)](q)=\mathbb{E}[Y_t^{(j)}(q) \mid X_t^{(j)}=x]$.

The aim is to construct, from $\boldsymbol{Z}_1,\dots,\boldsymbol{Z}_T$ and $\boldsymbol{X}_{T+1}$, a simultaneous prediction set $\mathcal{C}_{T,1-\alpha}(\boldsymbol{X}_{T+1}) \subseteq \mathcal{Y}$ for $\boldsymbol{Y}_{T+1}$, i.e.\ a set that covers the $p$ components jointly and over their whole domains, whose marginal coverage $\mathbb{P}(\boldsymbol{Y}_{T+1} \in \mathcal{C}_{T,1-\alpha}(\boldsymbol{X}_{T+1}))$ is close to the nominal coverage level $1-\alpha$, where $\alpha \in (0,1)$ is the miscoverage level and the probability refers to the joint distribution of $\boldsymbol{Z}_1,\dots,\boldsymbol{Z}_{T+1}$ and of the split (Section~\ref{sec:procedure}).

\subsection{Multivariate functional prediction band and nonconformity score}
\label{sec:score}

The bands of the three procedures are split conformal bands (\ref{app:primer}). The point predictors $\hat{\mu}^{(j)}_{\mathcal{I}_1}$ of $\mu^{(j)}$ are fitted on a training set $\{\boldsymbol{z}_h : h \in \mathcal{I}_1\}$, with $n_1=|\mathcal{I}_1|$, and the band is determined by the nonconformity scores of a calibration set (Section~\ref{sec:procedure}). We use the nonconformity measure of \citet{diquigiovanni_conformal_2021}, for which the prediction set is a band in closed form. For $\boldsymbol{z}=(\boldsymbol{x},\boldsymbol{y})$,
\begin{equation}
A(\{\boldsymbol{z}_h : h \in \mathcal{I}_1\},\boldsymbol{z}):=\max_{j=1,\dots,p}\ \esssup_{q \in \mathcal{Q}_j} \frac{\left|y^{(j)}(q)-[\hat{\mu}^{(j)}_{\mathcal{I}_1}(x^{(j)})](q)\right|}{s^{(j)}(q)},
\label{eq:score}
\end{equation}
where $[\hat{\mu}^{(j)}_{\mathcal{I}_1}(x^{(j)})](q)$, $q \in \mathcal{Q}_j$, is the point prediction of $[\mu^{(j)}(x^{(j)})](q)$ and the modulation function $s^{(j)}$ is a positive function on $\mathcal{Q}_j$, both computed from the training set; any regression method and any such function can be used. The finite-sample validity under exchangeability (Section~\ref{sec:guarantees}) and Propositions~\ref{prop:aci} and~\ref{prop:weighted} hold for any choice, and Proposition~\ref{prop:bound} holds under its conditions. The choice affects the size and shape of the band: a more accurate point predictor can be expected to produce smaller residuals and therefore a narrower band. The modulation function determines how the width of the band varies over $\mathcal{Q}_j$. A function that follows the variability of the residuals, instead of a constant one, reduces the size of the band when this variability changes over the domain, though not necessarily with small training sets \citep[Section~3.3]{diquigiovanni2021importance}. We use the standard deviation function of the training residuals,
\begin{equation}
\begin{split}
s^{(j)}(q)&:=\left(\frac{1}{n_1-1}\sum_{h \in \mathcal{I}_1}\left(r_h^{(j)}(q)-\bar{r}^{(j)}(q)\right)^2\right)^{1/2}, \qquad q \in \mathcal{Q}_j,\\
r_h^{(j)}(q)&:=y_h^{(j)}(q)-[\hat{\mu}^{(j)}_{\mathcal{I}_1}(x_h^{(j)})](q), \qquad \bar{r}^{(j)}(q):=\frac{1}{n_1}\sum_{h \in \mathcal{I}_1} r_h^{(j)}(q).
\end{split}
\label{eq:sd}
\end{equation}

For a candidate value $\boldsymbol{y} \in \mathcal{Y}$ of $\boldsymbol{Y}_{T+1}$, let $R_{T+1}(\boldsymbol{y}):=A(\{\boldsymbol{z}_h : h \in \mathcal{I}_1\},(\boldsymbol{x}_{T+1},\boldsymbol{y}))$ be its score. By \eqref{eq:score}, for a threshold $k \in [0,+\infty)$, $R_{T+1}(\boldsymbol{y}) \leq k$ if and only if $\boldsymbol{y}$ belongs to the band
\begin{equation}
\begin{split}
\Big\{\boldsymbol{y} \in \mathcal{Y}:\ & y^{(j)}(q) \in \left[\hat{y}^{(j)}_{T+1}(q)-k\, s^{(j)}(q),\ \hat{y}^{(j)}_{T+1}(q)+k\, s^{(j)}(q)\right]\\
& \text{for } j=1,\dots,p \text{ and all } q \in \mathcal{Q}_j\Big\},
\end{split}
\label{eq:band}
\end{equation}
where $\hat{y}^{(j)}_{T+1}(q):=[\hat{\mu}^{(j)}_{\mathcal{I}_1}(x^{(j)}_{T+1})](q)$, $q \in \mathcal{Q}_j$, is the point prediction. The three procedures differ in the way they determine $k$ from the calibration scores, i.e.\ in the way they calibrate the band.

The modulation functions make the residuals of components with different variability comparable in the maximum in \eqref{eq:score}. We measure the size of the band by $\sum_{j=1}^p \int_{\mathcal{Q}_j} 2k\,s^{(j)}(q)\,\mathrm{d}q$.

\subsection{Split conformal prediction with non-overlapping blocks}
\label{sec:procedure}

Procedure~1, block-split conformal bands (BSC), is the split version of the non-overlapping block scheme of \citet{chernozhukov2018exact} (\ref{app:remarks}).

\paragraph{Procedure~1 (BSC)} Figure~\ref{fig:scheme} illustrates the steps.
\begin{enumerate}
\item \textit{Split.} If $\boldsymbol{X}_t$ contains lagged values up to lag $t_0 \geq 0$, the first $t_0$ time points enter the analysis only as lagged covariates, and the indices $t_0+1,\dots,T$ are partitioned into the training index set $\mathcal{I}_1$ and the calibration index set $\mathcal{I}_2$, with $n_1=|\mathcal{I}_1|$, $n_2=|\mathcal{I}_2|$ and $T=t_0+n_1+n_2$. Any partition independent of the data can be used, e.g.\ a random split, with $\mathcal{I}_1$ drawn uniformly among the subsets of $\{t_0+1,\dots,T\}$ of size $n_1$, or a chronological split, with $\mathcal{I}_1=\{t_0+1,\dots,t_0+n_1\}$.
\item \textit{Point predictor.} The point predictors $\hat{\mu}^{(j)}_{\mathcal{I}_1}$ are fitted on the training set.
\item \textit{Blocks.} Let $\omega_d$ denote the $d$-th smallest element of $\mathcal{I}_2 \cup \{T+1\}$, $d=1,\dots,n_2+1$, so that $\omega_{n_2+1}=T+1$. Let $b \in \{1,\dots,n_2+1\}$ be the block length, with $K:=(n_2+1)/b$ an integer. The positions $1,\dots,n_2+1$ are partitioned into the $K$ blocks $\{(g-1)b+1,\dots,gb\}$, $g=1,\dots,K$. The block permutations $\pi_i:\{1,\dots,n_2+1\}\rightarrow\{1,\dots,n_2+1\}$, $i=1,\dots,K$, are the cyclic shifts by $(i-1)b$ positions,
\begin{equation*}
\pi_{i}(d)=\begin{cases}
d+(i-1)b & \text{if } 1 \leq d \leq n_2+1-(i-1)b,\\
d+(i-1)b-(n_2+1) & \text{if } n_2+1-(i-1)b < d \leq n_2+1,
\end{cases}
\end{equation*}
and $\Pi:=\{\pi_1,\dots,\pi_K\}$. Scores are computed at the images of the test position,
\begin{equation*}
\mathcal{D}_{\Pi}:=\{\pi_i(n_2+1): i=1,\dots,K\}=\{b,2b,\dots,(K-1)b,n_2+1\},
\end{equation*}
i.e.\ the last position of each block, so that $|\mathcal{D}_{\Pi}|=K$.
\item \textit{Nonconformity scores.} The scores of the $K-1$ selected calibration observations are $R_{\omega_d}:=A(\{\boldsymbol{z}_h : h \in \mathcal{I}_1\},\boldsymbol{z}_{\omega_d})$, $d \in \mathcal{D}_{\Pi}\setminus\{n_2+1\}$, and the test score of a candidate value $\boldsymbol{y} \in \mathcal{Y}$ is $R_{\omega_{n_2+1}}:=R_{T+1}(\boldsymbol{y})$.
\item \textit{Conformal p-value and prediction set.} The conformal p-value of $\boldsymbol{y}$ is
\begin{equation}
\begin{split}
\hat{p}(\boldsymbol{y})&:=\frac{1}{K}\left|\left\{d \in \mathcal{D}_{\Pi}: R_{\omega_d} \geq R_{T+1}(\boldsymbol{y})\right\}\right|\\
&\phantom{:}=\frac{1}{K}\left(1+\left|\left\{d \in \mathcal{D}_{\Pi}\setminus\{n_2+1\}: R_{\omega_d} \geq R_{T+1}(\boldsymbol{y})\right\}\right|\right),
\end{split}
\label{eq:pvalue}
\end{equation}
and the prediction set is $\mathcal{C}_{T,1-\alpha}(\boldsymbol{x}_{T+1}):=\{\boldsymbol{y} \in \mathcal{Y}: \hat{p}(\boldsymbol{y})>\alpha\}$. If $\alpha \geq 1/K$, it is the band \eqref{eq:band} with $k$ the ${\lceil K(1-\alpha) \rceil}$-th smallest of the $K-1$ calibration scores (derivation in \ref{app:remarks}). If $\alpha<1/K$, then $\hat{p}(\boldsymbol{y}) \geq 1/K>\alpha$ for every $\boldsymbol{y}$ and the prediction set is the whole space $\mathcal{Y}$.
\end{enumerate}

\begin{figure}[tbp]
\centering
\begin{tikzpicture}[
  font=\small,
  tc/.style={draw, minimum width=0.55cm, minimum height=0.55cm, inner sep=0pt, outer sep=0pt, font=\scriptsize},
  cell/.style={draw, minimum width=0.84cm, minimum height=0.6cm, inner sep=0pt, outer sep=0pt, anchor=west, font=\scriptsize},
  key/.style={draw, minimum width=0.4cm, minimum height=0.4cm, inner sep=0pt, outer sep=0pt, font=\scriptsize},
  keytext/.style={right=3pt, font=\footnotesize, inner sep=0pt},
  rowlabel/.style={anchor=south west, font=\small, inner sep=0pt},
  brace/.style={decorate, decoration={brace, amplitude=5pt}},
  mbrace/.style={decorate, decoration={brace, amplitude=5pt, mirror}}
]
\def\schemecell#1#2#3#4{%
  \ifcase#2
    \node[tc, densely dashed] at (#1*0.55,#4) {};
  \or
    \node[tc, fill=gray!20] at (#1*0.55,#4) {};
  \or
    \node[tc, fill=white] at (#1*0.55,#4) {#3};
  \or
    \node[tc, fill=gray!60] at (#1*0.55,#4) {#3};
  \or
    \node[tc, fill=black, text=white] at (#1*0.55,#4) {#3};
  \fi}
\node[key, densely dashed] (k0) at (0.55,2.75) {};
\node[keytext] (t0) at (k0.east) {lags only};
\node[key, fill=gray!20] (k1) at ($(t0.east)+(0.55,0)$) {};
\node[keytext] (t1) at (k1.east) {training ($\mathcal{I}_1$)};
\node[key, fill=white] (k2) at ($(t1.east)+(0.55,0)$) {$d$};
\node[keytext] (t2) at (k2.east) {calibration ($\mathcal{I}_2$)};
\node[key, fill=gray!60] (k3) at ($(t2.east)+(0.55,0)$) {$d$};
\node[keytext] (t3) at (k3.east) {scored};
\node[key, fill=black, text=white] (k4) at ($(t3.east)+(0.55,0)$) {$d$};
\node[keytext] (t4) at (k4.east) {test point};
\draw[-{Stealth[length=2.2mm]}] (0.275,1.95) -- node[above, pos=0.5, font=\footnotesize] {time} (12.375,1.95);
\foreach \t/\lab in {1/1, 3/t_0{+}1, 21/T, 22/T{+}1} { \node[font=\scriptsize, anchor=base] at (\t*0.55,1.5) {$\lab$}; }
\node[rowlabel] at (0.275,0.98) {(a) random split};
\foreach \t/\c/\d in {1/0/,2/0/,3/2/1,4/1/,5/2/2,6/3/3,7/1/,8/1/,9/2/4,10/2/5,11/1/,12/3/6,13/1/,14/2/7,15/2/8,16/1/,17/3/9,18/1/,19/2/10,20/1/,21/2/11,22/4/12}
  { \schemecell{\t}{\c}{\d}{0.7} }
\node[rowlabel] at (0.275,-0.27) {(b) chronological split};
\foreach \t/\c/\d in {1/0/,2/0/,3/1/,4/1/,5/1/,6/1/,7/1/,8/1/,9/1/,10/1/,11/2/1,12/2/2,13/3/3,14/2/4,15/2/5,16/3/6,17/2/7,18/2/8,19/3/9,20/2/10,21/2/11,22/4/12}
  { \schemecell{\t}{\c}{\d}{-0.55} }
\node[font=\small] at (6.325,-1.45) {$\omega_1<\omega_2<\dots<\omega_{n_2}<\omega_{n_2+1}=T+1$};
\foreach \d in {1,...,12} {
  \pgfmathtruncatemacro{\scored}{mod(\d,3)==0 ? 1 : 0}
  \ifnum\d=12
    \node[cell, fill=black, text=white] (c\d) at ($(1.285,-2.3)+(\d*0.84-0.84,0)$) {\d};
  \else\ifnum\scored=1
    \node[cell, fill=gray!60] (c\d) at ($(1.285,-2.3)+(\d*0.84-0.84,0)$) {\d};
  \else
    \node[cell, fill=white] (c\d) at ($(1.285,-2.3)+(\d*0.84-0.84,0)$) {\d};
  \fi\fi
}
\foreach \k in {3,6,9} { \draw[very thick] (c\k.north east) -- (c\k.south east); }
\draw[very thick] (c1.north west) rectangle (c12.south east);
\node[anchor=east, font=\small] at ($(c1.west)+(-0.15,0)$) {position $d$};
\node[below=2pt of c3.south east, anchor=north east] {$b$};
\node[below=2pt of c6.south east, anchor=north east] {$2b$};
\node[below=2pt of c9.south east, anchor=north east] {$(K{-}1)b$};
\node[below=2pt of c12.south east, anchor=north east] {$n_2{+}1$};
\draw[mbrace] ($(c1.south west)+(0.06,-0.72)$) -- node[below=5pt] {block $1$} ($(c3.south east)+(-0.06,-0.72)$);
\draw[mbrace] ($(c4.south west)+(0.06,-0.72)$) -- node[below=5pt] {block $2$} ($(c6.south east)+(-0.06,-0.72)$);
\draw[mbrace] ($(c7.south west)+(0.06,-0.72)$) -- node[below=5pt] {block $K{-}1$} ($(c9.south east)+(-0.06,-0.72)$);
\draw[mbrace] ($(c10.south west)+(0.06,-0.72)$) -- node[below=5pt] {block $K$} ($(c12.south east)+(-0.06,-0.72)$);
\end{tikzpicture}
\caption{Scheme of BSC with $t_0=2$, $n_1=8$, $n_2=11$ and $b=3$ ($T=21$, $K=4$), showing the time points $1,\dots,T+1$ under (a) a random and (b) a chronological split (top), each element of $\mathcal{I}_2 \cup \{T+1\}$ numbered by its position $d$, and the $K$ blocks of positions (bottom).}
\label{fig:scheme}
\end{figure}

When $b=1$, $K=n_2+1$ and BSC is split conformal prediction. The scored time points are at least $b$ apart (\ref{app:remarks}). For weakly dependent data, the dependence among the scores can be expected to decrease as $b$ increases, but so does their number $K$. Since the score \eqref{eq:score} depends on a single observation, BSC amounts to split conformal prediction on every $b$-th calibration score. Neither Proposition~\ref{prop:bound} nor the simulations of Section~\ref{sec:simulation} show a net benefit of $b>1$, and we keep it for the comparison only.

\subsection{Coverage guarantees of BSC}
\label{sec:guarantees}

\paragraph{Exchangeable data} If the pairs $\boldsymbol{Z}_t$, $t \in \mathcal{I}_1 \cup \mathcal{I}_2 \cup \{T+1\}$, are exchangeable, then, conditionally on the training set, the $K$ scores $R_{\omega_d}$, $d \in \mathcal{D}_{\Pi}$, with the test score evaluated at $\boldsymbol{y}=\boldsymbol{Y}_{T+1}$, are exchangeable for any $b$, which gives finite-sample validity (\ref{app:primer}):
\begin{equation*}
\mathbb{P}\left(\boldsymbol{Y}_{T+1} \in \mathcal{C}_{T,1-\alpha}(\boldsymbol{X}_{T+1})\right) \geq 1-\alpha
\end{equation*}
for all $\alpha \in (0,1)$ and all $T$, $n_1$, $n_2$ and $b$. If, in addition, the $K$ scores are almost surely distinct, the coverage is exactly $\lceil K(1-\alpha) \rceil/K$, which equals $1-\alpha$ when $\alpha K$ is an integer (\ref{app:primer}, with $K$ in place of $n_2+1$).

\paragraph{Dependent data} Following \citet{chernozhukov2018exact}, the coverage can be related to the accuracy with which the scores approximate oracle scores. Let $A^*$ be an oracle nonconformity measure, i.e.\ a real-valued measurable function of a single pair $\boldsymbol{z}$ that does not depend on the data, typically the population counterpart of $A$ (\ref{app:remarks}), and let $R^*_t:=A^*(\boldsymbol{Z}_t)$. In the rest of this section the scores are random variables, $R_{\omega_d}=A(\{\boldsymbol{Z}_h : h \in \mathcal{I}_1\},\boldsymbol{Z}_{\omega_d})$ for $d \in \mathcal{D}_{\Pi}$, with $R_{\omega_{n_2+1}}=R_{T+1}$ evaluated at $\boldsymbol{Y}_{T+1}$. Define, for $a \in \mathbb{R}$,
\begin{equation*}
\tilde{F}_K(a):=\frac{1}{K}\sum_{d \in \mathcal{D}_{\Pi}} \mathbbm{1}\{R^*_{\omega_d}<a\}, \qquad F(a):=\mathbb{P}(R^*_{T+1}<a).
\end{equation*}
Let $\delta_{1K}$ and $\zeta_{1K}$ be positive sequences converging to 0 as $K \rightarrow \infty$, and $\delta_{2n_1}$ and $\zeta_{2n_1}$ as $n_1 \rightarrow \infty$.

\begin{proposition}
\label{prop:bound}
Assume that the following conditions hold.
\begin{itemize}
\item[(E)] Approximate ergodicity: $\sup_{a \in \mathbb{R}} |\tilde{F}_K(a)-F(a)| \leq \delta_{1K}$ with probability at least $1-\zeta_{1K}$.
\item[(A)] Accuracy of the scores: with probability at least $1-\zeta_{2n_1}$, the two inequalities
\begin{equation*}
\frac{1}{K}\sum_{d \in \mathcal{D}_{\Pi}}\left(R_{\omega_d}-R^*_{\omega_d}\right)^2 \leq \delta_{2n_1}^2 \quad \text{and} \quad \left|R_{T+1}-R^*_{T+1}\right| \leq \delta_{2n_1}
\end{equation*}
hold jointly.
\item[(D)] Bounded density: $R^*_{T+1}$ has a density bounded by a constant $D$.
\end{itemize}
Then, for every $\alpha \in (0,1)$,
\begin{equation}
\begin{split}
\left|\mathbb{P}\left(\boldsymbol{Y}_{T+1} \in \mathcal{C}_{T,1-\alpha}(\boldsymbol{X}_{T+1})\right)-(1-\alpha)\right| &\leq 3\delta_{1K}+\delta_{2n_1}+D\left(\delta_{2n_1}+2\sqrt{\delta_{2n_1}}\right)\\
&\quad+\zeta_{1K}+\zeta_{2n_1}.
\end{split}
\label{eq:bound}
\end{equation}
\end{proposition}

The bound \eqref{eq:bound} holds for each $T$ and, provided that (E) and (A) hold with sequences that vanish, implies asymptotic exactness, i.e.\ marginal coverage converging to $1-\alpha$, when both $n_1 \rightarrow \infty$ and $K \rightarrow \infty$, e.g.\ as $T \rightarrow \infty$ with $n_1/T$, $n_2/T$ and $b$ fixed. For a fixed number of blocks, its right-hand side is bounded away from zero (\ref{app:remarks}). The terms in $\delta_{2n_1}$ and $\zeta_{2n_1}$ are governed by the training set size $n_1$, and those in $\delta_{1K}$ and $\zeta_{1K}$ by the number of blocks $K$ rather than by $n_2$. This indexing is only nominal: $\delta_{2n_1}$ in (A) can also depend on $K$ and on the split, through the average over the scored positions. Proposition~\ref{prop:bound} is the counterpart of Theorem~2 of \citet{chernozhukov2018exact} for the split setting (proof in \ref{app:proof}). Its conditions involve the unobservable oracle scores and cannot be verified from the data without further assumptions. Lemma~\ref{lem:example} gives an example, a correctly specified Gaussian autoregression, in which they hold with explicit rates.

The oracle scores are not the only route to a guarantee for dependent data. With a chronological split and $b=1$, Procedure~1 is split conformal prediction with training, calibration and test points ordered in time, and Theorem~4 of \citet{oliveira2024split} applies if the pairs $\boldsymbol{Z}_t$ are strictly stationary and $\beta$-mixing: for any point predictor, consistent or not, the coverage is at least $1-\alpha-\eta_{\beta}$, where $\eta_{\beta}$ depends on the mixing coefficients and on $n_2$, with no oracle score and no condition of type (A) or (D). Proposition~\ref{prop:bound} adds a two-sided bound for any block length and any split chosen independently of the data, at the cost of the high-level conditions (E), (A) and (D). Neither result comes with primitive conditions for the random split used in Sections~\ref{sec:simulation} and~\ref{sec:application}, apart from the example of Lemma~\ref{lem:example}.

\subsection{Adaptive conformal inference on a rolling window}
\label{sec:adaptive}

In operational use, bands are needed at the time points $t=T+1,\dots,T+H$, with $H \in \mathbb{N}^*$, and, when the band for time $t$ is computed, the responses are observed up to time $t-\tau-1$, where the delay $\tau \in \mathbb{N}=\{0,1,\dots\}$ is fixed ($\tau=0$ in Section~\ref{sec:sim_seq} and $\tau=7$ in Section~\ref{sec:application}). Procedure~1 can then be applied at each $t$ to the window of the $T$ most recent time points with observed response, $t-\tau-T,\dots,t-\tau-1$, with the point predictors refitted on each window and $\boldsymbol{x}_t$ observed when the band is computed (the series starts at time $1-\tau$). Over these time points, the frequency of misses of Procedure~1 can deviate from $\alpha$, in particular under a change in distribution. Procedure~2 addresses this with adaptive conformal inference (ACI) \citep{gibbs2021adaptive}, using the band of Procedure~1 with $b=1$, i.e.\ the split conformal band, as the base method and replacing $\alpha$ with a working miscoverage level $\alpha_t$ updated as the responses are observed.

\paragraph{Procedure~2 (ACI)} Fix $\alpha \in (0,1)$, a step size $\gamma>0$, $\tau$, $T$, $t_0$, $n_1$, $n_2$ and the type of split, and set $b=1$, so that $K=n_2+1$.
\begin{enumerate}
\item \textit{Initialisation.} Set $\alpha_t:=\alpha$ for $t=T+1,\dots,T+\tau+1$, and repeat Steps~2--4 for $t=T+1,\dots,{T+H}$.
\item \textit{Scores.} Apply Steps~1--4 of Procedure~1 with $b=1$ to the observations at times $t-\tau-T,\dots,t-\tau-1$, in the role of times $1,\dots,T$, with $\boldsymbol{x}_t$ in the role of $\boldsymbol{x}_{T+1}$ and a new random split or the chronological split of the window, to obtain the modulation functions $s^{(j)}_t$, the point predictions $\hat{y}^{(j)}_t$ and the $n_2$ calibration scores.
\item \textit{Band at level $\alpha_t$.} If $\alpha_t<1/(n_2+1)$, set $\mathcal{C}_{t-\tau-1,1-\alpha_t}(\boldsymbol{x}_t):=\mathcal{Y}$; if $\alpha_t \geq 1$, set $\mathcal{C}_{t-\tau-1,1-\alpha_t}(\boldsymbol{x}_t):=\emptyset$; otherwise, let $\mathcal{C}_{t-\tau-1,1-\alpha_t}(\boldsymbol{x}_t)$ be the band \eqref{eq:band} with $\hat{y}^{(j)}_{T+1}$, $s^{(j)}$ and $k$ replaced by $\hat{y}^{(j)}_t$, $s^{(j)}_t$ and $k_t$, where $k_t$ is the $\lceil (n_2+1)(1-\alpha_t)\rceil$-th smallest of the $n_2$ calibration scores.
\item \textit{Update.} If $t \geq T+\tau+1$, after observing $\boldsymbol{y}_{t-\tau}$, set
\begin{equation}
\alpha_{t+1}:=\alpha_t+\gamma\left(\alpha-\mathrm{err}_{t-\tau}\right),
\label{eq:aci_update}
\end{equation}
where $\mathrm{err}_s:=\mathbbm{1}\{\boldsymbol{y}_s \notin \mathcal{C}_{s-\tau-1,1-\alpha_s}(\boldsymbol{x}_s)\}$ is the miscoverage indicator of time $s$.
\end{enumerate}

\begin{proposition}
\label{prop:aci}
Let $\alpha \in (0,1)$, $\gamma>0$ and $\tau \in \mathbb{N}$. For every $H \in \mathbb{N}^*$ and every realisation of the time series $\boldsymbol{z}_{1-\tau},\dots,\boldsymbol{z}_{T+H}$ and of any auxiliary randomisation (e.g.\ random splits) used to compute the bands, the miscoverage indicators of Procedure~2 satisfy
\begin{equation}
\left|\frac{1}{H}\sum_{t=T+1}^{T+H}\mathrm{err}_t-\alpha\right| \leq \frac{\max\{\alpha,1-\alpha\}\,(1+\tau\gamma)+(\tau+1)\gamma}{\gamma H},
\label{eq:aci_bound}
\end{equation}
which, for $\tau=0$, is $(\max\{\alpha,1-\alpha\}+\gamma)/(\gamma H)$.
\end{proposition}

Proposition~\ref{prop:aci} is proved in \ref{app:proof_aci}. For $\tau=0$ it is Proposition~4.1 of \citet{gibbs2021adaptive} applied to the functional band; for $\tau>0$ it extends that result to the delayed update of \citet{hallberg2024adaptive} and \citet{wang2024online}.

Proposition~\ref{prop:aci} does not control the size of the bands, which can be infinite or empty (Step~3). The step size $\gamma$ trades adaptivity against stability \citep{zaffran2022adaptive}: a larger $\gamma$ gives a smaller bound \eqref{eq:aci_bound}, informative only when $H$ is large compared with $1/\gamma$ and $\tau$, and a faster reaction to changes in distribution, but a more variable working level and more frequent infinite bands.

\subsection{Weighted nonexchangeable conformal prediction}
\label{sec:weighted}

Under a gradual change in distribution, the calibration scores closest in time to the target time point can be expected to be the most informative. Procedure~3 therefore applies the split version of nonexchangeable conformal prediction \citep{barber2023beyond} to the split conformal band, i.e.\ to Procedure~1 with $b=1$, with fixed weights that decrease with the time distance.

\paragraph{Procedure~3 (WCP)} Fix $\alpha \in (0,1)$ and $\rho \in (0,1]$ before observing the data.
\begin{enumerate}
\item \textit{Scores.} Apply Steps~1--4 of Procedure~1 with $b=1$, which give the calibration scores $R_{\omega_d}$, $d \in \mathcal{D}_{\mathrm{cal}}:=\{1,\dots,n_2\}$.
\item \textit{Weights.} Set $w_d:=\rho^{\,T+1-\omega_d}$, $d \in \mathcal{D}_{\mathrm{cal}}$, where $T+1-\omega_d \geq 1$ is the time distance from the target time point $T+1$, and give the test point weight $1$. Let $W:=\sum_{d \in \mathcal{D}_{\mathrm{cal}}} w_d$, $\tilde{w}_d:=w_d/(W+1)$, $d \in \mathcal{D}_{\mathrm{cal}}$, and $\tilde{w}_{n_2+1}:=1/(W+1)$.
\item \textit{Band.} Let
\begin{equation}
k_{\rho}:=\inf\Big\{r \in \mathbb{R}: \sum_{d \in \mathcal{D}_{\mathrm{cal}}:\, R_{\omega_d} \leq r} \tilde{w}_d \geq 1-\alpha\Big\}, \qquad \inf\emptyset:=+\infty,
\label{eq:wquantile}
\end{equation}
and let the prediction set $\mathcal{C}^{\rho}_{T,1-\alpha}(\boldsymbol{x}_{T+1})$ be the band \eqref{eq:band} with $k$ replaced by $k_{\rho}$ if $k_{\rho}<+\infty$, and $\mathcal{Y}$ otherwise.
\end{enumerate}

For finite calibration scores, the band is the whole space if and only if $W<(1-\alpha)/\alpha$, an event that depends only on the split. For $\rho=1$, Procedure~3 is Procedure~1 with $b=1$ (\ref{app:remarks}). On a rolling window with delay $\tau$ (Section~\ref{sec:adaptive}), the target time point is $T+1+\tau$ in the notation of the window, and $w_d:=\rho^{\,T+1+\tau-\omega_d}$.

Let $\boldsymbol{R}:=(R_{\omega_1},\dots,R_{\omega_{n_2+1}})$ be the vector of the scores, with $R_{\omega_{n_2+1}}=R_{T+1}$ evaluated at $\boldsymbol{Y}_{T+1}$ as in Section~\ref{sec:guarantees}, and, for $d \in \mathcal{D}_{\mathrm{cal}}$, let $\boldsymbol{R}^{(d)}$ be $\boldsymbol{R}$ with the entries $d$ and $n_2+1$ exchanged.

\begin{proposition}
\label{prop:weighted}
Let $\alpha \in (0,1)$ and $\rho \in (0,1]$ be fixed, let the split be independent of the data, and let the scores be real-valued. Then, for any joint distribution of $\boldsymbol{Z}_1,\dots,\boldsymbol{Z}_{T+1}$,
\begin{equation}
\mathbb{P}\left(\boldsymbol{Y}_{T+1} \in \mathcal{C}^{\rho}_{T,1-\alpha}(\boldsymbol{X}_{T+1}) \mid \mathcal{I}_1,\mathcal{I}_2\right) \geq 1-\alpha-\sum_{d \in \mathcal{D}_{\mathrm{cal}}} \tilde{w}_d\, \mathsf{d}_{\mathsf{TV}}\big(\boldsymbol{R},\boldsymbol{R}^{(d)}\big),
\label{eq:weighted_bound}
\end{equation}
where $\mathsf{d}_{\mathsf{TV}}(\boldsymbol{R},\boldsymbol{R}^{(d)})$ is the total variation distance between the conditional distributions of $\boldsymbol{R}$ and $\boldsymbol{R}^{(d)}$ given the split. If, in addition, the $n_2+1$ scores are almost surely distinct, this conditional probability is smaller than $1-\alpha+\tilde{w}_{n_2+1}+\sum_{d \in \mathcal{D}_{\mathrm{cal}}} \tilde{w}_d\, \mathsf{d}_{\mathsf{TV}}(\boldsymbol{R},\boldsymbol{R}^{(d)})$.
\end{proposition}

Proposition~\ref{prop:weighted} is a corollary of the coverage bounds of \citet{barber2023beyond}, in their split version (\ref{app:remarks}). It holds for each band of a rolling-window deployment and, for $\rho=1$, for Procedure~1 with $b=1$, without the conditions of Proposition~\ref{prop:bound}. Under exchangeability of the pairs $\boldsymbol{Z}_t$, the total variation terms are zero and the coverage is at least $1-\alpha$ for every $\rho$. Since the test point keeps the mass $1/(W+1) \geq 1-\rho$ at $+\infty$ for every $n_2$, the band can overcover. Without exchangeability the total variation terms are unknown. Decreasing weights make their weighted sum small if the terms are small for recent calibration scores and grow slowly with the time distance, as under a slow drift of independent observations. Serial dependence can make them positive even without drift (\ref{app:remarks}). A smaller $\rho$ typically gives wider bands.

\FloatBarrier
\section{Simulation study}
\label{sec:simulation}

In Experiment~1 (Section~\ref{sec:sim_static}), Procedures~1 and~3, whose guarantees concern a single band, are applied once to each simulated series, and their marginal coverage is estimated by the fraction of replications in which the band contains $\boldsymbol{y}_{T+1}$. At comparable coverage, a smaller band is more informative \citep{lei2013distribution}. In Experiment~2 (Section~\ref{sec:sim_seq}), the three procedures, including Procedure~2, whose guarantee concerns a sequence of bands, are deployed on a rolling window, with stationary data and with a change in distribution. The point predictors include misspecified models, for which condition (A) of Proposition~\ref{prop:bound} fails, in general, for the oracle score defined through $\mu^{(j)}$ (Section~\ref{sec:guarantees}).

\subsection{Design}

For $p=1$, the response is $Y_t^{(1)}(q)=\boldsymbol{g}(q)^{\top}\boldsymbol{\xi}_t$, $q \in \mathcal{Q}_1=[0,1]$, where $\boldsymbol{g}(q)=(1,\sqrt{2}\sin(2\pi q),\sqrt{2}\cos(2\pi q))^{\top}$ and $\boldsymbol{\xi}_t \in \mathbb{R}^3$ follows a stable VAR(2) process driven by i.i.d.\ multivariate Student's $t$ innovations with $4$ degrees of freedom. For $p=2$, a VAR(2) process in $\mathbb{R}^6$ generates two dependent components with domains $\mathcal{Q}_1=[0,1]$ and $\mathcal{Q}_2=[0,2]$, and different scales (\ref{app:sim_design}).

The point predictors (\ref{app:sim_design}) use the $r$ most recent observations as covariates: the \textit{Oracle}, i.e.\ the regression function; \textit{VAR($r$)}, $r \in \{1,2,3\}$, a VAR model fitted to $\boldsymbol{\xi}_t$, with known basis functions \citep{chen2021review}; and \textit{FAR($r$)}, $r \in \{1,2,3\}$, the concurrent functional autoregressive model \citep{ramsay_functional_2005}. VAR(2) is correctly specified. VAR(1) omits a relevant lagged variable, VAR(3) includes an irrelevant one, and the FAR models omit relevant regressors \citep{rao1971some}. The Oracle has coverage exactly $0.75$ (Section~\ref{sec:guarantees}), which checks the implementation, and gives a reference band size.

We use the random split of Section~\ref{sec:procedure}, $\alpha=0.25$ (nominal coverage level $0.75$), $T \in \{25,50,100,1000\}$, $b \in \{1,3,6\}$ for Procedure~1, and $b=1$ and $\rho \in \{0.99,0.98,0.95\}$ for Procedure~3, on the same series, splits and point predictors, and choose $n_2$ so that $K$ and $\alpha K$ are integers and $\alpha \geq 1/K$ (\ref{app:sim_design}). For each design $(p,T,n_2)$, we generate $N=5000$ independent series of length $T+1$. The Monte Carlo standard error (SE) of the empirical coverage $c$ is $\{c(1-c)/N\}^{1/2}$, and we flag the cases in which $0.75$ lies outside the $99\%$ Monte Carlo interval $c \pm 2.576$ SE. The band size is summarised by its median and interquartile range (IQR).

\subsection{Experiment 1: a single band}
\label{sec:sim_static}

The coverage is between $0.729$ and $0.765$ in all the $139$ combinations of design, $b$ and point predictor (Tables~\ref{tab:cov_p1} and~\ref{tab:cov_p2} in \ref{app:sim_static}), also where condition (A) is not met (misspecified point predictors) or is questionable (short training sets). None of the $20$ Oracle cells is flagged, and $3$ of the $119$ cells of the estimated point predictors are, against about $1.2$ expected by chance (\ref{app:sim_static}).

\begin{figure}[tbp]
\centering
\includegraphics[width=0.8\textwidth]{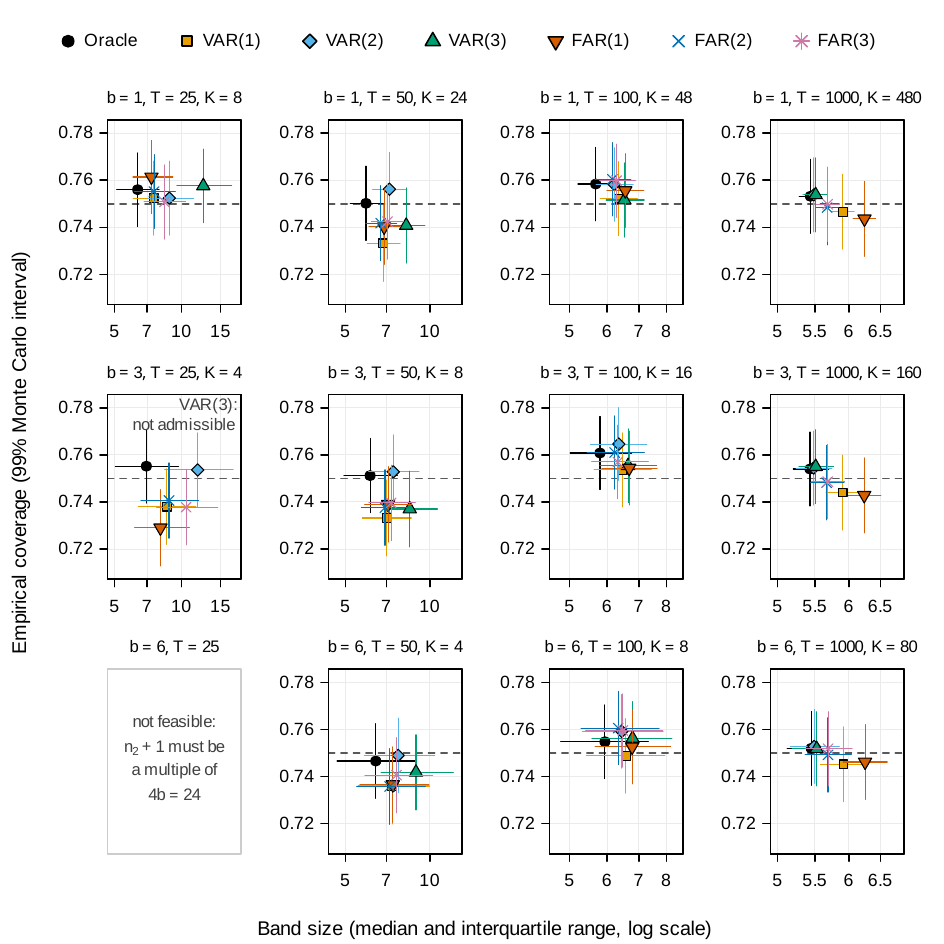}\\[2pt]
\includegraphics[width=0.92\textwidth]{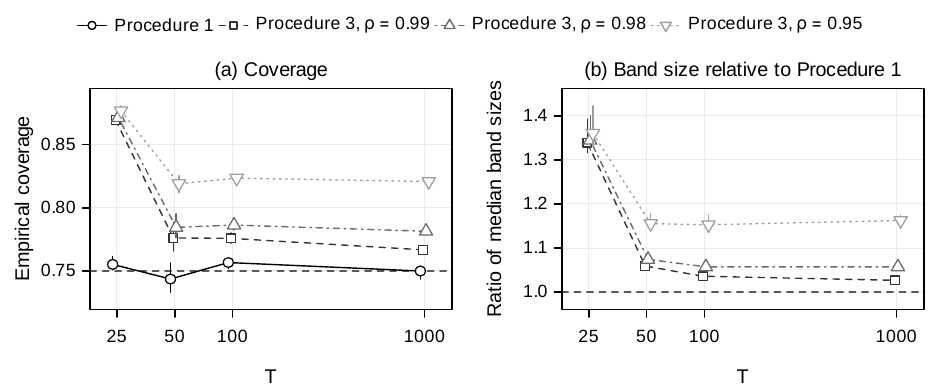}
\caption{Experiment~1, $p=1$. Top: empirical coverage and median band size (logarithmic scale) of Procedure~1; one panel for each $(b,T)$ and one marker for each point predictor; segments: IQR of the band size and $99\%$ Monte Carlo interval of the coverage. Bottom: Procedures~1 and~3 with $b=1$, against $T$; (a) empirical coverage; (b) median size of the finite bands of Procedure~3 divided by that of Procedure~1; markers: mean over the point predictors; segments: their range. Dashed lines: nominal coverage $0.75$ and ratio $1$.}
\label{fig:sim_cov_size_p1}
\end{figure}

For given $T$, the IQR of the band size increases with $b$, since fewer blocks yield more variable values of $k$, and so does the median for $T \in \{50,100\}$ (Figure~\ref{fig:sim_cov_size_p1}; Tables~\ref{tab:size_p1} and~\ref{tab:size_p2} in \ref{app:sim_static}). For $T=1000$, the median size exceeds that of the Oracle by $0.8\%$ to $1.6\%$ for VAR(2) and by $10.5\%$ to $14.9\%$ for FAR(1), for both values of $p$: with a long training set, misspecification affects the size of the bands more than their coverage.

Procedure~3 overcovers for every $T$ (Figure~\ref{fig:sim_cov_size_p1}, bottom; Table~\ref{tab:static_wcp}). Since $W<\rho/(1-\rho)$, the mass $1/(W+1)$ of the test point exceeds $1-\rho$. With $T=1000$, its mean is $0.021$, $0.041$ and $0.101$ for $\rho=0.99$, $0.98$ and $0.95$, and the coverage exceeds that of Procedure~1 with $b=1$ by $0.011$ to $0.021$, $0.023$ to $0.035$ and $0.061$ to $0.076$ (SE at most $0.005$). With $T=25$ ($n_2=7$), the quantile of Procedure~3 is a higher order statistic than that of Procedure~1 in $92\%$ to $97\%$ of the replications, and the coverage is $0.868$ to $0.881$ for the three values of $\rho$. The effective number of calibration scores, $(\sum_d w_d)^2/\sum_d w_d^2$, grows little with $n_2$: with $\rho=0.95$, its mean is between $15.9$ and $19.0$ for $T \geq 50$, against the $n_2=479$ calibration scores of Procedure~1 with $T=1000$. The median band sizes of Procedure~3 are $1.02$ to $1.42$ times those of Procedure~1, but at a different coverage they are not comparable.

\subsection{Experiment 2: sequential deployment}
\label{sec:sim_seq}

We use $\alpha=0.25$, $T=100$ and $n_2=47$, a setting close to that of Section~\ref{sec:application}. Procedure~1 uses $b \in \{1,3\}$ ($K=48$, $16$) for $p=1$ and $b=1$ for $p=2$, and Procedures~2 and~3 use $b=1$ ($K=48$). For each of $M=500$ independent trajectories and each test time $t=T+h$, $h=1,\dots,H=500$, every point predictor is refitted on the window of the $T$ most recent observations, with a new random split, and the three procedures use the same split and fitted point predictors. Procedure~2 uses $\gamma \in \{0.005,0.02,0.1\}$, for which the bound \eqref{eq:aci_bound} equals $0.302$, $0.077$ and $0.017$, and Procedure~3 uses $\rho \in \{0.99,0.98,0.95\}$. In scenario S1 the data are stationary. In scenario S2 the innovations are multiplied by $2$ from $h=251$ onwards, with unchanged dynamics, and the whole window follows the shift for $h>350$. We report the coverage over all $MH=250{,}000$ pairs (trajectory, $h$) and the long-run coverage $H^{-1}\sum_{t=T+1}^{T+H}(1-\mathrm{err}_t)$ of each trajectory, controlled by Proposition~\ref{prop:aci} for Procedure~2, with its standard deviation (SD) across trajectories (Tables~\ref{tab:seq}--\ref{tab:seq_b3} in \ref{app:sim_seq}).

\begin{figure}[tbp]
\centering
\includegraphics[width=0.97\textwidth]{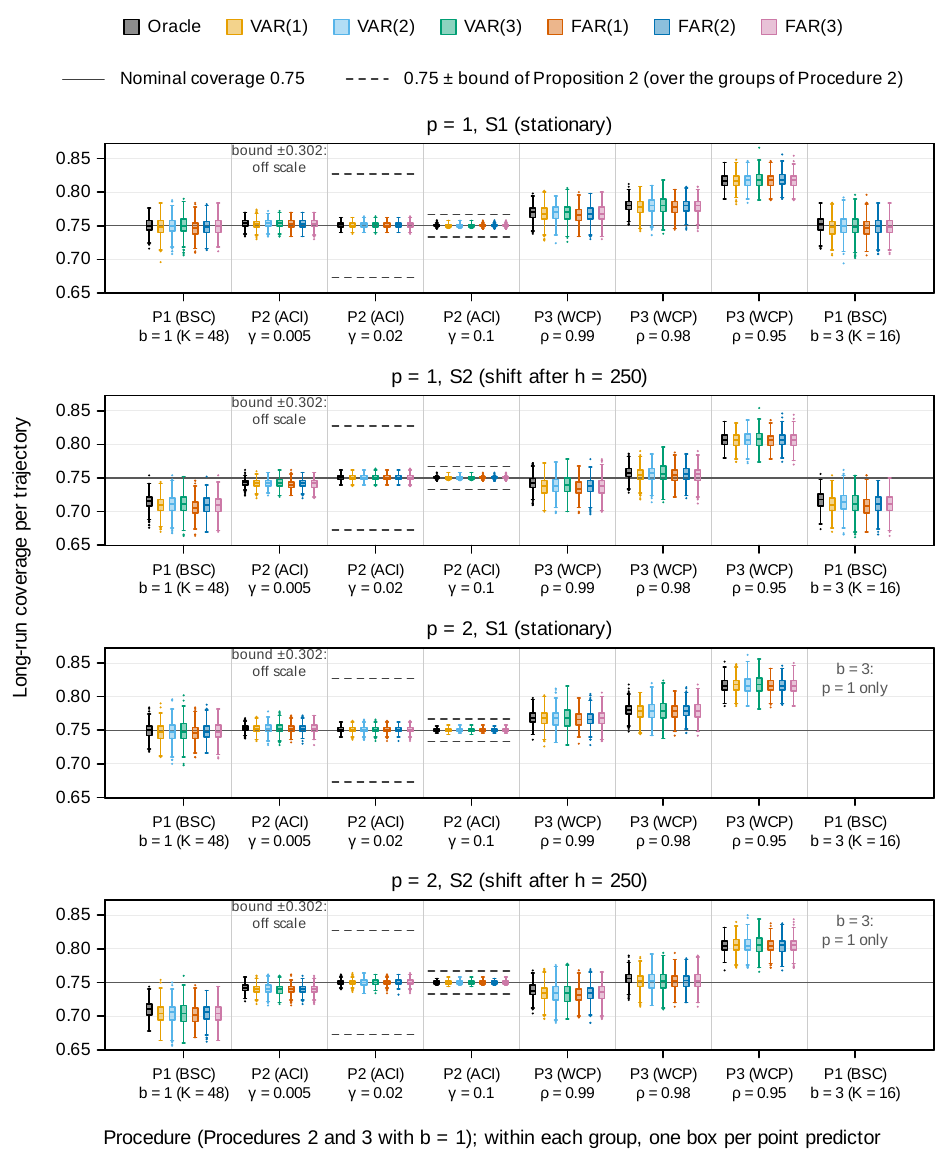}
\caption{Experiment~2: boxplots of the long-run coverage of the $M=500$ trajectories, one box per point predictor, by procedure (groups), $p$ and scenario (rows). Horizontal lines: nominal coverage $0.75$, and $0.75 \pm 0.077$ and $0.75 \pm 0.017$, the bounds of Proposition~\ref{prop:aci} for $\gamma=0.02$ and $0.1$.}
\label{fig:sim_seq_traj}
\end{figure}

The bound of Proposition~\ref{prop:aci} holds in all the $42{,}000$ trajectories of Procedure~2 (Figure~\ref{fig:sim_seq_traj}). In S1, Procedure~1 has coverage $0.746$--$0.751$ and an SD of $0.010$--$0.016$. Procedure~2 has coverage $0.750$--$0.754$ and an SD of $0.006$--$0.008$, $0.004$--$0.005$ and $0.002$ for $\gamma=0.005$, $0.02$ and $0.1$. For $\gamma \in \{0.005,0.02\}$, its median band size exceeds that of Procedure~1 by $0.4\%$--$1.0\%$, with coverage higher by at most $0.006$. For $\gamma=0.1$, $0.18\%$--$0.57\%$ of its bands are infinite, so the sizes are not comparable. Procedure~3 overcovers: for $\rho=0.99$, $0.98$ and $0.95$, its coverage is $0.766$--$0.770$, $0.777$--$0.780$ and $0.816$--$0.819$ (\ref{app:sim_seq}).

In S2, for $p=1$ and $b=1$, the coverage of Procedure~1 is between $0.666$ and $0.679$ after the shift. In the transition ($251 \leq h \leq 350$), the loss of coverage with respect to S1 is $0.177$--$0.199$ for Procedure~1, and $0.113$--$0.125$, $0.033$--$0.036$ and $0.002$--$0.003$ for Procedure~2 with $\gamma=0.005$, $0.02$ and $0.1$. Procedure~2 compensates these misses by overcovering afterwards, with bands wider than needed (\ref{app:sim_seq}). With $\gamma=0.1$, $0.9\%$--$1.2\%$ of its bands are infinite, against at most $5$ of the $250{,}000$ with $\gamma=0.02$. For Procedure~3, the loss is $0.138$--$0.156$, $0.108$--$0.121$ and $0.052$--$0.058$ for $\rho=0.99$, $0.98$ and $0.95$. With a fixed level, Procedure~3 does not compensate the misses: for $h>350$, its coverage is as in S1. The results for $b=3$ and for $p=2$ are similar (\ref{app:sim_seq}).

\paragraph{Practical guidance} In Experiment~2, $\gamma=0.02$ kept the long-run coverage within $0.75 \pm 0.017$ in at least $99.8\%$ of the trajectories, with almost no infinite bands. For the weakly dependent processes considered, $b=3$ changed the coverage of Procedure~1 at most slightly, as in \ref{app:sim_static}, and, by reducing $K$, increased the median size of the bands and the SD of the long-run coverage. A block length $b>1$ may be worth considering when the dependence among the scores is expected to be stronger. In S1, for each procedure, the misspecified VAR(1) and FAR models had coverage within $0.003$ of that of VAR(2) and median size $0.99$ to $1.06$ ($p=1$) and $0.90$ to $0.95$ ($p=2$) times that of VAR(2). For $p=1$, FAR(1) lost up to $0.010$ more coverage than VAR(2) in the transition (\ref{app:sim_seq}). Given $\alpha$, $n_2$ should make $\alpha K$ an integer and $K$ not too small, and $n_1$ should be large relative to the number of parameters of the point predictor, mainly to limit the size of the bands (Table~\ref{tab:size_p1}; in Section~\ref{sec:application}, $n_1=43$ and $K=40$).

\section{Application to the Italian gas market}
\label{sec:application}

\subsection{Market, data and curves}
\label{sec:app_data}

In the MGS, the pipeline manager, Snam S.p.A., and the participants admitted by the market operator, Gestore dei Mercati Energetici (GME), submit each day offers and bids for stored gas to an auction. The offer curve $y_t^{(1)}$ of day $t$ gives, for each quantity $q$ (MWh), the price (Euro/MWh) of the offer at which the cumulative quantity of the offers, sorted by increasing price, reaches $q$. The demand curve $y_t^{(2)}$ is defined in the same way from the bids, sorted by decreasing price. Both are positive, left-continuous step functions, non-decreasing and non-increasing, respectively, evaluated on a grid without smoothing or basis expansion. The clearing point $(Q_t,P_t)$, i.e.\ the quantity exchanged and the clearing price, lies at the intersection of the two curves, completed with vertical segments at their jumps. The offers and bids to its left are accepted at price $P_t$. Here $\boldsymbol{Y}_t=(Y_t^{(1)},Y_t^{(2)})$ and $p=2$.

\begin{figure}[tbp]
\centering
\includegraphics[width=12cm,height=8cm]{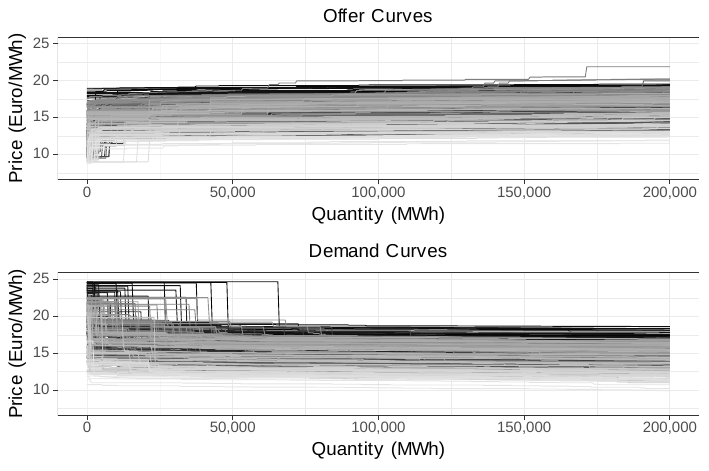}
\caption{Offer (top) and demand (bottom) curves of the 281 days from 26 April 2019 to 31 January 2020; older curves are darker.}
\label{fig:gasdata}
\end{figure}

We predict the curves of the 184 days between 1 August 2019 and 31 January 2020 (the evaluation period). The domains are $\mathcal{Q}_1=\mathcal{Q}_2=[0,2\cdot 10^5]$ MWh, on which the curves to be predicted and those of the 97 preceding days (Figure~\ref{fig:gasdata}) are defined without extension, since the total quantities offered and demanded exceed $2\cdot 10^5$ MWh on each of these days. On every day of the evaluation period, $Q_t$ lies in this interval (at most 116{,}734 MWh). All computations use the grid $\mathcal{G}=\{g_1,\dots,g_G\}$ of $G=401$ equispaced points with spacing 500 MWh, $g_1=0$ and $g_G=2\cdot 10^5$ (\ref{app:remarks}).

\subsection{Point predictor and implementation}
\label{sec:app_setup}

Due to the rules of the MGS, at the auction for day $t$ the most recent curves available to a participant are those of day $t-8$, and the most recent clearing price is $P_{t-2}$. We therefore use, for illustration, the simple concurrent functional autoregressive model
\begin{equation}
Y_t^{(j)}(q)=\beta_1^{(j)}(q)\,Y_{t-8}^{(j)}(q)+\beta_2^{(j)}(q)\,P_{t-2}+\varepsilon_t^{(j)}(q), \qquad q \in \mathcal{Q}_j,\ j=1,2,
\label{eq:gas_model}
\end{equation}
where $\varepsilon_t^{(j)}$ is a mean-zero, finite-variance error process, uncorrelated with the regressors, so that $X_t^{(j)}=(Y_{t-8}^{(j)},\allowbreak P_{t-2})$. The model is fitted on the training set by ordinary least squares at each grid point, separately for the two curves. Its predictions $\tilde{y}_t^{(j)}(q):=\hat{\beta}_1^{(j)}(q)\,y_{t-8}^{(j)}(q)+\hat{\beta}_2^{(j)}(q)\,P_{t-2}$, $q \in \mathcal{G}$, need not be monotone. The point predictions $\hat{y}_t^{(j)}(q):=[\hat{\mu}^{(j)}_{\mathcal{I}_1}(x_t^{(j)})](q)$ replace the parts of $\tilde{y}_t^{(j)}$ that violate monotonicity with a linear interpolation (\ref{app:app_extra}).

Sections~\ref{sec:app_setup}--\ref{sec:app_whatif} use Procedure~1 on a rolling window with delay $\tau=7$ (Section~\ref{sec:adaptive}): the band for day $t$, denoted $\mathcal{C}_{t-8,1-\alpha}(\boldsymbol{X}_t)$, is computed from the $T=90$ days $t-97,\dots,t-8$, of which the first $t_0=8$ serve only as lagged covariates; the training set consists of $n_1=43$ of the other 82 days, drawn at random anew for each $t$, and the calibration set of the remaining $n_2=39$ days (random split, Section~\ref{sec:procedure}). We set $b=1$, hence $K=40$, as suggested by the simulation study (Section~\ref{sec:sim_seq}). The region of Section~\ref{sec:app_region} and the what-if analysis of Section~\ref{sec:app_whatif}, illustrated with Procedure~1, can be computed in the same way from the bands of Procedures~2 and~3, which Section~\ref{sec:app_procedures} compares with Procedure~1.

We consider $\alpha=0.5$ and $\alpha=0.25$, for which $\alpha K$ is an integer (Section~\ref{sec:guarantees}) and $k_t$ is the 20th and the 30th smallest of the 39 calibration scores. Smaller values of $\alpha$ require a larger $K$ (Section~\ref{sec:procedure}; \ref{app:app_extra}).

\subsection{Prediction bands}
\label{sec:app_bands}

\begin{figure}[tbp]
\centering
\includegraphics[width=12cm,height=8cm]{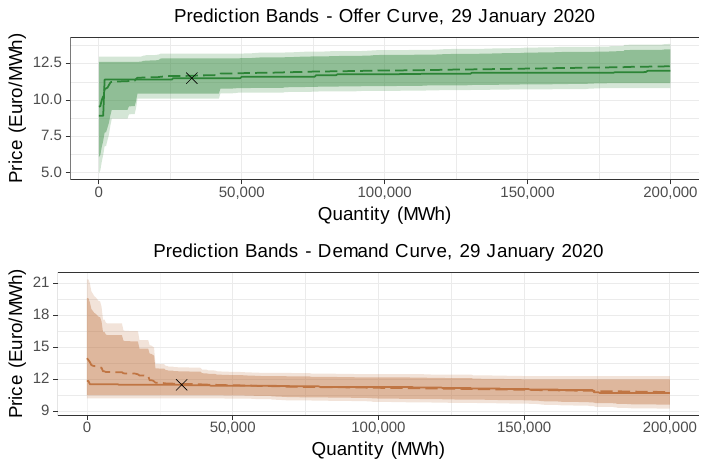}
\caption{Band of Procedure~1 on 29 January 2020, with $\alpha=0.5$ (darker) and $\alpha=0.25$ (lighter); observed curves (solid), point predictions (dashed) and observed $(Q_t,P_t)$ (cross).}
\label{fig:gaspredset}
\end{figure}

Since $\boldsymbol{Y}_t$ belongs almost surely to the set $\mathcal{M}$ of pairs $(y^{(1)},y^{(2)})$ with $y^{(1)}$ non-decreasing and $y^{(2)}$ non-increasing, whereas the bounds $\hat{y}_t^{(j)} \mp k_t\,s_t^{(j)}$ need not be monotone, the band is intersected with $\mathcal{M}$, which leaves its coverage unchanged. The set $\mathcal{C}_{t-8,1-\alpha}(\boldsymbol{X}_t) \cap \mathcal{M}$ consists of the pairs in $\mathcal{M}$ that lie, at all $g \in \mathcal{G}$, between the monotone bounds
\begin{align*}
\ell_t^{(1)}(g)&:=\max_{v \in \mathcal{G},\, v \leq g}\left\{\hat{y}_t^{(1)}(v)-k_t\,s_t^{(1)}(v)\right\},\\
u_t^{(1)}(g)&:=\min_{v \in \mathcal{G},\, v \geq g}\left\{\hat{y}_t^{(1)}(v)+k_t\,s_t^{(1)}(v)\right\},\\
\ell_t^{(2)}(g)&:=\max_{v \in \mathcal{G},\, v \geq g}\left\{\hat{y}_t^{(2)}(v)-k_t\,s_t^{(2)}(v)\right\},\\
u_t^{(2)}(g)&:=\min_{v \in \mathcal{G},\, v \leq g}\left\{\hat{y}_t^{(2)}(v)+k_t\,s_t^{(2)}(v)\right\}.
\end{align*}
Figure~\ref{fig:gaspredset} shows these bounds for 29 January 2020.

At each grid quantity, the band gives a range of prices for each curve, and the two curves lie in these ranges jointly with the coverage of the band. Hence, statements that hold whenever both curves lie in the band have probability at least this coverage, unlike those derived from intervals computed separately for each $q$.

For $T=90$, the observed curves lie in the band on 52.7\% of the days for $\alpha=0.5$ and on 75.5\% for $\alpha=0.25$. The window length $T=90$ was selected ex post among $T \in \{45,\allowbreak 60,\allowbreak 90,\allowbreak 180,\allowbreak 365\}$ as the value giving, for $\alpha=0.5$, the smallest median over the days of the width of the band (difference between the monotone bounds, averaged over the grid and the two curves; Table~\ref{tab:gas_T} in \ref{app:app_extra}).

The increase of the frequencies with $T$ (Table~\ref{tab:gas_T}) and the concentration of the misses in September and October 2019 (Section~\ref{sec:app_procedures}) suggest that the distribution of the curves changed over the period.

\subsection{Prediction region for the clearing point}
\label{sec:app_region}

\begin{figure}[tbp]
\centering
\includegraphics[width=12cm,height=6cm]{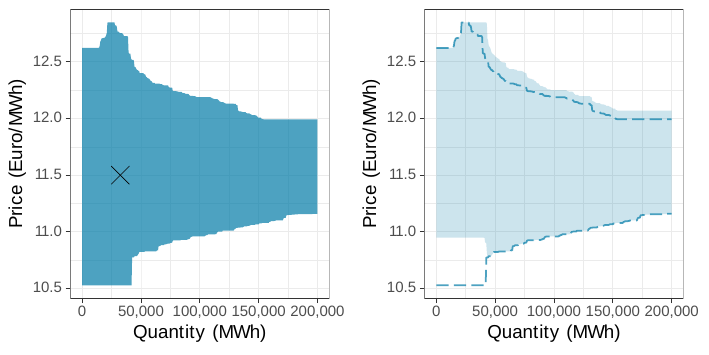}
\caption{Left: region $\mathcal{R}_t$ of Procedure~1 on 29 January 2020 ($\alpha=0.5$) and observed $(Q_t,P_t)$ (cross). Right: the same region (dashed boundary) and the region with an additional demand bid of 20{,}000 MWh at 12 Euro/MWh (lighter).}
\label{fig:extra_of}
\end{figure}

The bands induce a prediction region for $(Q_t,P_t)$: the set of points of the quantity--price plane at which the bands of the two curves overlap. For $q$ in a cell $[g_i,g_{i+1})$, $i=1,\dots,G-1$, let $J_t^{(1)}(q):=[\ell_t^{(1)}(g_i),u_t^{(1)}(g_{i+1})]$ and $J_t^{(2)}(q):=[\ell_t^{(2)}(g_{i+1}),u_t^{(2)}(g_i)]$, and define
\begin{equation*}
\mathcal{R}_t:=\left\{(q,c) \in [0,2\cdot 10^5) \times \mathbb{R}:\ c \in J_t^{(1)}(q) \cap J_t^{(2)}(q)\right\}
\end{equation*}
(left panel of Figure~\ref{fig:extra_of}).

Assume that the observed curves belong to the band, and hence, as they belong to $\mathcal{M}$, lie between the monotone bounds at all grid points, and that they intersect below $2\cdot 10^5$, with $g_i \leq Q_t<g_{i+1}$. With $y_t^{(j)}(q+)$ the right limit of the $j$-th curve at $q$,
\begin{align*}
&\ell_t^{(1)}(g_i) \leq y_t^{(1)}(g_i) \leq y_t^{(1)}(Q_t) \leq P_t \leq y_t^{(1)}(Q_t+) \leq y_t^{(1)}(g_{i+1}) \leq u_t^{(1)}(g_{i+1}),\\
&\ell_t^{(2)}(g_{i+1}) \leq y_t^{(2)}(g_{i+1}) \leq y_t^{(2)}(Q_t+) \leq P_t \leq y_t^{(2)}(Q_t) \leq y_t^{(2)}(g_i) \leq u_t^{(2)}(g_i),
\end{align*}
i.e.\ $(Q_t,P_t) \in \mathcal{R}_t$: the outer inequalities hold because the curves lie between the bounds, those involving $P_t$ because the clearing point lies on both completed curves, which are left-continuous, and the others by monotonicity. Hence, if the curves intersect below $2\cdot 10^5$ with probability one, $\mathbb{P}((Q_t,P_t) \in \mathcal{R}_t) \geq \mathbb{P}(\boldsymbol{Y}_t \in \mathcal{C}_{t-8,1-\alpha}(\boldsymbol{X}_t))$.

The observed clearing point lies in $\mathcal{R}_t$ on 92.4\% of the days for $\alpha=0.5$ and 97.8\% for $\alpha=0.25$, more often than the curves lie in the band (Section~\ref{sec:app_bands}).

\subsection{What-if analysis}
\label{sec:app_whatif}

A demand bid of quantity $\Delta$ at price $c$ is inserted in a demand curve as in Section~\ref{sec:app_data}: the curve is unchanged where its price is above $c$, equals $c$ on the next interval of length $\Delta$, and is shifted to the right by $\Delta$ beyond it; offers are inserted in the same way. The other participants' offers and bids are assumed to be unaffected (ceteris paribus).

In Figure~\ref{fig:extra_of} (right), the bid is inserted in the point prediction $\hat{y}_t^{(2)}$, and the band is recentred on the modified prediction with the same $k_t$ and $s_t^{(j)}$: a scenario analysis, to which the coverage guarantees do not extend in general. Under this assumption, inserting the bid in the bounds $\ell_t^{(2)}$ and $u_t^{(2)}$ instead gives a region whose coverage, under the same intersection condition, is at least that of the band (\ref{app:app_extra}). With the modified bounds, a demand bid priced above (below) all the prices in the modified region is accepted (rejected) whenever the clearing point lies in the region, i.e.\ with probability at least the coverage of the band.

\subsection{Daily deployment of the three procedures}
\label{sec:app_procedures}

\begin{figure}[tbp]
\centering
\includegraphics[width=\linewidth,height=0.72\textheight,keepaspectratio]{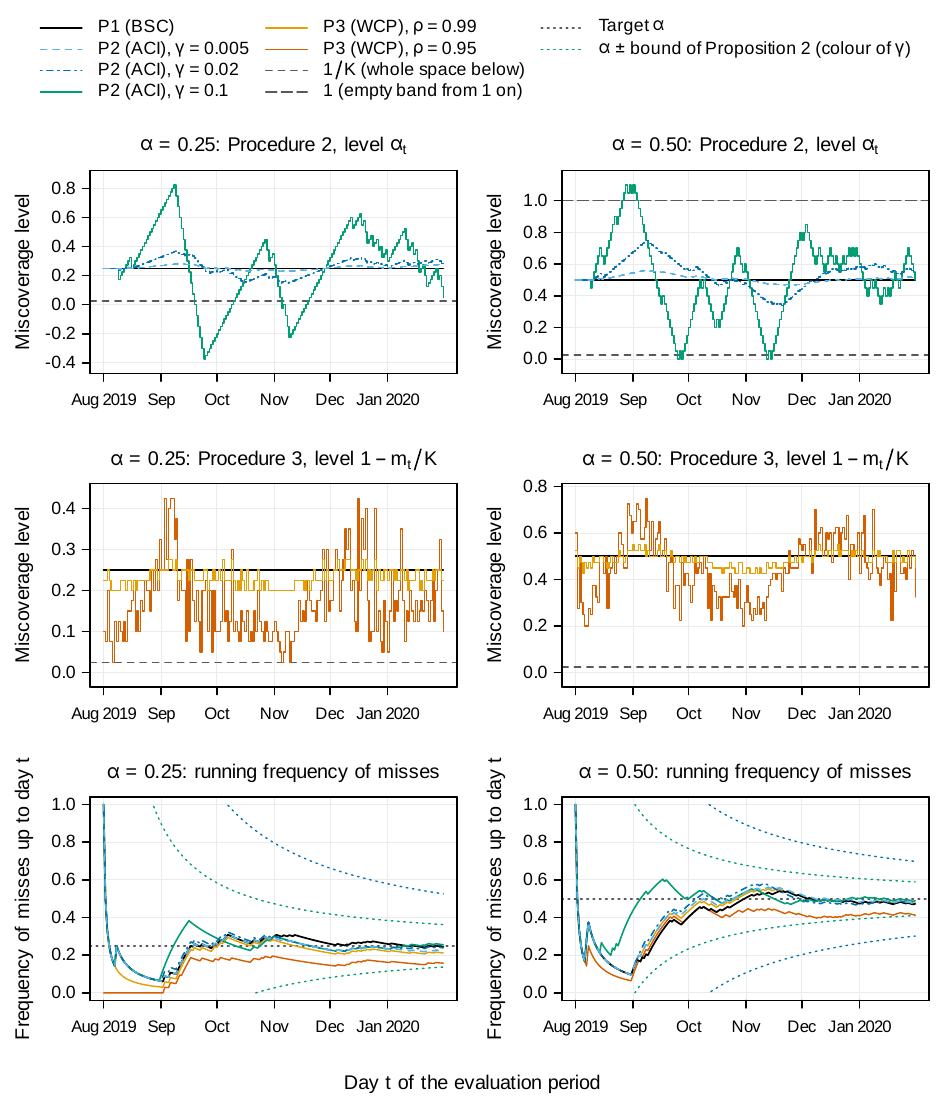}
\caption{Procedures~1, 2 and~3 on the 184 days, for $\alpha=0.25$ (left) and $\alpha=0.5$ (right). Top: working miscoverage level $\alpha_t$ of Procedure~2 ($\gamma=0.005$, $0.02$ and $0.1$; $\tau=7$), with $\alpha$ (Procedure~1), $1/K$ (whole space below) and $1$ (empty band from $1$ on). Middle: miscoverage level $1-m_t/K$ at which Procedure~1 gives the band of Procedure~3 ($\rho=0.99$ and $0.95$; $m_t$: position of $k_t$ among the sorted calibration scores), with $\alpha$ and $1/K$. Bottom: frequency of misses up to day $t$, with $\alpha$ and $\alpha \mp$ the bound~\eqref{eq:aci_bound} for $H$ the number of days up to $t$.}
\label{fig:gas_procedures}
\end{figure}

Procedures~2 and~3 are run on the same days as Procedure~1, with the same windows, splits and point predictor: Procedure~2 with $\gamma \in \{0.005,\allowbreak 0.02,\allowbreak 0.1\}$ and delay $\tau=7$; Procedure~3 with $\rho \in \{0.99,0.98,0.95\}$, as in Section~\ref{sec:sim_seq}. The deviations of the frequency of misses from $\alpha$ are not always smaller for Procedure~2 than for Procedure~1, and, for both, far below the bound~\eqref{eq:aci_bound} for $H=184$ (Table~\ref{tab:gas_procedures}). The misses are concentrated in September and October 2019, less so for Procedure~2 with $\gamma=0.1$ (Table~\ref{tab:gas_monthly}). Except with $\gamma=0.1$ and $\alpha=0.5$, Procedure~2 does not reduce the misses of September (Figure~\ref{fig:gas_procedures}; \ref{app:app_extra}). With $\gamma=0.1$, the band is the whole space or empty on some days, so that the widths of its finite, non-empty bands are not comparable with those of the other procedures.

With $\rho=0.99$, the curves lie in the band of Procedure~3 on 78.8\% and 52.2\% of the days ($\alpha=0.25$ and $0.5$). For $\alpha=0.5$, at a frequency similar to that of Procedure~1, its bands are slightly wider (Table~\ref{tab:gas_procedures}). With $\rho=0.95$, Procedure~3 overcovers, as in Section~\ref{sec:sim_seq}: the curves lie in the band on 84.2\% and 58.7\% of the days (Figure~\ref{fig:gas_procedures}).

At a frequency of coverage similar to that of Procedure~1, the width of the band of Procedure~2 with $\gamma=0.005$ and $\alpha=0.5$ is close to that of Procedure~1, and with $\gamma=0.02$ has a similar or smaller median and a larger IQR. With both step sizes, the clearing point lies in $\mathcal{R}_t$ on fewer days (Table~\ref{tab:gas_procedures}). With $\rho=0.95$, at a different frequency of coverage, the widths of Procedure~3 are not comparable with those of Procedure~1.

\begin{table}[tbp]
\centering
\caption{Procedures~1, 2 and~3 (P1, P2, P3) on the 184 days (P3 with weights $\rho^{\delta}$, $\delta$: number of days to the predicted day; $\rho=0.98$ in Table~\ref{tab:gas_procedures_full}). Frequency: percentage of days with the observed curves in the band and with $(Q_t,P_t)$ in $\mathcal{R}_t$; Width: median, IQR and mean over the days of the width of the finite, non-empty bands (Euro/MWh; Section~\ref{sec:app_bands}); Bands: number of bands equal to the whole space (Inf.) and empty (Empty); Dev.: $|H^{-1}\sum_t \mathrm{err}_t-\alpha|$; Bound: right-hand side of~\eqref{eq:aci_bound}, $H=184$ (--: not applicable).}
\label{tab:gas_procedures}
{\footnotesize\setlength{\tabcolsep}{4pt}
\begin{tabular}{lrrrrrrrrr}
\toprule
 & \multicolumn{2}{c}{Frequency (\%)} & \multicolumn{3}{c}{Width} & \multicolumn{2}{c}{Bands} & & \\
\cmidrule(lr){2-3}\cmidrule(lr){4-6}\cmidrule(lr){7-8}
 & Curves & $(Q_t,P_t)$ & Median & IQR & Mean & Inf. & Empty & Dev. & Bound \\
\midrule
\multicolumn{10}{l}{$\alpha=0.25$ (nominal level 0.75)} \\
P1 & 75.5 & 97.8 & 4.51 & 1.36 & 4.60 & 0 & 0 & 0.005 & -- \\
P2, $\gamma{=}0.005$ & 77.7 & 97.3 & 4.55 & 1.49 & 4.67 & 0 & 0 & 0.027 & 0.887 \\
P2, $\gamma{=}0.02$ & 76.1 & 97.3 & 4.50 & 1.92 & 4.74 & 0 & 0 & 0.011 & 0.276 \\
P2, $\gamma{=}0.1$ & 74.5 & 94.6 & 3.71 & 1.50 & 4.15 & 35 & 0 & 0.005 & 0.113 \\
P3, $\rho{=}0.99$ & 78.8 & 97.8 & 4.68 & 1.44 & 4.80 & 0 & 0 & 0.038 & -- \\
P3, $\rho{=}0.95$ & 84.2 & 97.3 & 5.82 & 2.57 & 5.90 & 0 & 0 & 0.092 & -- \\
\addlinespace
\multicolumn{10}{l}{$\alpha=0.5$ (nominal level 0.50)} \\
P1 & 52.7 & 92.4 & 3.16 & 0.65 & 3.20 & 0 & 0 & 0.027 & -- \\
P2, $\gamma{=}0.005$ & 52.2 & 91.3 & 3.14 & 0.72 & 3.21 & 0 & 0 & 0.022 & 0.606 \\
P2, $\gamma{=}0.02$ & 52.2 & 89.7 & 3.04 & 0.96 & 3.18 & 0 & 0 & 0.022 & 0.198 \\
P2, $\gamma{=}0.1$ & 51.1 & 85.9 & 3.07 & 1.10 & 3.61 & 4 & 9 & 0.011 & 0.090 \\
P3, $\rho{=}0.99$ & 52.2 & 92.4 & 3.20 & 0.75 & 3.27 & 0 & 0 & 0.022 & -- \\
P3, $\rho{=}0.95$ & 58.7 & 91.8 & 3.34 & 1.26 & 3.53 & 0 & 0 & 0.087 & -- \\
\bottomrule
\end{tabular}
}
\end{table}

\FloatBarrier
\section{Discussion and conclusions}
\label{sec:conclusion}

We have compared three procedures that turn any point predictor of the daily supply and demand curves of an auction market into simultaneous prediction bands. All three compute the same split conformal band and differ only in how they calibrate it.

Under stationarity, split conformal prediction (Procedure~1 with $b=1$) has coverage between $0.733$ and $0.761$ in all the designs of Experiment~1, also with misspecified point predictors. Larger blocks change the coverage little and make the band size more variable. Procedure~2 reduces the variability of the long-run coverage across trajectories, at the price of some infinite bands with $\gamma=0.1$. Procedure~3 overcovers for every length $T$ of the series, because the mass $1/(W+1)$ of the test point exceeds $1-\rho$, and its bands widen as $\rho$ decreases. After an abrupt change, Procedure~1 loses coverage until the window follows the change. Procedure~2 loses less, and less still for larger $\gamma$; with $\gamma \in \{0.005,0.02\}$ it then overcovers, with wider bands. Procedure~3 loses less than Procedure~1 but does not compensate the misses. On the gas market data, the frequencies of misses of Procedures~1 and~2 deviate from $\alpha$ by at most $5/184 \approx 0.027$. Procedure~3 overcovers with $\rho=0.95$, with wider bands, and deviates from $\alpha$ by up to $7/184 \approx 0.038$ with $\rho=0.99$.

For daily use, these results suggest Procedure~1 with $b=1$ as the default when the market is approximately stationary over the window: it gives a statement on the coverage probability of each day's band and uses all the calibration scores. Procedure~2 suits the repeated use of the bands over a long period, when the fraction of days on which the curves leave the band should be close to $\alpha$ whatever the data, including after abrupt changes. A delay in the feedback slows its reaction to misses. Procedure~3 is intended for gradual changes in distribution. This use rests on the form of Proposition~\ref{prop:weighted} alone, since our simulations consider only an abrupt change. Whether its band is the whole space depends only on the split, so it can be checked before deployment. With any of the three procedures, the bands give a prediction region for the clearing point and a what-if analysis of an additional bid or offer.

The analysis has limitations. First, the bounds \eqref{eq:bound} and \eqref{eq:weighted_bound} of Propositions~\ref{prop:bound} and~\ref{prop:weighted} involve the unknown distribution of the data and cannot be evaluated from the data without further assumptions (\ref{app:remarks}). Second, the procedures share the inefficiency of split conformal prediction: the point predictor is fitted on the $n_1$ training observations only, the band is computed from $K-1$ calibration scores, and, in Procedure~1 with $b>1$, $n_2+1-K$ calibration observations are not scored. Third, the theory does not indicate how to choose $T$, $n_2$, $b$, $\gamma$, $\rho$ and the domains $\mathcal{Q}_j$. The simulation study gives indications for $b$, $\gamma$, $\rho$, $n_1$ and $n_2$ (Section~\ref{sec:sim_seq}), while in the application $T$ and the domain were chosen using curves of the evaluation period, which is not possible in real time (Sections~\ref{sec:app_data} and~\ref{sec:app_bands}). Fourth, the point predictors are simple, and a predictor designed for this market can be expected to give narrower bands (Section~\ref{sec:score}).

As for further developments, Procedure~2 could be combined with the aggregation over several step sizes \citep{zaffran2022adaptive} or with their online tuning \citep{gibbs2024conformal}. Methods that do not rely on a single split, such as EnbPI \citep{xu2021conformal,xu2023conformal}, would use the data more efficiently, but their extension to multivariate functional responses, with closed-form simultaneous bands, remains to be studied. Finally, the use of the bands and of the prediction region for the clearing point within the optimisation problems that describe the decisions of market participants is left to future research.

\section*{Acknowledgements}
The authors thank Centrex Italia S.p.A. for a detailed overview of the Italian gas market.

\section*{Disclosure statement}
The authors report there are no competing interests to declare.

\section*{Declaration of generative AI use}
In the research and the preparation of this manuscript, the authors used Claude Opus 5.5 to proofread and condense the text, check the proofs in the appendices, validate the code and check the bibliographic references. The authors reviewed, verified and edited all outputs, and take full responsibility for the content of the manuscript.

\section*{Funding}
This work was supported by the Italian Space Agency (ASI) under Grant 2018-5-HH.0 (Accordo Quadro ASI-POLIMI ``Attivit\`a di Ricerca e Innovazione''; S.V.).

\section*{Data availability statement}
The offers and bids of the MGS are publicly available at \url{https://www.mercatoelettrico.org/en/}, the website of GME. The R code and the processed data are available at \codelocation. The code is distributed under the GNU General Public License, version~3 or later; the processed data, derived from the data published by GME (source: Gestore dei Mercati Energetici SpA), remain subject to the terms of use of GME. The data are the offer and demand curves of the MGS evaluated on a grid, with the clearing prices and quantities, for the gas days from 1~July~2017 to 31~August~2020. The code comprises the simulation study and the application, with their stored results, except the figures and the raw results of the simulation study, which the scripts regenerate. The code that builds the curves from the GME files is not included; for the analysis without the offers and bids of Snam (\ref{app:app_extra}), only its stored results are included, not the curves or the code that computed the bands from them.

\bibliographystyle{elsarticle-harv}
\bibliography{references}

\appendix
\section{A primer on conformal prediction}
\label{app:primer}

This appendix summarises the elements of split conformal prediction used in Section~\ref{sec:methods}, for a generic response; the general theory, whose terminology we follow in this appendix, is given by \citet{angelopoulos2024theoretical}, and the results stated below are in their Chapters~2--4.

\paragraph{Exchangeability} Random elements $V_1,\dots,V_m$ are exchangeable if the joint distribution of $(V_{\pi(1)},\dots,V_{\pi(m)})$ is that of $(V_1,\dots,V_m)$ for every permutation $\pi$ of $\{1,\dots,m\}$. Independent and identically distributed random elements are exchangeable; exchangeable random elements are identically distributed, but need not be independent. The observations of a time series with serial dependence, or with a distribution that changes over time, are in general not exchangeable, and the guarantee given below does not apply to them directly; Sections~\ref{sec:procedure}--\ref{sec:weighted} describe the three procedures that we use for such data.

\paragraph{Split conformal prediction} Let $\boldsymbol{Z}_i=(\boldsymbol{X}_i,\boldsymbol{Y}_i)$, $i=1,\dots,n+1$, be random elements of a measurable space $\mathcal{Z}$, of which $\boldsymbol{Z}_1,\dots,\boldsymbol{Z}_n$ and $\boldsymbol{X}_{n+1}$ are observed; $\boldsymbol{Z}_{n+1}$ is the test point. The indices $1,\dots,n$ are partitioned, independently of the data, into the index sets $\mathcal{I}_1$ of the pretraining data and $\mathcal{I}_2$ of the calibration data, with $n_1=|\mathcal{I}_1|$ and $n_2=|\mathcal{I}_2|$. A conformal score function is a measurable function $A(\{\boldsymbol{z}_h: h \in \mathcal{I}_1\},\cdot)$ from $\mathcal{Z}$ to $\bar{\mathbb{R}}=\mathbb{R}\cup\{-\infty,+\infty\}$, constructed from the pretraining data by any method; its value at $\boldsymbol{z}$ is the score of $\boldsymbol{z}$, and larger scores indicate that $\boldsymbol{z}$ does not conform to the pretraining data. For a real-valued response, a common example is the residual score $|y-\hat{\mu}(\boldsymbol{x})|$ of a predictive model $\hat{\mu}$ fitted on the pretraining data. The calibration scores are $R_i:=A(\{\boldsymbol{z}_h: h \in \mathcal{I}_1\},\boldsymbol{z}_i)$, $i \in \mathcal{I}_2$, and the score of a hypothesised value $\boldsymbol{y}$ of $\boldsymbol{Y}_{n+1}$ is $R_{n+1}(\boldsymbol{y}):=A(\{\boldsymbol{z}_h: h \in \mathcal{I}_1\},(\boldsymbol{x}_{n+1},\boldsymbol{y}))$. The rest of the paper uses the terms nonconformity measure, nonconformity score, training set, candidate value and point predictor for the score function, the score, the pretraining data, the hypothesised value and the predictive model, respectively.

\paragraph{Conformal quantile and prediction set} If $\alpha \geq 1/(n_2+1)$, the conformal quantile $k$ is the $\lceil (n_2+1)(1-\alpha)\rceil$-th smallest of the $n_2$ calibration scores, i.e.\ their empirical quantile of level $(1-\alpha)(1+1/n_2) \leq 1$. If $\alpha<1/(n_2+1)$, the index $\lceil (n_2+1)(1-\alpha)\rceil$ exceeds $n_2$, the level exceeds $1$ and, by convention, $k:=+\infty$. Equivalently, $k$ is the smallest $(1-\alpha)$-quantile of the distribution that gives mass $1/(n_2+1)$ to each calibration score and to $+\infty$; Procedure~3 replaces these equal masses with weights (Section~\ref{sec:weighted}). The prediction set is
\begin{equation*}
\mathcal{C}(\boldsymbol{x}_{n+1}):=\{\boldsymbol{y}: R_{n+1}(\boldsymbol{y}) \leq k\},
\end{equation*}
which is the whole response space if $k=+\infty$. The same set is $\{\boldsymbol{y}: \hat{p}(\boldsymbol{y})>\alpha\}$, where $\hat{p}(\boldsymbol{y}):=(1+|\{i \in \mathcal{I}_2: R_i \geq R_{n+1}(\boldsymbol{y})\}|)/(n_2+1)$ is the conformal p-value of $\boldsymbol{y}$; Section~\ref{sec:procedure} uses this form.

\paragraph{Coverage guarantee} If $\boldsymbol{Z}_1,\dots,\boldsymbol{Z}_{n+1}$ are exchangeable, then, since the split is independent of the data, the calibration data and the test point are exchangeable conditionally on the split and on the pretraining data, and so are the $n_2+1$ scores $R_i$, $i \in \mathcal{I}_2$, and $R_{n+1}(\boldsymbol{Y}_{n+1})$. Hence $R_{n+1}(\boldsymbol{Y}_{n+1})$ does not exceed $k$ with probability at least $\lceil (n_2+1)(1-\alpha)\rceil/(n_2+1)$, and
\begin{equation*}
\mathbb{P}\left(\boldsymbol{Y}_{n+1} \in \mathcal{C}(\boldsymbol{X}_{n+1})\right) \geq 1-\alpha
\end{equation*}
for every $\alpha \in (0,1)$ and all $n_1$ and $n_2$. If, in addition, the $n_2+1$ scores are almost surely distinct, the coverage is exactly $\lceil (n_2+1)(1-\alpha)\rceil/(n_2+1)=1-\lfloor (n_2+1)\alpha\rfloor/(n_2+1)$, a value in $[1-\alpha,1-\alpha+1/(n_2+1))$ that equals $1-\alpha$ when $(n_2+1)\alpha$ is an integer.

\paragraph{Marginal and conditional coverage} The probability in the guarantee refers to the joint distribution of the split, the pretraining and calibration data and the test point; the guarantee is therefore marginal. It implies neither training-conditional coverage, i.e.\ coverage conditional on the pretraining and calibration data, nor test-conditional coverage, i.e.\ coverage conditional on $\boldsymbol{X}_{n+1}$. For a real-valued response, if a procedure has test-conditional coverage at least $1-\alpha$, almost surely, for every distribution of independent and identically distributed data, then its prediction set has infinite expected length under every such distribution for which $\mathbb{P}(\boldsymbol{X}_{n+1}=\boldsymbol{x})=0$ for all $\boldsymbol{x}$. For exchangeable data, coverage conditional on $\boldsymbol{X}_{n+1}$ belonging to a bin of a fixed finite partition of the space in which $\boldsymbol{X}_{n+1}$ takes values can be guaranteed by computing the conformal quantile separately within each bin.

\section{Proof of Proposition~\ref{prop:bound}}
\label{app:proof}
\renewcommand{\thelemma}{\Alph{section}.\arabic{lemma}}
\renewcommand*{\theHlemma}{\thelemma}
\setcounter{lemma}{0}

Throughout the proof we write $\delta_1:=\delta_{1K}$ and $\delta_2:=\delta_{2n_1}$. All scores are random variables, and the test score $R_{\omega_{n_2+1}}=R_{T+1}$ is evaluated at the realised response $\boldsymbol{Y}_{T+1}$; similarly, $R^*_{\omega_{n_2+1}}=R^*_{T+1}$. The argument uses neither the specific form \eqref{eq:score} of the nonconformity measure nor the value of $b$: it only uses the $K$ scores $R_{\omega_d}$, $d \in \mathcal{D}_{\Pi}$, and the conditions (E), (A) and (D).
Let
\begin{equation*}
\hat{F}(a):=\frac{1}{K}\sum_{d \in \mathcal{D}_{\Pi}} \mathbbm{1}\{R_{\omega_d}<a\}, \qquad a \in \bar{\mathbb{R}},
\end{equation*}
be the empirical distribution function of the $K$ scores, which is defined on $\bar{\mathbb{R}}$ because the scores take values in $\bar{\mathbb{R}}$. By \eqref{eq:pvalue}, $\hat{p}(\boldsymbol{Y}_{T+1})=K^{-1}|\{d \in \mathcal{D}_{\Pi}: R_{\omega_d} \geq R_{T+1}\}|=1-\hat{F}(R_{T+1})$, so that
the coverage event $\mathcal{E}:=\{\boldsymbol{Y}_{T+1} \in \mathcal{C}_{T,1-\alpha}(\boldsymbol{X}_{T+1})\}$ satisfies
\begin{equation}
\mathcal{E}=\left\{\hat{p}(\boldsymbol{Y}_{T+1})>\alpha\right\}=\left\{\hat{F}(R_{T+1})<1-\alpha\right\}.
\label{eq:app_event}
\end{equation}
Let $f$ be the density of $R^*_{T+1}$ in (D). Then $F(a)=\int_{-\infty}^{a} f(u)\,\mathrm{d}u$, so that $F$ is continuous and $|F(a)-F(a')| \leq D|a-a'|$ for all $a,a' \in \mathbb{R}$. Since $F$ is the continuous distribution function of $R^*_{T+1}$, the random variable $U:=F(R^*_{T+1})$ is uniformly distributed on $(0,1)$, and therefore $\mathbb{P}(U<c)=\min\{\max\{c,0\},1\}$ for $c \in \mathbb{R}$; in particular,
\begin{equation}
\mathbb{P}(U<c) \leq c \ \text{ if } c \geq 0, \qquad \mathbb{P}(U<c) \geq c \ \text{ if } c \leq 1.
\label{eq:app_unif}
\end{equation}

Let $\Omega_1$ be the event in (E), $\Omega_2$ the event in (A) and $\Omega:=\Omega_1 \cap \Omega_2$. By the union bound, $\mathbb{P}(\Omega^c) \leq \zeta_{1K}+\zeta_{2n_1}$. Since $A^*$ is real-valued, on $\Omega_2$ all the scores $R_{\omega_d}$, $d \in \mathcal{D}_{\Pi}$, are finite.

\begin{lemma}
\label{lem:bound_edf}
On $\Omega$, $\sup_{a \in \mathbb{R}}|\hat{F}(a)-F(a)| \leq 3\delta_1+\delta_2+2D\sqrt{\delta_2}$.
\end{lemma}

\begin{proof}
Let $\mathcal{B}:=\{d \in \mathcal{D}_{\Pi}: |R_{\omega_d}-R^*_{\omega_d}| \geq \sqrt{\delta_2}\}$. By the first inequality in (A),
\begin{equation*}
|\mathcal{B}|\,\delta_2 \leq \sum_{d \in \mathcal{B}}\left(R_{\omega_d}-R^*_{\omega_d}\right)^2 \leq \sum_{d \in \mathcal{D}_{\Pi}}\left(R_{\omega_d}-R^*_{\omega_d}\right)^2 \leq K\delta_2^2,
\end{equation*}
hence $|\mathcal{B}|/K \leq \delta_2$.

We first compare $\hat{F}$ with $\tilde{F}_K$. Fix $a \in \mathbb{R}$ and $d \in \mathcal{D}_{\Pi}\setminus\mathcal{B}$ such that $\mathbbm{1}\{R_{\omega_d}<a\} \neq \mathbbm{1}\{R^*_{\omega_d}<a\}$. If $R_{\omega_d}<a \leq R^*_{\omega_d}$, then $a \leq R^*_{\omega_d}<R_{\omega_d}+\sqrt{\delta_2}<a+\sqrt{\delta_2}$; if $R^*_{\omega_d}<a \leq R_{\omega_d}$, then $a-\sqrt{\delta_2} \leq R_{\omega_d}-\sqrt{\delta_2}<R^*_{\omega_d}<a$. In both cases $a-\sqrt{\delta_2}<R^*_{\omega_d}<a+\sqrt{\delta_2}$. Therefore, since $|\mathcal{B}|/K \leq \delta_2$,
\begin{align*}
\left|\hat{F}(a)-\tilde{F}_K(a)\right| &\leq \frac{1}{K}\sum_{d \in \mathcal{D}_{\Pi}}\left|\mathbbm{1}\{R_{\omega_d}<a\}-\mathbbm{1}\{R^*_{\omega_d}<a\}\right| \\
&\leq \frac{|\mathcal{B}|}{K}+\frac{1}{K}\left|\left\{d \in \mathcal{D}_{\Pi}: |R^*_{\omega_d}-a|<\sqrt{\delta_2}\right\}\right| \\
&\leq \delta_2+\tilde{F}_K\left(a+\sqrt{\delta_2}\right)-\tilde{F}_K\left(a-\sqrt{\delta_2}\right),
\end{align*}
where the last inequality uses $K^{-1}|\{d \in \mathcal{D}_{\Pi}: R^*_{\omega_d} \leq a-\sqrt{\delta_2}\}| \geq \tilde{F}_K(a-\sqrt{\delta_2})$. By (E) and the Lipschitz property of $F$,
\begin{align*}
\tilde{F}_K\left(a+\sqrt{\delta_2}\right)-\tilde{F}_K\left(a-\sqrt{\delta_2}\right) &\leq F\left(a+\sqrt{\delta_2}\right)-F\left(a-\sqrt{\delta_2}\right)+2\delta_1\\
&\leq 2D\sqrt{\delta_2}+2\delta_1.
\end{align*}
Since $a$ is arbitrary, $\sup_{a \in \mathbb{R}}|\hat{F}(a)-\tilde{F}_K(a)| \leq 2\delta_1+\delta_2+2D\sqrt{\delta_2}$. Hence, by the triangle inequality and (E),
\begin{align*}
\sup_{a \in \mathbb{R}}\left|\hat{F}(a)-F(a)\right| &\leq \sup_{a \in \mathbb{R}}\left|\hat{F}(a)-\tilde{F}_K(a)\right|+\sup_{a \in \mathbb{R}}\left|\tilde{F}_K(a)-F(a)\right|\\
&\leq 3\delta_1+\delta_2+2D\sqrt{\delta_2},
\end{align*}
which proves the lemma.
\end{proof}

\paragraph{Proof of Proposition~\ref{prop:bound}} On $\Omega$, $R_{T+1}$ is finite, since $\Omega \subseteq \Omega_2$. Hence, by Lemma~\ref{lem:bound_edf}, the Lipschitz property of $F$ and the second inequality in (A),
\begin{align*}
\left|\hat{F}(R_{T+1})-U\right| &\leq \left|\hat{F}(R_{T+1})-F(R_{T+1})\right|+\left|F(R_{T+1})-F(R^*_{T+1})\right|\\
&\leq 3\delta_1+\delta_2+2D\sqrt{\delta_2}+D\delta_2=:\eta.
\end{align*}
By \eqref{eq:app_event} and this inequality, $\mathcal{E}\cap\Omega \subseteq \{U<1-\alpha+\eta\}$ and $\{U<1-\alpha-\eta\}\cap\Omega \subseteq \mathcal{E}$, so that
\begin{equation*}
\mathcal{E} \subseteq \{U<1-\alpha+\eta\} \cup \Omega^c \quad \text{and} \quad \{U<1-\alpha-\eta\} \subseteq \mathcal{E} \cup \Omega^c.
\end{equation*}
Hence, by \eqref{eq:app_unif} with $c=1-\alpha+\eta \geq 0$ and $c=1-\alpha-\eta \leq 1$,
\begin{align*}
\mathbb{P}(\mathcal{E}) &\leq \mathbb{P}(U<1-\alpha+\eta)+\mathbb{P}(\Omega^c) \leq 1-\alpha+\eta+\zeta_{1K}+\zeta_{2n_1},\\
\mathbb{P}(\mathcal{E}) &\geq \mathbb{P}(U<1-\alpha-\eta)-\mathbb{P}(\Omega^c) \geq 1-\alpha-\eta-\zeta_{1K}-\zeta_{2n_1}.
\end{align*}
Since $\eta+\zeta_{1K}+\zeta_{2n_1}=3\delta_{1K}+\delta_{2n_1}+D(\delta_{2n_1}+2\sqrt{\delta_{2n_1}})+\zeta_{1K}+\zeta_{2n_1}$, this proves \eqref{eq:bound}. \hfill$\square$

\paragraph{Role of the sample sizes} The training set enters the proof only through condition (A), i.e.\ through $\delta_2$ and $\zeta_{2n_1}$, which quantify the error of the scores computed with the point predictor $\hat{\mu}^{(j)}_{\mathcal{I}_1}$ and the modulation functions $s^{(j)}$ fitted on the $n_1$ training observations; the oracle scores do not depend on the training set. The calibration set enters only through the scores at the $K$ positions in $\mathcal{D}_{\Pi}$, i.e.\ through condition (E), a law-of-large-numbers requirement over them, and the empirical distribution functions $\hat{F}$ and $\tilde{F}_K$, which are averages of $K$ terms. Hence the terms $3\delta_{1K}+\zeta_{1K}$ of \eqref{eq:bound} are controlled by the number of blocks $K$, and the terms $\delta_{2n_1}+D(\delta_{2n_1}+2\sqrt{\delta_{2n_1}})+\zeta_{2n_1}$ by the training set size $n_1$; the $n_2+1-K$ calibration observations that are not scored do not enter the bound.

\paragraph{Relation with \citet{chernozhukov2018exact}} The argument follows the structure of the proof of their Theorem~2, specialised to the split setting, in which the scores are computed from a point predictor fitted once on the training set and the coverage event \eqref{eq:app_event} involves the empirical distribution function of the $K$ scores $R_{\omega_d}$, $d \in \mathcal{D}_{\Pi}$, only. Their bound, written with the symbols of Proposition~\ref{prop:bound}, $6\delta_{1n}+4\delta_{2n}+2D(\delta_{2n}+2\sqrt{\delta_{2n}})+\zeta_{1n}+\zeta_{2n}$, in which all the sequences are indexed by the number of permutations $n=|\Pi|$ (the number of blocks, for non-overlapping blocks), has the same form as \eqref{eq:bound} with larger numerical constants, so that it also holds in the present setting with the sequences indexed as in Proposition~\ref{prop:bound}. The difference in the constants has two sources. First, their proof bounds $|\hat{F}(R_{T+1})-U|$, in the present notation, by $3\delta_{1n}+2\delta_{2n}+D(\delta_{2n}+2\sqrt{\delta_{2n}})$, with $2\delta_{2n}$ in place of the term $\delta_2$ obtained in the proof of Lemma~\ref{lem:bound_edf}. Second, it bounds the coverage error by the probability that $U$ falls in an interval centred at $1-\alpha$ whose half-length is this bound, which doubles the constants; the last part of the proof of Proposition~\ref{prop:bound} avoids the doubling by comparing the coverage event separately from above and from below with events of the form $\{U<c\}$.

\paragraph{Relation with earlier statements of the bound} A bound of the same form, $6\delta_{1K}+2\delta_{2n_1}+2D(\delta_{2n_1}+2\sqrt{\delta_{2n_1}})+\zeta_{1K}+\zeta_{2n_1}$, with the sequences indexed by $K$ and $n_1$ as in Proposition~\ref{prop:bound}, is stated by \citet{diquigiovanni2021distribution}, who refer to the proof of Theorem~2 of \citet{chernozhukov2018exact} and give no proof; \citet[Theorem~1]{ajroldi2023conformal} report it from that source, for a single functional response defined on a two-dimensional domain, again without proof. Under conditions (E), (A) and (D), that bound follows from \eqref{eq:bound}, whose constants are smaller; the proof above is, to our knowledge, the first complete one for the split setting. The remark that the terms in $\delta_{2n_1}$ and $\zeta_{2n_1}$ depend on the training set and those in $\delta_{1K}$ and $\zeta_{1K}$ on the number of blocks is already made in these two works.

\section{Proof of Proposition~\ref{prop:aci}}
\label{app:proof_aci}
\setcounter{lemma}{0}

The proof adapts that of Proposition~4.1 of \citet{gibbs2021adaptive}, which corresponds to $\tau=0$. The realisation $\boldsymbol{z}_{1-\tau},\dots,\boldsymbol{z}_{T+H}$, and that of any auxiliary randomisation (e.g.\ random splits) used to compute the bands, is fixed, so that the working levels $\alpha_t$ and the miscoverage indicators $\mathrm{err}_t \in \{0,1\}$ are deterministic; let $\bar{\alpha}:=\max\{\alpha,1-\alpha\}$. The argument uses only the update \eqref{eq:aci_update} and two consequences of the conventions in Step~3 of Procedure~2: if $\alpha_t<0$, then $\alpha_t<1/(n_2+1)$, the band is the whole space $\mathcal{Y}$ and $\mathrm{err}_t=0$; if $\alpha_t>1$, the band is empty and $\mathrm{err}_t=1$. By Steps~1 and~4, $\alpha_{t+1}-\alpha_t=0$ for $T+1 \leq t \leq T+\tau$ and $\alpha_{t+1}-\alpha_t \in \{\gamma\alpha,-\gamma(1-\alpha)\}$ for $T+\tau+1 \leq t \leq T+H$; hence, for $T+1 \leq s \leq t \leq T+H+1$,
\begin{equation}
-(t-s)\,\gamma(1-\alpha) \leq \alpha_t-\alpha_s \leq (t-s)\,\gamma\alpha.
\label{eq:app_incr}
\end{equation}

\begin{lemma}
\label{lem:aci_levels}
For $T+1 \leq t \leq T+H+1$, $-(\tau+1)\gamma \leq \alpha_t \leq 1+(\tau+1)\gamma$.
\end{lemma}

\begin{proof}
Suppose that the upper bound fails, let $u \leq T+H+1$ be the smallest index such that $\alpha_u>1+(\tau+1)\gamma$, and let $t:=u-1$. Since $\alpha_s=\alpha<1$ for $T+1 \leq s \leq T+\tau+1$, we have $T+\tau+1 \leq t \leq T+H$, hence $t-\tau \geq T+1$, and $\alpha_t \leq 1+(\tau+1)\gamma$ by the definition of $u$. By \eqref{eq:app_incr}, $\alpha_t \geq \alpha_{t+1}-\gamma\alpha>1+\tau\gamma+\gamma(1-\alpha)$ and $\alpha_{t-\tau} \geq \alpha_t-\tau\gamma\alpha>1+(\tau+1)\gamma(1-\alpha)>1$. Hence the band at time $t-\tau$ is empty, $\mathrm{err}_{t-\tau}=1$ and, by \eqref{eq:aci_update}, $\alpha_{t+1}=\alpha_t-\gamma(1-\alpha)<\alpha_t \leq 1+(\tau+1)\gamma$, which contradicts the definition of $u$. Similarly, if the lower bound fails, let $u \leq T+H+1$ be the smallest index such that $\alpha_u<-(\tau+1)\gamma$, and let $t:=u-1$. Then $T+\tau+1 \leq t \leq T+H$, $\alpha_t \geq -(\tau+1)\gamma$ and, by \eqref{eq:app_incr}, $\alpha_t \leq \alpha_{t+1}+\gamma(1-\alpha)<-\tau\gamma-\gamma\alpha$ and $\alpha_{t-\tau} \leq \alpha_t+\tau\gamma(1-\alpha)<-(\tau+1)\gamma\alpha<0$. Hence the band at time $t-\tau$ is the whole space, $\mathrm{err}_{t-\tau}=0$ and $\alpha_{t+1}=\alpha_t+\gamma\alpha>\alpha_t \geq -(\tau+1)\gamma$, a contradiction.
\end{proof}

\paragraph{Proof of Proposition~\ref{prop:aci}} Summing \eqref{eq:aci_update} over $t=T+\tau+1,\dots,T+H$ and using $\alpha_{T+\tau+1}=\alpha$ gives, for every $H \in \mathbb{N}^*$,
\begin{equation*}
\gamma\sum_{t=T+1}^{T+H-\tau}\left(\alpha-\mathrm{err}_t\right)=\alpha_{T+H+1}-\alpha,
\end{equation*}
where both sides are zero if $H \leq \tau$. By Lemma~\ref{lem:aci_levels}, $-(\tau+1)\gamma-\alpha \leq \alpha_{T+H+1}-\alpha \leq 1-\alpha+(\tau+1)\gamma$, so that the sum on the left-hand side is at most $(\bar{\alpha}+(\tau+1)\gamma)/\gamma$ in absolute value. The remaining terms of $\sum_{t=T+1}^{T+H}(\alpha-\mathrm{err}_t)$, those with $t>T+H-\tau$, are at most $\tau$ in number, and each is at most $\bar{\alpha}$ in absolute value. Hence
\begin{equation*}
\left|\sum_{t=T+1}^{T+H}\left(\mathrm{err}_t-\alpha\right)\right| \leq \frac{\bar{\alpha}+(\tau+1)\gamma}{\gamma}+\tau\bar{\alpha}=\frac{\bar{\alpha}(1+\tau\gamma)+(\tau+1)\gamma}{\gamma},
\end{equation*}
and dividing by $H$ gives \eqref{eq:aci_bound}. \hfill$\square$

\paragraph{Remarks} The delay enters the bound twice: in Lemma~\ref{lem:aci_levels}, through the $\tau$ updates between times $t-\tau$ and $t$, and through the last $\tau$ indicators, which do not enter $\alpha_{T+H+1}$. Applying the argument for $\tau=0$ separately to the $\tau+1$ subsequences of times that are congruent modulo $\tau+1$, each updated as $\alpha_{t+\tau+1}:=\alpha_t+\gamma(\alpha-\mathrm{err}_t)$, a different scheme, studied by \citet{elhalabi2026adaptive}, gives the bound $(\tau+1)(\bar{\alpha}+\gamma)/(\gamma H)$, which, for $\gamma<1$, exceeds \eqref{eq:aci_bound} by $\tau\bar{\alpha}(1-\gamma)/(\gamma H)$.

\paragraph{Relation with multi-step adaptive conformal inference} \citet{gibbs2021adaptive} consider responses revealed at each time point and mention delayed responses as an open problem. For $\tau>0$, the update \eqref{eq:aci_update} coincides with the updates proposed, for scalar series and forecasts $\tau+1$ steps ahead, by \citet{hallberg2024adaptive} and by \citet{wang2024online}. \citet{hallberg2024adaptive} states for it, without proof, the bound of \citet{gibbs2021adaptive} with no delay term; that bound rests on working levels in $[-\gamma,1+\gamma]$, whereas with a delay the levels can leave this interval, and Lemma~\ref{lem:aci_levels} gives the range $[-(\tau+1)\gamma,1+(\tau+1)\gamma]$. \citet{wang2024online} prove bounds on the long-run frequency of miscoverage, with an additive term of order $(\tau+1)/H$ due to the delay, for updates of the quantile with bounded scores, and none for the update of the level. Proposition~\ref{prop:aci} gives a bound for the update of the level, with an explicit dependence on $\tau$ and no condition on the scores.

The conventions at the extremes are the only link between the working level and the indicators used in the proof, and the bound need not hold without them: if the band at levels $\alpha_t \geq 1$ were not empty, a sequence of responses lying in all the bands would give $\mathrm{err}_t=0$ for all $t$, and $\alpha_t$ would increase without bound; symmetrically, if the band at negative levels were not the whole space, a sequence of misses would give $\mathrm{err}_t=1$ for all $t$. The argument does not involve the form \eqref{eq:score} of the nonconformity measure, the dependence structure of the data, the point predictors, the split, the overlap of the windows, or the calibration set size $n_2$. The nesting of the band in the level is not used in the bound; it ensures that the update moves the band in the intended direction (\ref{app:remarks}).

The bound concerns the realised indicators $\mathrm{err}_t$ and implies no statement on the probability $\mathbb{P}(\boldsymbol{Y}_t \notin \mathcal{C}_{t-\tau-1,1-\alpha_t}(\boldsymbol{X}_t))$ at a given time $t$: when the data are random, taking expectations in \eqref{eq:aci_bound} shows that the average of these probabilities over $t=T+1,\dots,T+H$, not each of them, is within the same distance of $\alpha$.

\section{Further remarks on the procedures}
\label{app:remarks}

This appendix collects remarks on the procedures of Section~\ref{sec:methods} that are not needed to apply them.

\paragraph{Functions and grids} Functions that coincide almost everywhere with respect to a reference measure on $\mathcal{Q}_j$ (e.g.\ the Lebesgue measure) are identified; the essential supremum in \eqref{eq:score} is taken with respect to this measure, and the conditions defining a band hold almost everywhere. In applications, the functions are evaluated on a finite grid of points of $\mathcal{Q}_j$, on which the essential supremum is a maximum and the conditions defining a band are checked.

\paragraph{Closed form of the band} If $\alpha \geq 1/K$, then, by \eqref{eq:pvalue}, $\hat{p}(\boldsymbol{y})>\alpha$ if and only if at least $\lfloor \alpha K \rfloor$ calibration scores are not smaller than $R_{T+1}(\boldsymbol{y})$. Since $K-\lfloor \alpha K \rfloor={\lceil K(1-\alpha) \rceil}$, this holds if and only if $R_{T+1}(\boldsymbol{y}) \leq k$, which, by \eqref{eq:score}, is equivalent to the conditions in \eqref{eq:band}. Hence $\mathcal{C}_{T,1-\alpha}(\boldsymbol{x}_{T+1})$ is the band \eqref{eq:band}.

\paragraph{Bands, split and blocks} Bands are simple to interpret and can be visualised in parallel coordinates \citep{inselberg1985plane}; their properties are discussed by \citet{diquigiovanni2021importance}. As in split conformal prediction, $\mathcal{C}_{T,1-\alpha}(\boldsymbol{x}_{T+1})$ contains the candidate values whose score is not exceptionally large compared with the scored calibration ones. With a chronological split and $b=1$, BSC is split conformal prediction with the sequential split of \citet{wisniewski2020application}. With a random split, consecutive positions need not be consecutive time points, but, since $\omega_{d+b}-\omega_d \geq b$, the scored time points are at least $b$ apart (exactly $b$ with a chronological split).

\paragraph{Relation with \citet{chernozhukov2018exact}} BSC is the split (inductive) version of their conformal inference with non-overlapping blocks. In their Section~2.4, they note that inductive conformal prediction is a special case of their permutation framework, obtained by restricting the permutations to the indices of the calibration set, and that their results can be used to justify it for dependent data. They do not write this special case out. It was first written out by \citet{diquigiovanni2021distribution}, for multivariate functional responses and with a score of the form \eqref{eq:score}, which yields a band in closed form, together with a coverage bound stated without proof. \citet{ajroldi2023conformal} apply the same block-split scheme to a single functional response defined on a two-dimensional domain, with point predictors for such data, and report the procedure and the coverage bound from that source (their Section~2 and Theorem~1). With respect to \citet{ajroldi2023conformal}, the present paper considers $p \geq 1$ responses, gives a complete proof of the coverage bound, with smaller numerical constants (\ref{app:proof}), and an example of primitive conditions (Lemma~\ref{lem:example}), adds Procedures~2 and~3, with the bound for delayed feedback (Proposition~\ref{prop:aci}), and compares the three procedures. The numerical constants of the bound \eqref{eq:bound} are compared with those of Theorem~2 of \citet{chernozhukov2018exact} and of these earlier statements at the end of \ref{app:proof}.

\paragraph{Verifiability of the conditions of Proposition~\ref{prop:bound}} The conditions involve the oracle scores, which are not observable, and sequences that depend on the unknown data generating process; hence they cannot be verified, nor the right-hand side of \eqref{eq:bound} evaluated, from the observed data without further assumptions. Proposition~\ref{prop:bound} describes when the marginal coverage can be expected to be close to $1-\alpha$: scores close to the oracle scores (A), oracle scores that are weakly dependent across blocks and stable in distribution over time (E), and a bounded density of the oracle test score (D).

\paragraph{Conditions of Proposition~\ref{prop:bound}} An example of oracle nonconformity measure is the score \eqref{eq:score} with $\hat{\mu}^{(j)}_{\mathcal{I}_1}$ replaced by $\mu^{(j)}$ and the modulation function $s^{(j)}$ by a population counterpart. If $\hat{\mu}^{(j)}_{\mathcal{I}_1}$ is not consistent for $\mu^{(j)}$, condition (A) may still hold for an oracle score defined through the limit of the misspecified predictor, if such a limit exists. By Lemma~1 of \citet{chernozhukov2018exact}, condition (E) holds, for suitable sequences $\delta_{1K}$ and $\zeta_{1K}$, when the oracle scores at the $K$ time points $\omega_d$, $d \in \mathcal{D}_{\Pi}$, which are equally spaced with a chronological split, form a stationary strongly mixing sequence with summable mixing coefficients; with a random split, the lemma does not apply directly. Primitive conditions for (A) and (D) with the score \eqref{eq:score}, which involves a supremum over $q$ and estimated modulation functions, are given below in a Gaussian example only (Lemma~\ref{lem:example}); in general, (A) requires the point predictor and the modulation function to be consistent uniformly over $\mathcal{Q}_j$ and the modulation function to be bounded away from zero, which is more than the $L^2$ consistency usually available for functional autoregressive estimators. Under (D), $F$ is continuous, whereas $\tilde{F}_K$ is a step function with jumps of size at least $1/K$; hence $\sup_{a}|\tilde{F}_K(a)-F(a)| \geq 1/(2K)$ for every realisation, and (E) can hold with $\zeta_{1K}<1$ only if $\delta_{1K} \geq 1/(2K)$: for a fixed number of blocks, the right-hand side of \eqref{eq:bound} is bounded away from zero.

\paragraph{Primitive conditions in an example}
\setcounter{lemma}{0}%
Conditions (E), (A) and (D) can be verified in a correctly specified Gaussian example. Let $m,r \in \mathbb{N}^*$ and let $\{\boldsymbol{\xi}_t\}_{t \in \mathbb{Z}}$ be the stationary solution of $\boldsymbol{\xi}_t=\sum_{i=1}^{r}\Psi_i\boldsymbol{\xi}_{t-i}+\boldsymbol{\varepsilon}_t$, where the $\boldsymbol{\varepsilon}_t \in \mathbb{R}^m$ are i.i.d.\ Gaussian with mean zero and positive definite covariance matrix $\Sigma$, and $\det(I_m-\sum_{i=1}^{r}\Psi_i z^i) \neq 0$ for all complex $z$ with $|z| \leq 1$; then $\boldsymbol{\varepsilon}_t$ is independent of $\boldsymbol{\xi}_{t-1},\boldsymbol{\xi}_{t-2},\dots$, and the companion matrix of the process has spectral radius smaller than one \citep[Section~2.1]{lutkepohl2005new}. For $j=1,\dots,p$, let $Y_t^{(j)}(q)=\boldsymbol{g}_j(q)^{\top}\boldsymbol{\xi}_t$, $q \in \mathcal{Q}_j$, where $\boldsymbol{g}_j:\mathcal{Q}_j \rightarrow \mathbb{R}^m\setminus\{\boldsymbol{0}\}$ is known and continuous and the Gram matrix $\sum_{j=1}^{p}\int_{\mathcal{Q}_j}\boldsymbol{g}_j(q)\boldsymbol{g}_j(q)^{\top}\mathrm{d}q$, in which the integrals are taken with respect to the reference measure on $\mathcal{Q}_j$, assumed finite, is nonsingular, so that $\boldsymbol{\xi}_t$ is a linear function of $\boldsymbol{Y}_t$, also when functions that coincide almost everywhere are identified. For every $j$, the covariate $X_t^{(j)}$ consists of the whole responses $\boldsymbol{Y}_{t-1},\dots,\boldsymbol{Y}_{t-r}$ ($t_0=r$); it is therefore in linear one-to-one correspondence with $\boldsymbol{V}_t:=(\boldsymbol{\xi}_{t-1}^{\top},\dots,\boldsymbol{\xi}_{t-r}^{\top})^{\top} \in \mathbb{R}^{mr}$, of which $\boldsymbol{\varepsilon}_t$ is independent, and, with $\Psi:=[\Psi_1,\dots,\Psi_r]$, the regression map is $[\mu^{(j)}(X_t^{(j)})](q)=\boldsymbol{g}_j(q)^{\top}\Psi\boldsymbol{V}_t$. The point predictor replaces $\Psi$ by the least-squares estimator $\hat{\Psi}:=(\sum_{h \in \mathcal{I}_1}\boldsymbol{\xi}_h\boldsymbol{V}_h^{\top})(\sum_{h \in \mathcal{I}_1}\boldsymbol{V}_h\boldsymbol{V}_h^{\top})^{-1}$ (arbitrary if the inverse does not exist), the score is \eqref{eq:score} with \eqref{eq:sd}, $n_1 \geq 2$ (set to $+\infty$ if some $s^{(j)}$ is not strictly positive on $\mathcal{Q}_j$), and the oracle nonconformity measure is built on the true regression map,
\begin{equation*}
A^*(\boldsymbol{z}):=\max_{j=1,\dots,p}\ \esssup_{q \in \mathcal{Q}_j}\frac{\left|y^{(j)}(q)-[\mu^{(j)}(x^{(j)})](q)\right|}{\sigma_j(q)}, \qquad \sigma_j(q)^2:=\boldsymbol{g}_j(q)^{\top}\Sigma\,\boldsymbol{g}_j(q),
\end{equation*}
where $\sigma_j(q)^2$ is the variance of the oracle residual $\boldsymbol{g}_j(q)^{\top}\boldsymbol{\varepsilon}_t$; $A^*$ is real-valued, since $\sigma_j$ is continuous and positive on $\mathcal{Q}_j$. Let $\Lambda(l):=\mathbb{E}[\boldsymbol{V}_t\boldsymbol{V}_{t+l}^{\top}]$, $\Lambda:=\Lambda(0)$, let $\lambda_V$ and $\lambda_{\Sigma}$ be the smallest eigenvalues of $\Lambda$ and $\Sigma$, $\kappa:=\mathrm{tr}(\Lambda)/\lambda_V$, $c_{\Lambda}:=(mr+1)\sum_{l \in \mathbb{Z}}\|\Lambda(l)\|_F^2$,
\begin{equation*}
L_1:=\mathrm{tr}(\Sigma)\,(9+12\kappa) \qquad \text{and} \qquad L_2:=\left\{\frac{8\,\mathrm{tr}(\Sigma)}{\lambda_{\Sigma}}\left(4\kappa^2+\frac{L_1^2}{\lambda_{\Sigma}^2}\right)\right\}^{1/2}.
\end{equation*}

\begin{lemma}
\label{lem:example}
In this setting, $\lambda_V>0$, $c_{\Lambda}<\infty$ and, for any block length $b$ allowed in Step~3 of Procedure~1 and any split chosen independently of the data, including the random one:
\begin{itemize}
\item[(D)] holds with $D=m\sqrt{2/\pi}$;
\item[(E)] holds with $\delta_{1K}=\{\log(2K)/(2K)\}^{1/2}$ and $\zeta_{1K}=1/K$, the $K$ oracle scores $R^*_{\omega_d}$, $d \in \mathcal{D}_{\Pi}$, being i.i.d.\ given the split;
\item[(A)] holds with $\delta_{2n_1}=L_2\,n_1^{-1/3}$ and $\zeta_{2n_1}=5n_1^{-1/6}+4c_{\Lambda}/(\lambda_V^2 n_1)$, if $n_1^{5/6} \geq 4L_1^2/\lambda_{\Sigma}^2$.
\end{itemize}
Hence, if $n_1^{5/6} \geq 4L_1^2/\lambda_{\Sigma}^2$, the left-hand side of \eqref{eq:bound} is at most
\begin{equation*}
3\left\{\frac{\log(2K)}{2K}\right\}^{1/2}+\frac{1}{K}+(1+D)L_2\,n_1^{-1/3}+\left(2D\sqrt{L_2}+5\right)n_1^{-1/6}+\frac{4c_{\Lambda}}{\lambda_V^2 n_1}.
\end{equation*}
\end{lemma}

\begin{proof}
Condition on the split; no bound below depends on it. Let $|\cdot|$, $\|\cdot\|$ and $\|\cdot\|_F$ denote the Euclidean, spectral and Frobenius norms.

\emph{Preliminaries.} If $\boldsymbol{v}^{\top}\Lambda\boldsymbol{v}=0$ for some $\boldsymbol{v}=(\boldsymbol{v}_1^{\top},\dots,\boldsymbol{v}_r^{\top})^{\top} \neq \boldsymbol{0}$, and $i$ is the smallest index with $\boldsymbol{v}_i \neq \boldsymbol{0}$, then $\sum_{i' \geq i}\boldsymbol{v}_{i'}^{\top}\boldsymbol{\xi}_{t-i'}=0$ almost surely, so that $\boldsymbol{v}_i^{\top}\boldsymbol{\varepsilon}_{t-i}$ is almost surely equal to a function of $\boldsymbol{\xi}_{t-i-1},\boldsymbol{\xi}_{t-i-2},\dots$, of which it is independent; hence $\boldsymbol{v}_i^{\top}\Sigma\boldsymbol{v}_i=0$, a contradiction, and $\lambda_V>0$. Since $\boldsymbol{V}_{t+1}$ is the companion matrix times $\boldsymbol{V}_t$ plus a vector independent of $\boldsymbol{V}_t$, for $l \geq 0$ the matrix $\Lambda(l)$ is $\Lambda$ times the $l$-th power of the transpose of the companion matrix, and $\Lambda(-l)=\Lambda(l)^{\top}$; hence $\|\Lambda(l)\|_F$ decays geometrically and $c_{\Lambda}<\infty$.

\emph{Condition (D).} Let $\boldsymbol{a}_j(q):=\Sigma^{1/2}\boldsymbol{g}_j(q)/\sigma_j(q)$, a unit vector, and $\|\boldsymbol{v}\|_*:=\max_{j}\esssup_{q \in \mathcal{Q}_j}|\boldsymbol{a}_j(q)^{\top}\boldsymbol{v}|$ for $\boldsymbol{v} \in \mathbb{R}^m$, so that $R^*_t=\|\Sigma^{-1/2}\boldsymbol{\varepsilon}_t\|_*$. This is a norm: if $\|\boldsymbol{v}\|_*=0$, then $\boldsymbol{g}_j(q)^{\top}\Sigma^{1/2}\boldsymbol{v}=0$ for almost every $q$ and all $j$, and $\boldsymbol{v}=\boldsymbol{0}$ because the Gram matrix is nonsingular. Hence $\mathcal{S}:=\{\boldsymbol{v}:\|\boldsymbol{v}\|_* \leq 1\}$ is compact. Moreover, if $q_0$ belongs to the support of the reference measure on some $\mathcal{Q}_j$ (one exists, the Gram matrix being nonzero), then $\|\boldsymbol{v}\|_* \geq |\boldsymbol{a}_j(q_0)^{\top}\boldsymbol{v}|$ by continuity. Let $\boldsymbol{\chi}:=\Sigma^{-1/2}\boldsymbol{\varepsilon}_{T+1}$, with density $\vartheta(\boldsymbol{v}):=(2\pi)^{-m/2}\exp(-|\boldsymbol{v}|^2/2)$, and, for $u>0$,
\begin{equation*}
\Phi_*(u):=\mathbb{P}(R^*_{T+1} \leq u)=\int_{u\mathcal{S}}\vartheta(\boldsymbol{v})\,\mathrm{d}\boldsymbol{v}=\int_{\mathcal{S}}u^m\vartheta(u\boldsymbol{v})\,\mathrm{d}\boldsymbol{v}.
\end{equation*}
The last integrand is continuously differentiable in $u$, with derivative bounded on $\mathcal{S}\times[u_0,u_1]$ for $0<u_0<u_1$; differentiating under the integral sign,
\begin{equation*}
0 \leq \Phi_*'(u)=\int_{\mathcal{S}}\left(m u^{m-1}-u^{m+1}|\boldsymbol{v}|^2\right)\vartheta(u\boldsymbol{v})\,\mathrm{d}\boldsymbol{v} \leq \frac{m}{u}\,\Phi_*(u) \leq m\sqrt{2/\pi},
\end{equation*}
because $\Phi_*(u) \leq \mathbb{P}(|\boldsymbol{a}_j(q_0)^{\top}\boldsymbol{\chi}| \leq u) \leq u\sqrt{2/\pi}$, the variable $\boldsymbol{a}_j(q_0)^{\top}\boldsymbol{\chi}$ being standard Gaussian. The same inequality gives $\mathbb{P}(R^*_{T+1} \leq 0)=0$, so that $R^*_{T+1}$ has the density $\Phi_*'\mathbbm{1}_{(0,\infty)} \leq D$.

\emph{Condition (E).} The $K$ time points $\omega_d$, $d \in \mathcal{D}_{\Pi}$, are distinct and $R^*_t$ is a function of $\boldsymbol{\varepsilon}_t$ only; hence the $K$ oracle scores are i.i.d.\ with the continuous distribution function $\Phi_*$ (equal to $0$ on $(-\infty,0]$), and $F=\Phi_*$. If $\hat{\Phi}_K$ denotes their right-continuous empirical distribution function, then $\tilde{F}_K(a)=\hat{\Phi}_K(a-)$, so that $\sup_a|\tilde{F}_K(a)-F(a)| \leq \sup_a|\hat{\Phi}_K(a)-\Phi_*(a)|$, which exceeds $x>0$ with probability at most $2\exp(-2Kx^2)$ \citep{massart1990tight}; this equals $1/K$ for $x=\delta_{1K}$.

\emph{Condition (A): algebra.} Let $\hat{\Lambda}$, $\hat{\Xi}$, $\tilde{\Sigma}$ and $\bar{\boldsymbol{\varepsilon}}$ be the averages over $h \in \mathcal{I}_1$ of $\boldsymbol{V}_h\boldsymbol{V}_h^{\top}$, $\boldsymbol{\varepsilon}_h\boldsymbol{V}_h^{\top}$, $\boldsymbol{\varepsilon}_h\boldsymbol{\varepsilon}_h^{\top}$ and $\boldsymbol{\varepsilon}_h$. Since $\boldsymbol{\xi}_h=\Psi\boldsymbol{V}_h+\boldsymbol{\varepsilon}_h$, when $\hat{\Lambda}$ is invertible $\hat{\Theta}:=\hat{\Psi}-\Psi=\hat{\Xi}\hat{\Lambda}^{-1}$, the training residuals are $r_h^{(j)}(q)=\boldsymbol{g}_j(q)^{\top}\hat{\boldsymbol{\varepsilon}}_h$ with $\hat{\boldsymbol{\varepsilon}}_h:=\boldsymbol{\varepsilon}_h-\hat{\Theta}\boldsymbol{V}_h$, and $s^{(j)}(q)^2=\boldsymbol{g}_j(q)^{\top}\hat{\Sigma}\,\boldsymbol{g}_j(q)$, where $\hat{\Sigma}$ is the sample covariance matrix, with divisor $n_1-1$, of the $\hat{\boldsymbol{\varepsilon}}_h$. Their average $\bar{\boldsymbol{e}}$ and the average of the $\hat{\boldsymbol{\varepsilon}}_h\hat{\boldsymbol{\varepsilon}}_h^{\top}$ satisfy
\begin{equation*}
\frac{1}{n_1}\sum_{h \in \mathcal{I}_1}\hat{\boldsymbol{\varepsilon}}_h\hat{\boldsymbol{\varepsilon}}_h^{\top}=\tilde{\Sigma}-\hat{\Xi}\hat{\Lambda}^{-1}\hat{\Xi}^{\top}, \qquad |\bar{\boldsymbol{e}}|^2 \leq 2|\bar{\boldsymbol{\varepsilon}}|^2+\frac{2}{n_1}\sum_{h \in \mathcal{I}_1}|\hat{\Theta}\boldsymbol{V}_h|^2=2|\bar{\boldsymbol{\varepsilon}}|^2+2\,\mathrm{tr}(\hat{\Xi}\hat{\Lambda}^{-1}\hat{\Xi}^{\top}),
\end{equation*}
and $\hat{\Sigma}-\Sigma=\frac{n_1}{n_1-1}\{(\tilde{\Sigma}-\Sigma)-\hat{\Xi}\hat{\Lambda}^{-1}\hat{\Xi}^{\top}-\bar{\boldsymbol{e}}\bar{\boldsymbol{e}}^{\top}\}+\frac{1}{n_1-1}\Sigma$. Hence, as $n_1/(n_1-1) \leq 2$ and $\mathrm{tr}(\hat{\Xi}\hat{\Lambda}^{-1}\hat{\Xi}^{\top}) \leq \|\hat{\Lambda}^{-1}\|\,\|\hat{\Xi}\|_F^2$,
\begin{equation}
\|\hat{\Sigma}-\Sigma\| \leq 2\|\tilde{\Sigma}-\Sigma\|_F+6\|\hat{\Lambda}^{-1}\|\,\|\hat{\Xi}\|_F^2+4|\bar{\boldsymbol{\varepsilon}}|^2+\frac{\|\Sigma\|}{n_1-1}.
\label{eq:ex_S}
\end{equation}
Let $\hat{\theta}:=\|\hat{\Sigma}-\Sigma\|/\lambda_{\Sigma}$ and suppose that $\hat{\theta} \leq 1/2$. For all $j$ and $q$, which are omitted from the notation, $\sigma^2 \geq \lambda_{\Sigma}|\boldsymbol{g}|^2$ and $|s^2-\sigma^2| \leq \|\hat{\Sigma}-\Sigma\|\,|\boldsymbol{g}|^2 \leq \hat{\theta}\sigma^2$, so that $s \geq \sigma/\sqrt{2}>0$ and $|\sigma/s-1|=|\sigma^2-s^2|/\{s(\sigma+s)\} \leq \sqrt{2}\,\hat{\theta}$. For any time point $t$, the residual of the fitted predictor is $\boldsymbol{g}^{\top}(\boldsymbol{\varepsilon}_t-\hat{\Theta}\boldsymbol{V}_t)$, and
\begin{equation*}
\left|\frac{|\boldsymbol{g}^{\top}(\boldsymbol{\varepsilon}_t-\hat{\Theta}\boldsymbol{V}_t)|}{s}-\frac{|\boldsymbol{g}^{\top}\boldsymbol{\varepsilon}_t|}{\sigma}\right| \leq \frac{|\boldsymbol{g}^{\top}\hat{\Theta}\boldsymbol{V}_t|}{s}+\frac{|\boldsymbol{g}^{\top}\boldsymbol{\varepsilon}_t|}{\sigma}\left|\frac{\sigma}{s}-1\right| \leq \sqrt{\frac{2}{\lambda_{\Sigma}}}\left(\|\hat{\Theta}\|_F|\boldsymbol{V}_t|+\hat{\theta}|\boldsymbol{\varepsilon}_t|\right),
\end{equation*}
since $|\boldsymbol{g}^{\top}\boldsymbol{v}|/\sigma \leq |\boldsymbol{v}|/\sqrt{\lambda_{\Sigma}}$ for $\boldsymbol{v} \in \mathbb{R}^m$. Two maxima of essential suprema differ by at most the supremum of the absolute difference; therefore
\begin{equation}
(R_t-R^*_t)^2 \leq \frac{4}{\lambda_{\Sigma}}\left(\|\hat{\Theta}\|_F^2|\boldsymbol{V}_t|^2+\hat{\theta}^2|\boldsymbol{\varepsilon}_t|^2\right).
\label{eq:ex_R}
\end{equation}

\emph{Condition (A): probabilities.} Let $\varsigma_V:=K^{-1}\sum_{d \in \mathcal{D}_{\Pi}}|\boldsymbol{V}_{\omega_d}|^2+|\boldsymbol{V}_{T+1}|^2$, and define $\varsigma_{\varepsilon}$ likewise from the $\boldsymbol{\varepsilon}_t$; by stationarity, $\mathbb{E}\varsigma_V=2\,\mathrm{tr}(\Lambda)$ and $\mathbb{E}\varsigma_{\varepsilon}=2\,\mathrm{tr}(\Sigma)$. For $h<h'$, $\boldsymbol{\varepsilon}_{h'}$ has mean zero and is independent of $(\boldsymbol{\varepsilon}_h,\boldsymbol{V}_h,\boldsymbol{V}_{h'})$, and $\boldsymbol{\varepsilon}_h$ is independent of $\boldsymbol{V}_h$; hence $\mathbb{E}\|\hat{\Xi}\|_F^2=\mathrm{tr}(\Sigma)\mathrm{tr}(\Lambda)/n_1$. Moreover, $\mathbb{E}|\bar{\boldsymbol{\varepsilon}}|^2=\mathrm{tr}(\Sigma)/n_1$ and, the innovations being Gaussian, $\mathbb{E}\|\tilde{\Sigma}-\Sigma\|_F^2=[\{\mathrm{tr}(\Sigma)\}^2+\mathrm{tr}(\Sigma^2)]/n_1 \leq 2\{\mathrm{tr}(\Sigma)\}^2/n_1$. By Isserlis' formula for the fourth moments of the zero-mean Gaussian process $\{\boldsymbol{V}_t\}$, for $h,h' \in \mathcal{I}_1$ and $l=h'-h$,
\begin{equation*}
\mathbb{E}\,\mathrm{tr}\{(\boldsymbol{V}_h\boldsymbol{V}_h^{\top}-\Lambda)(\boldsymbol{V}_{h'}\boldsymbol{V}_{h'}^{\top}-\Lambda)\}=\{\mathrm{tr}\,\Lambda(l)\}^2+\mathrm{tr}\{\Lambda(l)^2\} \leq (mr+1)\|\Lambda(l)\|_F^2,
\end{equation*}
and, for each $h$, the lags $h'-h$, $h' \in \mathcal{I}_1$, are distinct, so that $\mathbb{E}\|\hat{\Lambda}-\Lambda\|_F^2 \leq c_{\Lambda}/n_1$. By Chebyshev's inequality for $\|\hat{\Lambda}-\Lambda\|_F$, Markov's inequality for the other five variables and the union bound, with probability at least $1-\zeta_{2n_1}$,
\begin{gather*}
\|\hat{\Lambda}-\Lambda\|_F \leq \frac{\lambda_V}{2}, \qquad \|\hat{\Xi}\|_F^2 \leq \frac{\mathrm{tr}(\Sigma)\mathrm{tr}(\Lambda)}{n_1^{5/6}}, \qquad \|\tilde{\Sigma}-\Sigma\|_F^2 \leq \frac{2\{\mathrm{tr}(\Sigma)\}^2}{n_1^{5/6}},\\
|\bar{\boldsymbol{\varepsilon}}|^2 \leq \frac{\mathrm{tr}(\Sigma)}{n_1^{5/6}}, \qquad \varsigma_V \leq 2\,\mathrm{tr}(\Lambda)\,n_1^{1/6}, \qquad \varsigma_{\varepsilon} \leq 2\,\mathrm{tr}(\Sigma)\,n_1^{1/6}.
\end{gather*}
On this event, the smallest eigenvalue of $\hat{\Lambda}$ is at least $\lambda_V/2$ by Weyl's inequality, so that $\|\hat{\Lambda}^{-1}\| \leq 2/\lambda_V$ and $\|\hat{\Theta}\|_F^2 \leq \|\hat{\Xi}\|_F^2\|\hat{\Lambda}^{-1}\|^2 \leq 4\,\mathrm{tr}(\Sigma)\kappa^2/\{\mathrm{tr}(\Lambda)\,n_1^{5/6}\}$. By \eqref{eq:ex_S}, $\|\Sigma\| \leq \mathrm{tr}(\Sigma)$ and $1/(n_1-1) \leq 2/n_1$,
\begin{equation*}
\|\hat{\Sigma}-\Sigma\| \leq \mathrm{tr}(\Sigma)\left(2\sqrt{2}+12\kappa+4+2\right)n_1^{-5/12} \leq L_1 n_1^{-5/12}, \qquad \text{hence} \qquad \hat{\theta} \leq \frac{L_1}{\lambda_{\Sigma}\,n_1^{5/12}} \leq \frac{1}{2}.
\end{equation*}
Thus $s^{(j)}>0$ on $\mathcal{Q}_j$ and, averaging \eqref{eq:ex_R} over $t=\omega_d$, $d \in \mathcal{D}_{\Pi}$, or taking $t=T+1$, both $K^{-1}\sum_{d \in \mathcal{D}_{\Pi}}(R_{\omega_d}-R^*_{\omega_d})^2$ and $(R_{T+1}-R^*_{T+1})^2$ are at most
\begin{equation*}
\frac{4}{\lambda_{\Sigma}}\left(\|\hat{\Theta}\|_F^2\varsigma_V+\hat{\theta}^2\varsigma_{\varepsilon}\right) \leq \frac{4}{\lambda_{\Sigma}}\left(8\,\mathrm{tr}(\Sigma)\kappa^2+\frac{2\,\mathrm{tr}(\Sigma)L_1^2}{\lambda_{\Sigma}^2}\right)n_1^{-2/3}=L_2^2\,n_1^{-2/3}=\delta_{2n_1}^2,
\end{equation*}
which is (A). The last statement of the lemma follows by substitution in \eqref{eq:bound}.
\end{proof}

Lemma~\ref{lem:example} has a limited scope. The point predictor is a correctly specified VAR with Gaussian innovations: up to the initialisation of the recursion, the designs of \ref{app:sim_design} with the VAR(2) predictor have this form, except that their innovations are Student's $t$ with four degrees of freedom (with scale matrix, not covariance matrix, denoted there by $\Sigma$), which have no finite fourth moment and are not covered. The oracle score is built on the true regression map, so that the oracle scores are functions of the i.i.d.\ innovations: (E) then follows from the Dvoretzky--Kiefer--Wolfowitz inequality for any split chosen independently of the data, the random one included, with no mixing condition, and the serial dependence enters only through (A). An oracle defined through the limit of a misspecified predictor would require a mixing argument and is not covered. The rate $n_1^{-1/6}$ is obtained from Markov's and Chebyshev's inequalities only and is not claimed to be sharp, and the constants are large: the bound is smaller than one only if $n_1>5^6$. The lemma shows that the conditions of Proposition~\ref{prop:bound} can hold with explicit rates, not that the coverage error is small at the sample sizes of Sections~\ref{sec:simulation} and~\ref{sec:application}.

\paragraph{A guarantee without condition (A)} With a chronological split and $b=1$, BSC is split conformal prediction with training, calibration and test points ordered in time. If the pairs $\boldsymbol{Z}_t$ are strictly stationary and $\beta$-mixing, Theorem~4 of \citet{oliveira2024split} applies: for any real-valued score computed from the training set, hence for any point predictor, consistent or not, the coverage of their set at level $1-\alpha$ is at least $1-\alpha-\eta_{\beta}$, where $\eta_{\beta}$ depends on the mixing coefficients and on $n_2$ and not on the accuracy of the predictor; their set uses the $\lceil n_2(1-\alpha)\rceil$-th smallest calibration score, which is not larger than the $\lceil (n_2+1)(1-\alpha)\rceil$-th smallest used by BSC, so that the lower bound holds for BSC. If, in addition, the score of an independent copy of the test point has a continuous distribution given the training set, their two-sided bound, applied at the level $\lceil (n_2+1)(1-\alpha)\rceil/n_2$ (smaller than 1 if $\alpha \geq 2/(n_2+1)$; for smaller $\alpha$ the upper bound below exceeds 1 and holds trivially), for which their set is the band of BSC, gives a coverage of at most $1-\alpha+(2-\alpha)/n_2+\eta_{\beta}$. This result requires neither an oracle score nor conditions (A) and (D), and its bound is explicit in the mixing coefficients. Proposition~\ref{prop:bound} covers every split chosen independently of the data and every $b$, under the high-level conditions (E), (A) and (D), with strong mixing, which is weaker than $\beta$-mixing, behind (E) for a chronological split. Apart from the Gaussian example of Lemma~\ref{lem:example}, in which (E) holds for every split because the oracle scores are independent, neither result comes with primitive conditions for the random split, which is used in all the experiments of Sections~\ref{sec:simulation} and~\ref{sec:application}; these experiments, with Student's $t$ innovations and, for most point predictors, misspecification, or with real data, are outside the scope of the available primitive conditions.

\paragraph{Modulation functions and computational cost} The factor $1/(n_1-1)$ in \eqref{eq:sd} does not affect the band, since multiplying all $s^{(j)}$ by a positive constant divides $k$ by it. The computational cost of BSC is mainly that of fitting the point predictor once: since the band \eqref{eq:band} is in closed form, no refitting is needed for candidate values of $\boldsymbol{Y}_{T+1}$, and, on a fixed grid, computing $s^{(j)}$, the scores and $k$ requires $O(T+K\log K)$ operations.

\paragraph{Conventions and nesting in Procedure~2} The conventions in Step~3 of Procedure~2 follow from the definition $\{\boldsymbol{y} \in \mathcal{Y}: \hat{p}(\boldsymbol{y})>\alpha_t\}$ of Section~\ref{sec:procedure}, with $b=1$, applied to any real level, since $1/(n_2+1) \leq \hat{p}(\boldsymbol{y}) \leq 1$; for $\alpha_t \in [1/(n_2+1),1)$, this set is the band of Step~3 by the argument given above for the closed form of the band. For a given window, the band is nested in the level, since $k_t$ is non-decreasing in $\lceil (n_2+1)(1-\alpha_t)\rceil$: as intended in ACI, a miss does not narrow the band, and a covered outcome does not widen it.

\paragraph{Procedure~3: prediction set} Let
\begin{equation*}
\hat{p}_{\mathrm{w}}(\boldsymbol{y}):=\tilde{w}_{n_2+1}+\sum_{d \in \mathcal{D}_{\mathrm{cal}}:\, R_{\omega_d} \geq R_{T+1}(\boldsymbol{y})}\tilde{w}_d, \qquad \hat{F}_{\mathrm{w}}(r):=\sum_{d \in \mathcal{D}_{\mathrm{cal}}:\, R_{\omega_d} \leq r}\tilde{w}_d, \quad r \in \mathbb{R},
\end{equation*}
be the weighted counterpart of the p-value \eqref{eq:pvalue} and the weighted distribution function of the calibration scores, a non-decreasing, right-continuous step function with finitely many jumps, so that $k_{\rho}$ in \eqref{eq:wquantile} equals $\inf\{r \in \mathbb{R}: \hat{F}_{\mathrm{w}}(r) \geq 1-\alpha\}$. Thus $k_{\rho}$ is the smallest $(1-\alpha)$-quantile of the distribution with mass $\tilde{w}_d$ at $R_{\omega_d}$, $d \in \mathcal{D}_{\mathrm{cal}}$, and $\tilde{w}_{n_2+1}$ at $+\infty$, i.e.\ the threshold of the nonexchangeable split conformal prediction of \citet{barber2023beyond}, whose prediction set, for a general nonconformity score, is $\{\boldsymbol{y} \in \mathcal{Y}: R_{T+1}(\boldsymbol{y}) \leq k_{\rho}\}$. For $R \in [0,+\infty]$, $R \leq k_{\rho}$ if and only if $\hat{F}_{\mathrm{w}}(r)<1-\alpha$ for all $r<R$, i.e.\ if and only if $\sum_{d \in \mathcal{D}_{\mathrm{cal}}:\, R_{\omega_d}<R}\tilde{w}_d<1-\alpha$; since the normalised weights $\tilde{w}_d$, $d=1,\dots,n_2+1$, sum to one, this is equivalent to $\tilde{w}_{n_2+1}+\sum_{d \in \mathcal{D}_{\mathrm{cal}}:\, R_{\omega_d} \geq R}\tilde{w}_d>\alpha$. With $R=R_{T+1}(\boldsymbol{y})$, which is non-negative by \eqref{eq:score},
\begin{equation*}
\{\boldsymbol{y} \in \mathcal{Y}: \hat{p}_{\mathrm{w}}(\boldsymbol{y})>\alpha\}=\{\boldsymbol{y} \in \mathcal{Y}: R_{T+1}(\boldsymbol{y}) \leq k_{\rho}\}.
\end{equation*}
If $k_{\rho}<+\infty$, then, by \eqref{eq:score}, $R_{T+1}(\boldsymbol{y}) \leq k_{\rho}$ is equivalent to the conditions in \eqref{eq:band} with $k_{\rho}$ in place of $k$; if $k_{\rho}=+\infty$, it holds for every $\boldsymbol{y}$. Hence the prediction set of nonexchangeable split conformal prediction is $\mathcal{C}^{\rho}_{T,1-\alpha}(\boldsymbol{x}_{T+1})$, which is nested in the level, since $k_{\rho}$ is non-decreasing in $1-\alpha$; as in Procedure~1, a candidate value whose score equals $k_{\rho}$ is in the band. Moreover, $k_{\rho}=+\infty$ if and only if $\hat{F}_{\mathrm{w}}(r)<1-\alpha$ for all $r$, which, if the calibration scores are finite, is equivalent to $W/(W+1)<1-\alpha$, or $W<(1-\alpha)/\alpha$; otherwise, the infimum is attained at a calibration score.

\paragraph{Procedure~3 with $\rho=1$} If $\rho=1$, then $w_d=1$, $W=n_2$ and $\tilde{w}_d=1/(n_2+1)$ for $d=1,\dots,n_2+1$, so that $\hat{p}_{\mathrm{w}}$ is the p-value \eqref{eq:pvalue} with $b=1$, and $\mathcal{C}^{1}_{T,1-\alpha}(\boldsymbol{x}_{T+1})$ is the band of Procedure~1 with $b=1$. Directly, $\hat{F}_{\mathrm{w}}(r) \geq 1-\alpha$ if and only if at least $\lceil (n_2+1)(1-\alpha)\rceil$ calibration scores are not larger than $r$, so that $k_1$ is the $\lceil (n_2+1)(1-\alpha)\rceil$-th smallest of the $n_2$ calibration scores if $\lceil (n_2+1)(1-\alpha)\rceil \leq n_2$, i.e.\ if $\alpha \geq 1/(n_2+1)$, and $k_1=+\infty$ otherwise, in agreement with the rule $W<(1-\alpha)/\alpha$, which, for $W=n_2$, reads $\alpha<1/(n_2+1)$. In the code, the cumulative sums of the unnormalised weights are compared with $(1-\alpha)(W+1)$, since summing the normalised weights $1/(n_2+1)$ in floating-point arithmetic can select a different order statistic.

\paragraph{Derivation of Proposition~\ref{prop:weighted}} Condition on the split: since it is independent of the data, the joint distribution of $\boldsymbol{Z}_1,\dots,\boldsymbol{Z}_{T+1}$ is unchanged, and the positions $\omega_d$ and the weights $w_d \in (0,1]$ are fixed. Consider the $n_2+1$ data points whose responses are the scores $R_{\omega_d}$, $d \in \mathcal{D}_{\mathrm{cal}}$, and, for the test point, $R_{T+1}$, with trivial covariates, and the model that predicts $0$, which does not depend on any data and is therefore pre-fitted in the sense of \citet{barber2023beyond}. The absolute residuals are the scores, which are non-negative; the nonexchangeable split conformal interval with weights $w_d$ and weight $1$ for the test point is $[-k_{\rho},k_{\rho}]$, whose coverage event $\{R_{T+1} \leq k_{\rho}\}$ is the event $\{\boldsymbol{Y}_{T+1} \in \mathcal{C}^{\rho}_{T,1-\alpha}(\boldsymbol{X}_{T+1})\}$ by the previous paragraphs; the vector of residuals is $\boldsymbol{R}$, and its version after swapping the test point with the point at position $d$ is $\boldsymbol{R}^{(d)}$. Their coverage bounds, stated for full conformal prediction and, as they note, valid for split conformal prediction, require fixed weights in $[0,1]$ and no assumption on the joint distribution of the data points; the lower bound gives \eqref{eq:weighted_bound}, and the upper bound, which requires residuals that are distinct with probability one, gives the second statement of Proposition~\ref{prop:weighted}. The joint distribution of the scores is arbitrary: it may reflect serial dependence and the dependence of all scores on the training set through $\hat{\mu}^{(j)}_{\mathcal{I}_1}$ and $s^{(j)}$. Since these are fitted on the training set only, $\boldsymbol{R}^{(d)}$ is the vector of scores computed from the data in which $\boldsymbol{Z}_{\omega_d}$ and $\boldsymbol{Z}_{T+1}$ exchange their positions. With a random split, averaging over the split gives the unconditional lower bound $1-\alpha-\mathbb{E}[\sum_{d \in \mathcal{D}_{\mathrm{cal}}}\tilde{w}_d\,\mathsf{d}_{\mathsf{TV}}(\boldsymbol{R},\boldsymbol{R}^{(d)} \mid \mathcal{I}_1,\mathcal{I}_2)]$, in which the total variation distances are between conditional distributions given the split and the expectation is over the split. For $\rho=1$, Proposition~\ref{prop:weighted} gives their bound for the unweighted method applied to Procedure~1 with $b=1$, $\mathbb{P}(\boldsymbol{Y}_{T+1} \in \mathcal{C}_{T,1-\alpha}(\boldsymbol{X}_{T+1}) \mid \mathcal{I}_1,\mathcal{I}_2) \geq 1-\alpha-(n_2+1)^{-1}\sum_{d \in \mathcal{D}_{\mathrm{cal}}}\mathsf{d}_{\mathsf{TV}}(\boldsymbol{R},\boldsymbol{R}^{(d)})$, which requires none of the conditions of Proposition~\ref{prop:bound} but, like them, involves unknown quantities.

\paragraph{Weights of Procedure~3} The weights must be fixed before the data are observed: a value of $\rho$ (or of $\alpha$) chosen after inspecting the coverage obtained on the same data is not covered by Proposition~\ref{prop:weighted}; data-dependent weights would require the conditional version of the bound discussed by \citet[Section~4.5]{barber2023beyond}. Since the weight of the test point is fixed to $1$, the normalised weights change if all $w_d$ are multiplied by a constant; hence, on a rolling window with delay $\tau$, measuring the time distance from time $T+1$ of the window, $T+1-\omega_d$, instead of from the target time point, $T+1+\tau-\omega_d$, would give a different procedure. The time distances are distinct positive integers, so that $W \leq \sum_{h=1}^{\infty}\rho^h=\rho/(1-\rho)$ for $\rho<1$ (99, 49 and 19 for $\rho=0.99$, $0.98$ and $0.95$) whatever $n_2$ and $T$, and $\tilde{w}_{n_2+1}=1/(W+1) \geq 1-\rho$: unlike for Procedure~1, the upper bound of Proposition~\ref{prop:weighted} does not approach $1-\alpha$ as $n_2 \rightarrow \infty$, even under exchangeability. The effective number of calibration scores reported in \ref{app:sim_seq} and \ref{app:app_extra}, $(\sum_{d} w_d)^2/\sum_{d} w_d^2$, equals $n_2$ for $\rho=1$ and depends only on the split; the version that includes the test point, $(W+1)^2/(\sum_{d} w_d^2+1)$, which these authors use in their Section~5.3, equals $n_2+1$ for $\rho=1$.

\paragraph{Procedure~3 for dependent data} The explicit bounds of \citet{barber2023beyond} on the total variation terms under a bounded drift and after a change point assume independent data points; they do not apply here, since $\boldsymbol{X}_t$ contains lagged responses, so that the observations, and the scores even given the training set, are dependent. Even for strictly stationary data, the total variation terms can be positive: exchanging the test score with a calibration score changes its position relative to the other scores and to the training observations on which it depends, and, with a random split, a calibration observation can be followed by training observations whose lagged values contain its response, which may make its score smaller than that of the test point. These discrepancies need not grow with the time distance, so that weights decreasing with it do not remove them; a chronological split removes the discrepancy due to the random split. Summing the lower bounds \eqref{eq:weighted_bound} of the bands of a deployment period gives a bound on the expected frequency of misses only, which involves the unknown total variation terms of all the bands; unlike Proposition~\ref{prop:aci}, no statement holds for every sequence of data.
\section{Design details and additional results of the simulation study}
\setcounter{figure}{0}\setcounter{table}{0}
\renewcommand*{\theHfigure}{app.\thefigure}\renewcommand*{\theHtable}{app.\thetable}

This appendix gives the details of the design of the simulation study of Section~\ref{sec:simulation} (\ref{app:sim_design}) and the complete results of Experiment~1 (\ref{app:sim_static}) and Experiment~2 (\ref{app:sim_seq}).

\subsection{Design}
\label{app:sim_design}

\paragraph{Data generating processes} For $p=1$, the functional response is $Y_t^{(1)}(q)=\boldsymbol{g}(q)^{\top}\boldsymbol{\xi}_t$, $q \in \mathcal{Q}_1=[0,1]$, where $\boldsymbol{g}(q)=(1,\sqrt{2}\sin(2\pi q),\sqrt{2}\cos(2\pi q))^{\top}$ and the coefficient vector $\boldsymbol{\xi}_t \in \mathbb{R}^3$ follows the VAR(2) process
\begin{equation*}
\boldsymbol{\xi}_t=\Psi_1\boldsymbol{\xi}_{t-1}+\Psi_2\boldsymbol{\xi}_{t-2}+\boldsymbol{\varepsilon}_t,
\end{equation*}
with $\Psi_i=\Upsilon_i/(2\|\Upsilon_i\|_F)$, $i=1,2$, where $\|\cdot\|_F$ is the Frobenius norm,
\begin{equation*}
\Upsilon_1=
\begin{bmatrix}
0.8 & 0.3 & 0.3\\
0.3 & 0.8 & 0.3\\
0.3 & 0.3 & 0.8
\end{bmatrix},
\qquad
\Upsilon_2=
\begin{bmatrix}
0.5 & 0.1 & 0.1\\
0.1 & 0.5 & 0.1\\
0.1 & 0.1 & 0.5
\end{bmatrix},
\end{equation*}
and the innovations $\boldsymbol{\varepsilon}_t$ are i.i.d.\ multivariate Student's $t$ random vectors with $\nu=4$ degrees of freedom and scale matrix
\begin{equation*}
\Sigma=
\begin{bmatrix}
0.5 & 0.3 & 0.3\\
0.3 & 0.5 & 0.3\\
0.3 & 0.3 & 0.5
\end{bmatrix},
\end{equation*}
i.e.\ with covariance matrix $2\Sigma$ and heavy tails. The largest modulus of the eigenvalues of the companion matrix is $0.885$, so that the process is stable. The recursion is started from $\boldsymbol{\xi}_{-B-1}=\boldsymbol{\xi}_{-B}=\boldsymbol{0}$, and the first $B=200$ generated vectors are discarded as a burn-in, so that the effect of the initial values on the series used in the study is negligible.

\begin{figure}[tbp]
\centering
\includegraphics[width=\textwidth]{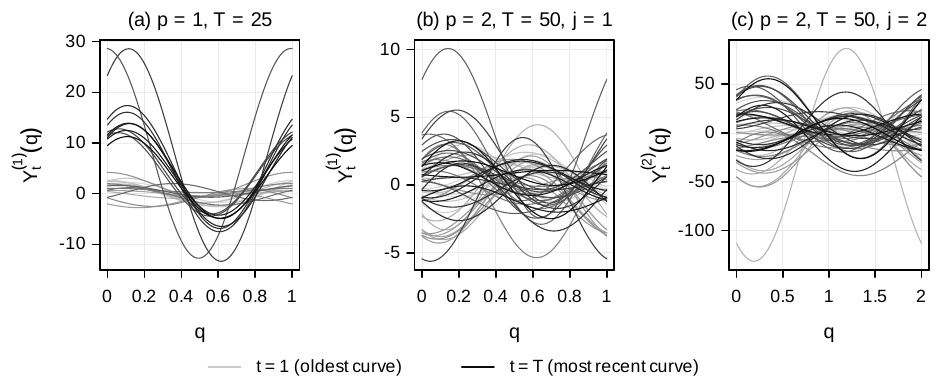}
\caption{Simulated series (first replication of the designs with $(p,T,n_2)=(1,25,7)$ and $(2,50,23)$); older curves are lighter. (a) $p=1$: curves $Y_t^{(1)}(q)$, $t=1,\dots,T$; the curves from $t=16$ onwards are large because of two large innovations, at $t=16$ and $t=19$ (heavy tails), propagated by the VAR dynamics. (b) and (c) $p=2$: first component, $q \in [0,1]$, and second component, $q \in [0,2]$.}
\label{fig:sim_data}
\end{figure}

For $p=2$, $\boldsymbol{\xi}_t=(\boldsymbol{\xi}_t^{(1)\top},\boldsymbol{\xi}_t^{(2)\top})^{\top} \in \mathbb{R}^6$ follows a VAR(2) process of the same form, with $\Psi_i=U_i/(2\|U_i\|_F)$, $U_i=C \otimes \Upsilon_i$, and innovations with scale matrix $C \otimes \Sigma$, where $\otimes$ denotes the Kronecker product and
\begin{equation*}
C=
\begin{bmatrix}
1 & 0.5\\
0.5 & 1
\end{bmatrix};
\end{equation*}
the largest modulus of the eigenvalues of the companion matrix is $0.855$, and the initial values and the burn-in are as for $p=1$. The two components of the response,
\begin{align*}
Y_t^{(1)}(q)&=\boldsymbol{g}(q)^{\top}\boldsymbol{\xi}_t^{(1)}, & q &\in \mathcal{Q}_1=[0,1],\\
Y_t^{(2)}(q)&=10\,(1,\sqrt{2}\sin(\pi q),\sqrt{2}\cos(\pi q))\,\boldsymbol{\xi}_t^{(2)}, & q &\in \mathcal{Q}_2=[0,2],
\end{align*}
are dependent, as the offer and demand curves of Section~\ref{sec:application}, and differ in domain and in scale, as allowed in Section~\ref{sec:setting} (Figure~\ref{fig:sim_data}).

\paragraph{Point predictors} The point predictors use the $r$ most recent observations as covariates:
\begin{itemize}
\item \textit{Oracle} ($r=2$): the regression function, e.g.\ $\boldsymbol{g}(q)^{\top}(\Psi_1\boldsymbol{\xi}_{t-1}+\Psi_2\boldsymbol{\xi}_{t-2})$ for $p=1$.
\item \textit{VAR($r$)}, $r \in \{1,2,3\}$: with known basis functions, a VAR($r$) model without intercept ($3pr$ coefficients per equation) is fitted to $\boldsymbol{\xi}_t$ by least squares, and the predicted coefficients are mapped to curves \citep{chen2021review}.
\item \textit{FAR($r$)}, $r \in \{1,2,3\}$: the concurrent functional autoregressive model \citep{ramsay_functional_2005} $Y_t^{(j)}(q)=\sum_{i=1}^{r}\beta_i^{(j)}(q)\,Y_{t-i}^{(j)}(q)+a_t^{(j)}(q)$, $q \in \mathcal{Q}_j$, with zero-mean error $a_t^{(j)}$, fitted by least squares at each grid point; for $p=2$, each component is regressed on its own lags.
\end{itemize}
VAR(2) is correctly specified; VAR(1) omits a relevant lagged variable and VAR(3) includes an irrelevant one, while the FAR models omit relevant regressors, namely the previous curves at points other than $q$ and, for $p=2$, the other component \citep{rao1971some}. The residuals of the Oracle are functions of the i.i.d.\ innovations and the split does not depend on the data, so that its $K$ scores are i.i.d.\ given the split and $s^{(1)},\dots,s^{(p)}$, and its coverage is exactly $\lceil K(1-\alpha)\rceil/K=0.75$ (Section~\ref{sec:guarantees}): the Oracle checks the implementation and gives a reference band size, but it is not informative on the effect of dependence.

\paragraph{Split and designs} We use the random split of Section~\ref{sec:procedure} with common random numbers: in each replication, we draw $U_1,\dots,U_T$ i.i.d.\ uniform on $(0,1)$, shared by all point predictors and values of $b$, and, for a point predictor with $r$ lags ($t_0=r$), $\mathcal{I}_1$ consists of the $n_1=T-n_2-r$ indices in $\{r+1,\dots,T\}$ with the smallest $U_t$, and $\mathcal{I}_2$ of the other $n_2$ indices. The curves are evaluated on $201$ equispaced points of each $\mathcal{Q}_j$, on which the essential supremum in \eqref{eq:score} is computed as a maximum and the integrals by the trapezoidal rule. We consider $T \in \{25,50,100,1000\}$ and $b \in \{1,3,6\}$, and choose $n_2$ so that (i) $K=(n_2+1)/b$ is an integer; (ii) $\alpha K$ is an integer, so that the coverage under exchangeability is exactly $1-\alpha$ and deviations cannot be due to the discreteness of the conformal quantile; (iii) $\alpha \geq 1/K$, without which the band is the whole space; (iv) where possible, every point predictor is admissible, i.e.\ has at least three residual degrees of freedom per equation or grid point (next paragraph). With $\alpha=0.25$, (i)--(iii) hold if and only if $n_2+1$ is a multiple of $4b$. For $T \in \{50,100,1000\}$, we use the same $n_2$ for all $b$, i.e.\ $n_2+1$ a multiple of $24$: $n_2=23$, $47$ and $479$. For $T=25$, $n_2=23$ would leave $n_1 \leq 1$, so that $(T,b)=(25,6)$ is not feasible. For $b=3$, we use $n_2=11$, the only feasible value, with which VAR(3) ($n_1=11$, $9$ coefficients per equation) is not admissible; for $b=1$, we use $n_2=7$, with which all point predictors are admissible. As $n_1$ also differs, the two designs with $T=25$ are not compared across $b$. With $K=4$, $k$ is the largest of three calibration scores. For $p=2$, $T=25$ is excluded, since the VAR(3) model has $18$ coefficients per equation and $n_1 \leq 19$.

\paragraph{Admissibility of the point predictors} For the Oracle and the VAR models, $s^{(j)}(q)^2$ is a quadratic form in the basis functions at $q$, with matrix $\tilde{E}_j^{\top}\tilde{E}_j/(n_1-1)$, $\tilde{E}_j$ being the $n_1 \times 3$ matrix of centred training residuals of the basis coefficients; for VAR($r$), $\tilde{E}_j$ has rank at most $n_1-3pr$ and, if this is smaller than $3$, $s^{(j)}(q)$ can be close to zero and the band very wide. We therefore fit VAR($r$) only if $n_1-3pr \geq 3$, and FAR($r$), for which $s^{(j)}(q)>0$ almost surely, only if $n_1-r \geq 3$; this excludes only VAR(3) for $(p,T,b)=(1,25,3)$, which, in the first $1000$ replications of its design, would have median band size $45.9$ ($11.5$ for VAR(2)) and median ratio $\min_q s^{(1)}(q)/\max_q s^{(1)}(q)$ of $0.109$ ($0.29$ to $0.32$ for the admissible models). A grid of $2001$ instead of $201$ points changes the median band size of every admissible model by less than $0.07\%$ (first $1000$ replications, $(p,T) \in \{(1,25),(2,50)\}$).

\paragraph{Miscoverage level} The miscoverage level of the bands is $\alpha=0.25$, i.e.\ the nominal coverage level is $0.75$. Conditions (ii) and (iii) require $n_2+1 \geq b/\alpha$, so that a smaller $\alpha$ requires longer calibration sets: for $b=6$, $\alpha=0.1$ is not possible with $T=50$, nor $\alpha=0.05$ with $T=100$. With $\alpha=0.25$, all values of $b$ can be studied with $T \geq 50$; $0.25$ is also one of the levels used in Section~\ref{sec:application}. The coverage at other levels follows from the computed p-values (Section~\ref{sec:procedure}): for $T=1000$, it is between $0.893$ and $0.905$ for $\alpha=0.1$ and between $0.945$ and $0.951$ for $\alpha=0.05$ (SE at most $0.0044$), levels at which $\alpha K$ is an integer for all $b$.

\paragraph{Replications and performance measures} For each design $(p,T,n_2)$, we generate $N=5000$ independent series of length $T+1$, each used for all point predictors, values of $b$ and, for Procedure~3, values of $\rho$. We report the empirical coverage $c$ and its Monte Carlo standard error (SE) $\{c(1-c)/N\}^{1/2}$, between $0.0060$ and $0.0063$, and flag the cases in which $0.75$ lies outside the $99\%$ Monte Carlo confidence interval $c \pm 2.576$ SE. The level $99\%$, which refers to this interval and not to the nominal coverage of the bands, limits the cases flagged by chance: if all coverages were $0.75$, about $1.4$ of the $139$ combinations of design, $b$ and point predictor would be flagged (about $7$ at $95\%$). The SE, not the standard deviation $\{c(1-c)\}^{1/2}\approx 0.43$ of the coverage indicators, which describes single replications, measures the precision of $c$. The band size $\sum_{j=1}^p \int_{\mathcal{Q}_j} 2k\,s^{(j)}(q)\,\mathrm{d}q$ is right-skewed and is summarised by its median and interquartile range (IQR).

The study is implemented in R \citep{R_cit}, version 4.3.3, with one random-number substream per replication. On four cores, Experiment~1 takes about 8 minutes for Procedure~1 and 7 minutes for Procedure~3, and Experiment~2, with the three procedures, about 23 ($p=1$) and 38 ($p=2$) minutes. The code is available at \codelocation.

\subsection{Experiment~1}
\label{app:sim_static}

\begin{figure}[tbp]
\centering
\includegraphics[width=0.84\textwidth]{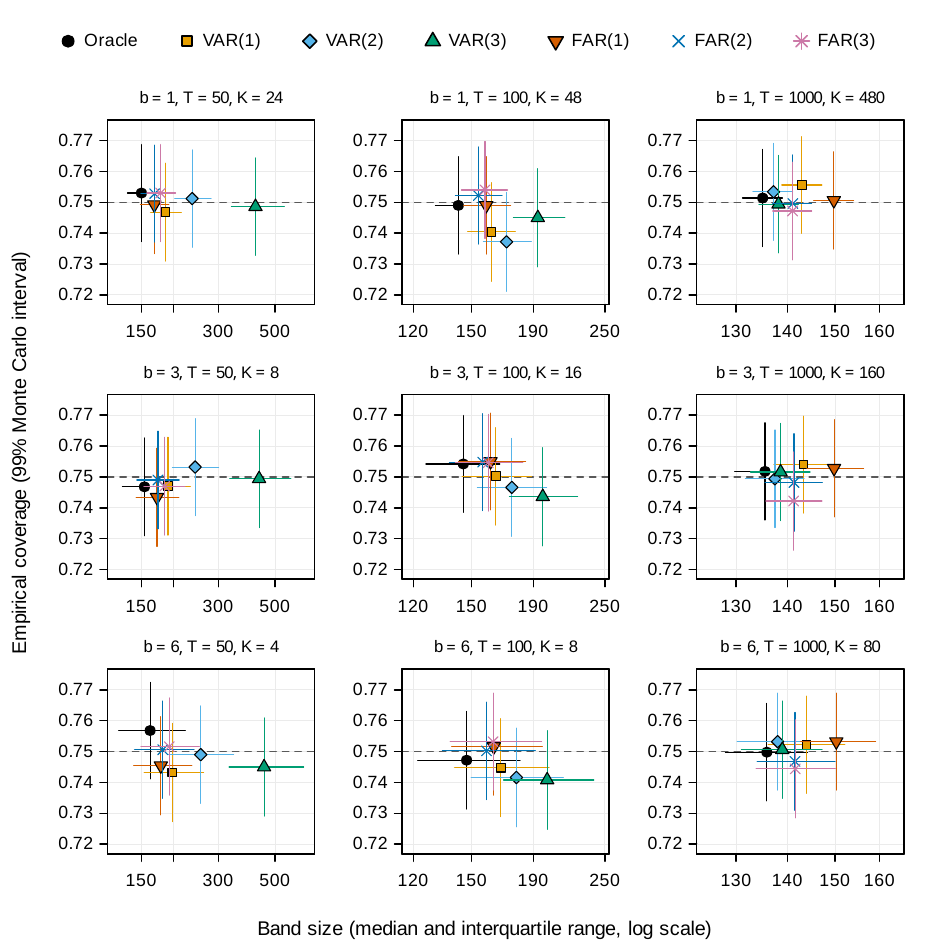}\\[2pt]
\includegraphics[width=0.92\textwidth]{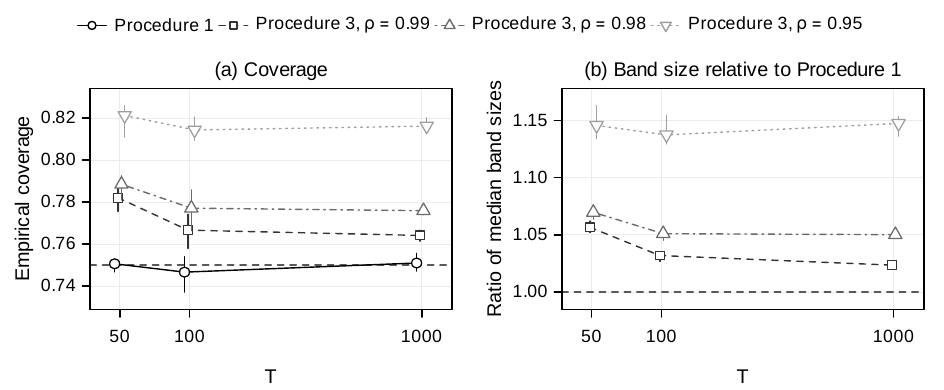}
\caption{Experiment~1, $p=2$. Top: empirical coverage and band size of Procedure~1; bottom: Procedures~1 and~3 with $b=1$, against $T$; panels, markers and segments as in Figure~\ref{fig:sim_cov_size_p1}.}
\label{fig:sim_cov_size_p2}
\end{figure}

\begin{figure}[tbp]
\centering
\includegraphics[width=\textwidth]{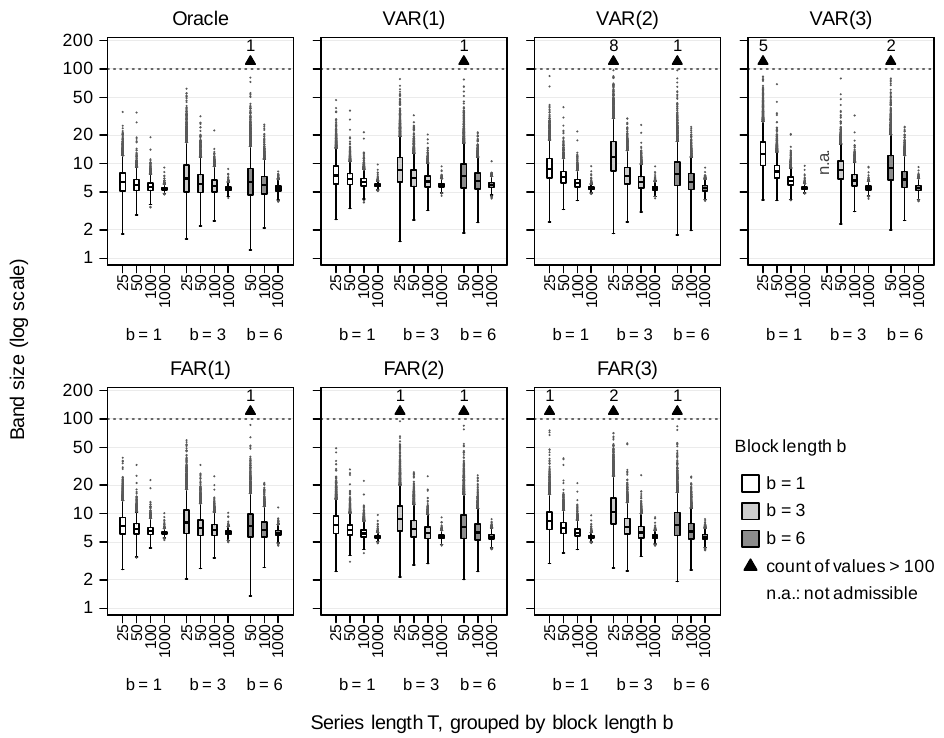}
\caption{Experiment~1: boxplots of the $N=5000$ band sizes of Procedure~1, $p=1$, one panel per point predictor; boxes grouped by $b$ and ordered by $T$; whiskers: most extreme values within $1.5$ IQR of the box. Values are truncated at $100$ (dotted line); the number above a triangle is the number of sizes larger than $100$. n.a.: not admissible.}
\label{fig:sim_size_p1}
\end{figure}

Tables~\ref{tab:cov_p1} and~\ref{tab:cov_p2} report the empirical coverage of Procedure~1 and Tables~\ref{tab:size_p1} and~\ref{tab:size_p2} the median and the IQR of the band size; Figure~\ref{fig:sim_cov_size_p2} shows coverage and band size jointly for $p=2$, as Figure~\ref{fig:sim_cov_size_p1} for $p=1$, and Figure~\ref{fig:sim_size_p1} the distribution of the band size for $p=1$. For $T \geq 50$, $b \in \{3,6\}$ changes the coverage obtained with $b=1$ by $-0.007$ to $0.010$ for the estimated point predictors and by $-0.006$ to $0.005$ for the Oracle, whose coverage is exactly $0.75$. Table~\ref{tab:static_wcp} reports the results of Procedure~3, computed in the same replications as Procedure~1 with $b=1$, whose coverages and median band sizes it reproduces; the bottom panels of Figures~\ref{fig:sim_cov_size_p1} and~\ref{fig:sim_cov_size_p2} show them against $T$. With $T=25$ and $\rho=0.95$, up to $1.5\%$ of its bands are the whole space ($W<(1-\alpha)/\alpha=3$); for $T \geq 50$ and for $\rho \in \{0.99,0.98\}$, none is.

\begin{table}[tbp]
\centering
\caption{Experiment~1: empirical coverage of Procedure~1, $p=1$, $\alpha=0.25$ (nominal coverage $0.75$), $N=5000$ replications per design. Columns $b$, $T$, $n_2$, $K$: block length, length of the series, calibration set size, number of blocks; $n_1=T-n_2-r$, with $r=2$ for the Oracle and $r$ equal to the order of the other models. Each cell: empirical coverage $c$, i.e.\ the fraction of the $N$ replications in which the band contains $\boldsymbol{y}_{T+1}$; its Monte Carlo SE $\{c(1-c)/N\}^{1/2}$ is between $0.0060$ and $0.0063$ in every cell. Bold: $0.75$ outside the $99\%$ Monte Carlo interval $c \pm 2.576$ SE. ---: model not admissible. The design $(T,b)=(25,6)$ is not feasible.}
\label{tab:cov_p1}
{\footnotesize\setlength{\tabcolsep}{4pt}
\begin{tabular}{rrrrccccccc}
\toprule
$b$ & $T$ & $n_2$ & $K$ & Oracle & VAR(1) & VAR(2) & VAR(3) & FAR(1) & FAR(2) & FAR(3) \\
\midrule
1 & 25 & 7 & 8 & 0.756 & 0.752 & 0.752 & 0.758 & 0.761 & 0.755 & 0.751 \\
 & 50 & 23 & 24 & 0.750 & \textbf{0.733} & 0.756 & 0.741 & 0.740 & 0.742 & 0.742 \\
 & 100 & 47 & 48 & 0.758 & 0.752 & 0.758 & 0.752 & 0.756 & 0.760 & 0.760 \\
 & 1000 & 479 & 480 & 0.753 & 0.747 & 0.754 & 0.754 & 0.744 & 0.748 & 0.750 \\
\midrule
3 & 25 & 11 & 4 & 0.755 & 0.738 & 0.754 & --- & \textbf{0.729} & 0.741 & 0.738 \\
 & 50 & 23 & 8 & 0.751 & \textbf{0.733} & 0.753 & 0.737 & 0.739 & 0.738 & 0.740 \\
 & 100 & 47 & 16 & 0.761 & 0.754 & 0.765 & 0.755 & 0.754 & 0.761 & 0.757 \\
 & 1000 & 479 & 160 & 0.754 & 0.744 & 0.754 & 0.755 & 0.743 & 0.748 & 0.749 \\
\midrule
6 & 50 & 23 & 4 & 0.747 & 0.736 & 0.749 & 0.742 & 0.737 & 0.736 & 0.741 \\
 & 100 & 47 & 8 & 0.755 & 0.749 & 0.759 & 0.756 & 0.753 & 0.761 & 0.760 \\
 & 1000 & 479 & 80 & 0.752 & 0.745 & 0.753 & 0.752 & 0.746 & 0.749 & 0.752 \\
\bottomrule
\end{tabular}
}
\end{table}

\begin{table}[tbp]
\centering
\caption{Experiment~1: empirical coverage of Procedure~1, $p=2$. Columns, entries and bold face as in Table~\ref{tab:cov_p1}.}
\label{tab:cov_p2}
{\footnotesize\setlength{\tabcolsep}{4pt}
\begin{tabular}{rrrrccccccc}
\toprule
$b$ & $T$ & $n_2$ & $K$ & Oracle & VAR(1) & VAR(2) & VAR(3) & FAR(1) & FAR(2) & FAR(3) \\
\midrule
1 & 50 & 23 & 24 & 0.753 & 0.747 & 0.751 & 0.749 & 0.749 & 0.753 & 0.753 \\
 & 100 & 47 & 48 & 0.749 & 0.740 & 0.737 & 0.745 & 0.749 & 0.752 & 0.754 \\
 & 1000 & 479 & 480 & 0.751 & 0.756 & 0.753 & 0.749 & 0.751 & 0.750 & 0.747 \\
\midrule
3 & 50 & 23 & 8 & 0.747 & 0.747 & 0.753 & 0.749 & 0.743 & 0.749 & 0.747 \\
 & 100 & 47 & 16 & 0.754 & 0.750 & 0.747 & 0.744 & 0.755 & 0.755 & 0.755 \\
 & 1000 & 479 & 160 & 0.752 & 0.754 & 0.749 & 0.752 & 0.753 & 0.748 & 0.742 \\
\midrule
6 & 50 & 23 & 4 & 0.757 & 0.743 & 0.749 & 0.745 & 0.745 & 0.751 & 0.752 \\
 & 100 & 47 & 8 & 0.747 & 0.745 & 0.742 & 0.741 & 0.752 & 0.750 & 0.753 \\
 & 1000 & 479 & 80 & 0.750 & 0.752 & 0.753 & 0.751 & 0.753 & 0.747 & 0.744 \\
\bottomrule
\end{tabular}
}
\end{table}

\paragraph{Coverage} The three flagged cells are those of VAR(1) with $p=1$, $T=50$ and $b \in \{1,3\}$ ($0.733$ for both values of $b$, which use the same series) and of FAR(1) with $p=1$, $T=25$ and $b=3$ ($0.729$); no cell is flagged for $T \geq 100$ or $p=2$. The coverage of the Oracle is between $0.747$ and $0.761$ and, for $T \geq 100$ or $p=2$, that of every point predictor is between $0.737$ and $0.765$. For $p=1$ and $T \leq 50$, the coverage of the correctly specified VAR(2) is between $0.749$ and $0.756$; that of the other estimated point predictors is between $0.751$ and $0.761$ for $(T,b)=(25,1)$, and between $0.729$ and $0.742$, slightly below $0.75$, in the other $19$ cells. These deviations are not explained by the number of estimated coefficients: with $9$ and $18$ coefficients per equation and $n_1=15$ and $24$, VAR(3) has coverage $0.758$ for $(p,T,b)=(1,25,1)$ and $0.745$ to $0.749$ for $p=2$ and $T=50$.

\paragraph{Band size} For $T=1000$, the median sizes differ by less than $1\%$ across $b$. For $T \leq 100$, a FAR model gives the smallest median size among the estimated point predictors (Tables~\ref{tab:size_p1} and~\ref{tab:size_p2}). The training set size has an effect of its own (Table~\ref{tab:size_p1}, Figure~\ref{fig:sim_size_p1}): in the designs with $p=1$ and $K=8$ ($(T,b)=(25,1)$, $(50,3)$, $(100,6)$), the median size of VAR(2) decreases from $8.82$ to $7.41$ and $6.44$ as $n_1$ increases from $16$ to $25$ and $51$, and that of the Oracle only from $6.36$ to $6.13$ and $5.93$. For $p=2$ (Table~\ref{tab:size_p2}, Figure~\ref{fig:sim_cov_size_p2}), the band size is dominated by the second component (median contributions $6.42$ and $128.6$ for the Oracle with $T=1000$, $b=1$), and, with $T=50$, the median size of VAR(3) is $2.8$ times that of the Oracle.

\begin{figure}[tbp]
\centering
\includegraphics[width=\textwidth]{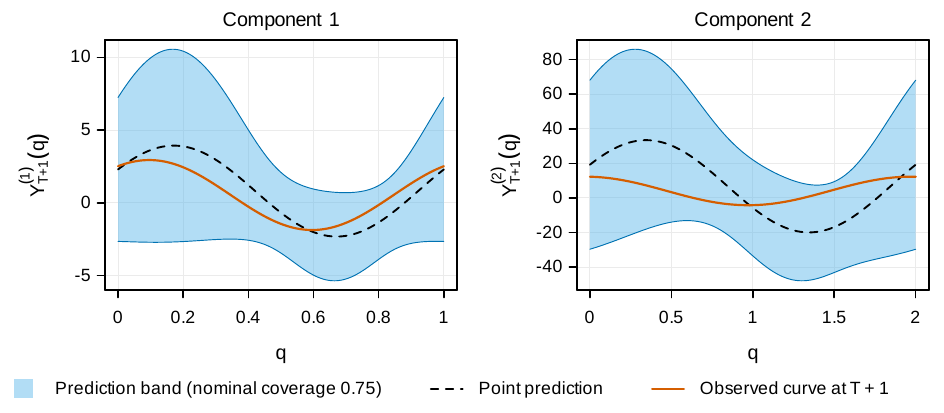}
\caption{Experiment~1: band of Procedure~1 for $p=2$ (shaded), nominal coverage $0.75$, first replication of the design with $T=100$, $n_2=47$, $b=1$, VAR(2) point predictor. Left: first component; right: second component. Dashed lines: point predictions $\hat{y}^{(j)}_{T+1}(q)$; solid lines: observed curves $y^{(j)}_{T+1}(q)$.}
\label{fig:sim_band_p2}
\end{figure}

Figure~\ref{fig:sim_band_p2} shows a band for $p=2$ ($k=3.152$, the $36$th smallest of $47$ calibration scores; test score $2.622$). Its half-width $k\,s^{(j)}(q)$ is smaller where the training residuals are less dispersed; the second component, with longer domain and larger scale, accounts for $147.4$ of the size $155.7$.

\begin{table}[tbp]
\centering
\caption{Experiment~1: band size of Procedure~1, $p=1$, $\alpha=0.25$, $N=5000$ replications per design. Columns as in Table~\ref{tab:cov_p1}. Each cell: median of the band size $\sum_{j=1}^{p}\int_{\mathcal{Q}_j} 2k\,s^{(j)}(q)\,\mathrm{d}q$ and, beneath it, its IQR. ---: model not admissible.}
\label{tab:size_p1}
{\footnotesize\setlength{\tabcolsep}{4pt}
\begin{tabular}{rrrrccccccc}
\toprule
$b$ & $T$ & $n_2$ & $K$ & Oracle & VAR(1) & VAR(2) & VAR(3) & FAR(1) & FAR(2) & FAR(3) \\
\midrule
1 & 25 & 7 & 8 & 6.36 & 7.50 & 8.82 & 12.5 & 7.33 & 7.53 & 8.38 \\
 & & & & (2.82) & (3.41) & (4.26) & (7.28) & (3.09) & (3.24) & (3.65) \\
\addlinespace[2pt]
 & 50 & 23 & 24 & 5.92 & 6.82 & 7.18 & 8.24 & 6.87 & 6.68 & 7.07 \\
 & & & & (1.59) & (1.84) & (2.04) & (2.51) & (1.76) & (1.70) & (1.86) \\
\addlinespace[2pt]
 & 100 & 47 & 48 & 5.67 & 6.34 & 6.21 & 6.53 & 6.55 & 6.16 & 6.28 \\
 & & & & (1.03) & (1.15) & (1.12) & (1.21) & (1.18) & (1.08) & (1.10) \\
\addlinespace[2pt]
 & 1000 & 479 & 480 & 5.44 & 5.91 & 5.49 & 5.51 & 6.24 & 5.68 & 5.68 \\
 & & & & (0.31) & (0.34) & (0.31) & (0.32) & (0.36) & (0.32) & (0.32) \\
\midrule
3 & 25 & 11 & 4 & 6.96 & 8.59 & 11.8 & --- & 8.05 & 8.80 & 10.5 \\
 & & & & (4.66) & (5.22) & (8.68) &  & (4.71) & (5.38) & (6.84) \\
\addlinespace[2pt]
 & 50 & 23 & 8 & 6.13 & 7.03 & 7.41 & 8.48 & 7.10 & 6.92 & 7.28 \\
 & & & & (2.72) & (2.84) & (3.09) & (3.75) & (2.76) & (2.73) & (2.82) \\
\addlinespace[2pt]
 & 100 & 47 & 16 & 5.79 & 6.47 & 6.35 & 6.64 & 6.68 & 6.22 & 6.32 \\
 & & & & (1.75) & (1.80) & (1.75) & (1.84) & (1.80) & (1.75) & (1.77) \\
\addlinespace[2pt]
 & 1000 & 479 & 160 & 5.43 & 5.90 & 5.48 & 5.51 & 6.24 & 5.67 & 5.68 \\
 & & & & (0.49) & (0.51) & (0.50) & (0.49) & (0.52) & (0.48) & (0.48) \\
\midrule
6 & 50 & 23 & 4 & 6.42 & 7.34 & 7.71 & 8.92 & 7.38 & 7.19 & 7.61 \\
 & & & & (4.22) & (4.41) & (4.54) & (5.42) & (4.27) & (4.16) & (4.36) \\
\addlinespace[2pt]
 & 100 & 47 & 8 & 5.93 & 6.57 & 6.44 & 6.79 & 6.78 & 6.33 & 6.48 \\
 & & & & (2.55) & (2.52) & (2.55) & (2.64) & (2.50) & (2.44) & (2.46) \\
\addlinespace[2pt]
 & 1000 & 479 & 80 & 5.45 & 5.92 & 5.49 & 5.53 & 6.25 & 5.69 & 5.69 \\
 & & & & (0.69) & (0.71) & (0.69) & (0.70) & (0.70) & (0.67) & (0.68) \\
\bottomrule
\end{tabular}
}
\end{table}

\begin{table}[tbp]
\centering
\caption{Experiment~1: band size of Procedure~1, $p=2$. Columns and entries as in Table~\ref{tab:size_p1}.}
\label{tab:size_p2}
{\footnotesize\setlength{\tabcolsep}{4pt}
\begin{tabular}{rrrrccccccc}
\toprule
$b$ & $T$ & $n_2$ & $K$ & Oracle & VAR(1) & VAR(2) & VAR(3) & FAR(1) & FAR(2) & FAR(3) \\
\midrule
1 & 50 & 23 & 24 & 150 & 186 & 237 & 419 & 169 & 169 & 178 \\
 & & & & (39.7) & (52.1) & (78.5) & (207) & (43.5) & (43.6) & (46.5) \\
\addlinespace[2pt]
 & 100 & 47 & 48 & 143 & 162 & 172 & 193 & 159 & 154 & 158 \\
 & & & & (26.1) & (29.7) & (31.5) & (38.3) & (28.7) & (27.6) & (27.8) \\
\addlinespace[2pt]
 & 1000 & 479 & 480 & 135 & 143 & 137 & 138 & 150 & 141 & 141 \\
 & & & & (7.80) & (8.30) & (7.88) & (7.89) & (8.66) & (7.96) & (7.94) \\
\midrule
3 & 50 & 23 & 8 & 154 & 191 & 244 & 433 & 173 & 174 & 185 \\
 & & & & (64.5) & (74.5) & (102) & (244) & (67.0) & (66.8) & (69.0) \\
\addlinespace[2pt]
 & 100 & 47 & 16 & 145 & 165 & 175 & 197 & 161 & 157 & 160 \\
 & & & & (41.2) & (44.1) & (46.5) & (51.7) & (41.7) & (40.9) & (41.5) \\
\addlinespace[2pt]
 & 1000 & 479 & 160 & 136 & 143 & 138 & 139 & 150 & 141 & 141 \\
 & & & & (11.6) & (12.2) & (11.7) & (11.9) & (12.3) & (11.8) & (11.5) \\
\midrule
6 & 50 & 23 & 4 & 162 & 198 & 256 & 453 & 179 & 182 & 193 \\
 & & & & (101) & (109) & (153) & (317) & (96.7) & (100) & (104) \\
\addlinespace[2pt]
 & 100 & 47 & 8 & 147 & 168 & 178 & 200 & 163 & 159 & 163 \\
 & & & & (58.6) & (61.3) & (63.4) & (69.9) & (57.7) & (57.3) & (57.9) \\
\addlinespace[2pt]
 & 1000 & 479 & 80 & 136 & 144 & 138 & 139 & 150 & 142 & 142 \\
 & & & & (16.2) & (16.0) & (16.2) & (16.3) & (16.4) & (16.0) & (16.3) \\
\bottomrule
\end{tabular}
}
\end{table}

\begin{table}[tbp]
\centering
\caption{Experiment~1: Procedure~3 with $b=1$, compared with Procedure~1 with $b=1$ in the same replications, $\alpha=0.25$ (nominal coverage $0.75$), $N=5000$ replications per design. $1/(W+1)$: mean mass of the test point; $n_{\mathrm{eff}}$: mean effective number of calibration scores $(\sum_d w_d)^2/\sum_d w_d^2$ ($n_2$ for Procedure~1); ranges over the seven point predictors of the empirical coverage of Procedure~3 (Coverage), of its difference from the coverage of Procedure~1 (Difference; SE at most $0.005$), and of the ratio of the median size of its finite bands to that of Procedure~1 (Size ratio).}
\label{tab:static_wcp}
{\footnotesize\setlength{\tabcolsep}{4pt}
\begin{tabular}{rrrrrrccc}
\toprule
$p$ & $T$ & $n_2$ & $\rho$ & $1/(W+1)$ & $n_{\mathrm{eff}}$ & Coverage & Difference & Size ratio \\
\midrule
1 & 25 & 7 & 0.99 & 0.139 & 7.0 & 0.868--0.874 & 0.106--0.123 & 1.316--1.392 \\
 &  &  & 0.98 & 0.153 & 6.9 & 0.869--0.876 & 0.108--0.125 & 1.321--1.400 \\
 &  &  & 0.95 & 0.201 & 6.4 & 0.875--0.881 & 0.113--0.130 & 1.333--1.423 \\
\addlinespace[2pt]
 & 50 & 23 & 0.99 & 0.052 & 22.6 & 0.766--0.788 & 0.028--0.036 & 1.055--1.070 \\
 &  &  & 0.98 & 0.064 & 21.4 & 0.776--0.795 & 0.035--0.047 & 1.069--1.085 \\
 &  &  & 0.95 & 0.108 & 15.9 & 0.812--0.825 & 0.069--0.086 & 1.141--1.177 \\
\addlinespace[2pt]
 & 100 & 47 & 0.99 & 0.033 & 43.6 & 0.772--0.780 & 0.016--0.023 & 1.033--1.042 \\
 &  &  & 0.98 & 0.047 & 36.0 & 0.782--0.792 & 0.026--0.032 & 1.053--1.064 \\
 &  &  & 0.95 & 0.101 & 18.6 & 0.819--0.826 & 0.060--0.073 & 1.143--1.175 \\
\addlinespace[2pt]
 & 1000 & 479 & 0.99 & 0.021 & 95.6 & 0.763--0.770 & 0.014--0.021 & 1.023--1.029 \\
 &  &  & 0.98 & 0.041 & 47.7 & 0.778--0.784 & 0.028--0.035 & 1.051--1.063 \\
 &  &  & 0.95 & 0.101 & 19.0 & 0.818--0.823 & 0.067--0.076 & 1.144--1.175 \\
\midrule
2 & 50 & 23 & 0.99 & 0.052 & 22.6 & 0.776--0.788 & 0.027--0.035 & 1.052--1.062 \\
 &  &  & 0.98 & 0.064 & 21.4 & 0.781--0.793 & 0.032--0.041 & 1.064--1.076 \\
 &  &  & 0.95 & 0.108 & 15.9 & 0.811--0.826 & 0.062--0.073 & 1.134--1.163 \\
\addlinespace[2pt]
 & 100 & 47 & 0.99 & 0.033 & 43.6 & 0.758--0.774 & 0.016--0.022 & 1.027--1.035 \\
 &  &  & 0.98 & 0.047 & 36.1 & 0.766--0.786 & 0.028--0.035 & 1.045--1.055 \\
 &  &  & 0.95 & 0.101 & 18.6 & 0.809--0.820 & 0.064--0.072 & 1.129--1.154 \\
\addlinespace[2pt]
 & 1000 & 479 & 0.99 & 0.021 & 95.7 & 0.761--0.767 & 0.011--0.014 & 1.021--1.026 \\
 &  &  & 0.98 & 0.041 & 47.8 & 0.774--0.778 & 0.023--0.028 & 1.047--1.052 \\
 &  &  & 0.95 & 0.101 & 19.0 & 0.814--0.820 & 0.061--0.068 & 1.136--1.153 \\
\bottomrule
\end{tabular}}
\end{table}

\subsection{Experiment~2}
\label{app:sim_seq}

\begin{figure}[tbp]
\centering
\includegraphics[width=\textwidth]{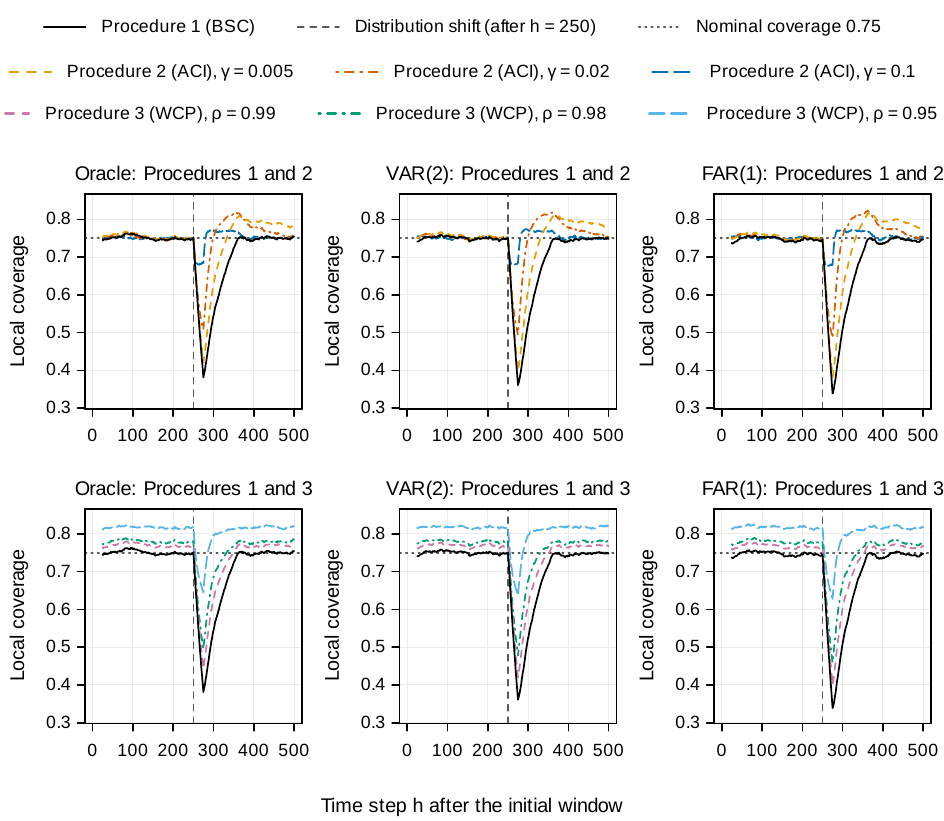}
\caption{Experiment~2, scenario S2, $p=1$, $b=1$: local coverage against $h$; columns: Oracle, VAR(2) and FAR(1). Top: Procedure~1 (solid) and Procedure~2 with $\gamma=0.005$, $0.02$ and $0.1$ (dashed, dash-dotted and long-dashed); bottom: Procedure~1 and Procedure~3 with $\rho=0.99$, $0.98$ and $0.95$ (same line types). The local coverage at $h$, computed over $h-24,\dots,h$, reacts to the shift with a delay of up to $24$ steps. Dashed vertical line: shift; dotted horizontal line: nominal coverage $0.75$.}
\label{fig:sim_seq}
\end{figure}

\paragraph{Design and performance measures} In each window, the random split is drawn as in Experiment~1, and scenario S2 uses the same random numbers as S1, including those of the splits. In S2, we distinguish the periods $h \leq 250$ (before the shift), $251 \leq h \leq 350$ (transition: the window contains observations that precede the shift, which can be training or calibration observations for $h \leq 350-r$), and $h>350$ (the whole window follows the shift). The loss of coverage of a procedure in the transition is its coverage in S1, over all pairs, minus its coverage in S2 for $251 \leq h \leq 350$. The weights of Procedure~3 are computed in each window from the times $\omega_d \in \{1,\dots,T\}$ of the calibration observations, the target time point being $T+1$ ($\tau=0$); they change with the split of each window, and $W$, the effective number of calibration scores and the occurrence of whole-space bands depend only on the split, and are therefore the same in S1 and S2. As a check of the implementation, Procedure~3 with $\rho=1$ was also computed in every window and gave the same coverage indicator and band size as Procedure~1 with $b=1$, and the band of Procedure~3 was infinite exactly when $W<(1-\alpha)/\alpha$. The SD of the long-run coverage across trajectories, divided by $M^{1/2}$, gives the SE of the coverage over all pairs, at most $0.00073$ (the pairs of a trajectory are dependent). The local coverage at $h$ is the frequency of coverage over the $25$ most recent test times, averaged over trajectories (Figure~\ref{fig:sim_seq}). In all trajectories of Procedure~2, $\alpha_t \in [-0.05,0.825]$, so that no band is empty. Tables~\ref{tab:seq} and~\ref{tab:seq_p2} report the results for $(p,b)=(1,1)$ and $(2,1)$, and Table~\ref{tab:seq_b3} those of Procedure~1 for $(p,b)=(1,3)$.

\begin{table}[tbp]
\centering
\caption{Experiment~2, $p=1$, $b=1$ ($K=48$), $T=100$, $\alpha=0.25$, $M=H=500$; P1, P2, P3: Procedures~1, 2 (step size $\gamma$) and~3 (weights $\rho^{\,T+1-\omega_d}$). Cov.: coverage over all pairs (trajectory, $h$); SD: standard deviation of the long-run coverage; Size: median size of the finite bands; Inf.: percentage of infinite bands (${<}0.1$: at most $0.002$); S2: innovations doubled from $h=251$.}
\label{tab:seq}
{\scriptsize\setlength{\tabcolsep}{4pt}\setlength{\defaultaddspace}{2pt}\renewcommand{\arraystretch}{0.9}
\begin{tabular}{llcccccccccc}
\toprule
 & & \multicolumn{4}{c}{S1 (stationary)} & \multicolumn{6}{c}{S2 (shift after $h = 250$)} \\
\cmidrule(lr){3-6} \cmidrule(lr){7-12}
 & & & & & & \multicolumn{3}{c}{Coverage, $h$} & \multicolumn{2}{c}{Size, $h$} & \\
\cmidrule(lr){7-9} \cmidrule(lr){10-11}
Model & Method & Cov. & SD & Size & Inf. & ${\le}250$ & ${>}250$ & ${>}350$ & ${\le}250$ & ${>}250$ & Inf. \\
\midrule
Oracle & P1 & 0.751 & 0.010 & 5.69 & 0 & 0.751 & 0.679 & 0.750 & 5.70 & 10.6 & 0 \\
 & P2, $\gamma{=}0.005$ & 0.754 & 0.006 & 5.73 & 0 & 0.755 & 0.731 & 0.791 & 5.76 & 11.7 & 0 \\
 & P2, $\gamma{=}0.02$ & 0.751 & 0.004 & 5.72 & 0 & 0.752 & 0.750 & 0.772 & 5.73 & 11.8 & 0 \\
 & P2, $\gamma{=}0.1$ & 0.750 & 0.002 & 5.74 & 0.2 & 0.750 & 0.750 & 0.752 & 5.75 & 11.5 & 0.9 \\
 & P3, $\rho{=}0.99$ & 0.769 & 0.010 & 5.91 & 0 & 0.770 & 0.714 & 0.769 & 5.92 & 11.2 & 0 \\
 & P3, $\rho{=}0.98$ & 0.780 & 0.010 & 6.04 & 0 & 0.780 & 0.736 & 0.779 & 6.06 & 11.5 & 0 \\
 & P3, $\rho{=}0.95$ & 0.817 & 0.010 & 6.64 & 0 & 0.817 & 0.796 & 0.818 & 6.67 & 12.9 & 0 \\
\addlinespace
VAR(1) & P1 & 0.748 & 0.013 & 6.34 & 0 & 0.748 & 0.671 & 0.746 & 6.36 & 11.8 & 0 \\
 & P2, $\gamma{=}0.005$ & 0.753 & 0.006 & 6.40 & 0 & 0.755 & 0.728 & 0.793 & 6.43 & 13.0 & 0 \\
 & P2, $\gamma{=}0.02$ & 0.751 & 0.004 & 6.39 & 0 & 0.751 & 0.750 & 0.772 & 6.40 & 13.2 & 0 \\
 & P2, $\gamma{=}0.1$ & 0.750 & 0.002 & 6.36 & 0.3 & 0.751 & 0.750 & 0.751 & 6.38 & 12.7 & 1.1 \\
 & P3, $\rho{=}0.99$ & 0.767 & 0.012 & 6.57 & 0 & 0.768 & 0.707 & 0.766 & 6.59 & 12.4 & 0 \\
 & P3, $\rho{=}0.98$ & 0.778 & 0.011 & 6.70 & 0 & 0.778 & 0.731 & 0.777 & 6.73 & 12.8 & 0 \\
 & P3, $\rho{=}0.95$ & 0.817 & 0.010 & 7.30 & 0 & 0.817 & 0.794 & 0.818 & 7.33 & 14.2 & 0 \\
\addlinespace
VAR(2) & P1 & 0.749 & 0.013 & 6.21 & 0 & 0.750 & 0.672 & 0.747 & 6.23 & 11.6 & 0 \\
 & P2, $\gamma{=}0.005$ & 0.753 & 0.006 & 6.26 & 0 & 0.756 & 0.728 & 0.792 & 6.30 & 12.8 & 0 \\
 & P2, $\gamma{=}0.02$ & 0.751 & 0.004 & 6.24 & 0 & 0.752 & 0.749 & 0.771 & 6.27 & 12.9 & 0 \\
 & P2, $\gamma{=}0.1$ & 0.750 & 0.002 & 6.22 & 0.3 & 0.751 & 0.750 & 0.751 & 6.24 & 12.5 & 1.1 \\
 & P3, $\rho{=}0.99$ & 0.769 & 0.012 & 6.43 & 0 & 0.769 & 0.709 & 0.768 & 6.45 & 12.2 & 0 \\
 & P3, $\rho{=}0.98$ & 0.780 & 0.012 & 6.57 & 0 & 0.780 & 0.734 & 0.780 & 6.59 & 12.6 & 0 \\
 & P3, $\rho{=}0.95$ & 0.818 & 0.011 & 7.17 & 0 & 0.818 & 0.797 & 0.818 & 7.20 & 14.0 & 0 \\
\addlinespace
VAR(3) & P1 & 0.750 & 0.015 & 6.54 & 0 & 0.749 & 0.674 & 0.750 & 6.56 & 12.3 & 0 \\
 & P2, $\gamma{=}0.005$ & 0.754 & 0.007 & 6.59 & 0 & 0.756 & 0.729 & 0.793 & 6.64 & 13.5 & 0 \\
 & P2, $\gamma{=}0.02$ & 0.751 & 0.004 & 6.57 & 0 & 0.752 & 0.749 & 0.772 & 6.61 & 13.6 & ${<}0.1$ \\
 & P2, $\gamma{=}0.1$ & 0.750 & 0.002 & 6.52 & 0.3 & 0.751 & 0.750 & 0.752 & 6.56 & 13.1 & 1.2 \\
 & P3, $\rho{=}0.99$ & 0.770 & 0.014 & 6.77 & 0 & 0.769 & 0.710 & 0.769 & 6.80 & 12.9 & 0 \\
 & P3, $\rho{=}0.98$ & 0.780 & 0.014 & 6.91 & 0 & 0.780 & 0.734 & 0.780 & 6.94 & 13.3 & 0 \\
 & P3, $\rho{=}0.95$ & 0.819 & 0.012 & 7.50 & 0 & 0.819 & 0.797 & 0.819 & 7.54 & 14.7 & 0 \\
\addlinespace
FAR(1) & P1 & 0.747 & 0.012 & 6.56 & 0 & 0.747 & 0.666 & 0.745 & 6.58 & 12.2 & 0 \\
 & P2, $\gamma{=}0.005$ & 0.752 & 0.006 & 6.62 & 0 & 0.754 & 0.727 & 0.794 & 6.66 & 13.5 & 0 \\
 & P2, $\gamma{=}0.02$ & 0.751 & 0.004 & 6.60 & 0 & 0.752 & 0.749 & 0.773 & 6.62 & 13.6 & ${<}0.1$ \\
 & P2, $\gamma{=}0.1$ & 0.750 & 0.002 & 6.54 & 0.3 & 0.751 & 0.750 & 0.752 & 6.56 & 13.1 & 1.2 \\
 & P3, $\rho{=}0.99$ & 0.767 & 0.011 & 6.77 & 0 & 0.767 & 0.703 & 0.765 & 6.80 & 12.8 & 0 \\
 & P3, $\rho{=}0.98$ & 0.778 & 0.011 & 6.91 & 0 & 0.779 & 0.729 & 0.776 & 6.94 & 13.2 & 0 \\
 & P3, $\rho{=}0.95$ & 0.817 & 0.010 & 7.49 & 0 & 0.818 & 0.794 & 0.817 & 7.53 & 14.6 & 0 \\
\addlinespace
FAR(2) & P1 & 0.748 & 0.012 & 6.14 & 0 & 0.748 & 0.670 & 0.747 & 6.16 & 11.5 & 0 \\
 & P2, $\gamma{=}0.005$ & 0.753 & 0.006 & 6.20 & 0 & 0.755 & 0.728 & 0.792 & 6.23 & 12.7 & 0 \\
 & P2, $\gamma{=}0.02$ & 0.751 & 0.004 & 6.18 & 0 & 0.752 & 0.749 & 0.772 & 6.20 & 12.8 & ${<}0.1$ \\
 & P2, $\gamma{=}0.1$ & 0.750 & 0.002 & 6.14 & 0.3 & 0.751 & 0.750 & 0.752 & 6.16 & 12.3 & 1.1 \\
 & P3, $\rho{=}0.99$ & 0.768 & 0.012 & 6.36 & 0 & 0.768 & 0.707 & 0.767 & 6.38 & 12.1 & 0 \\
 & P3, $\rho{=}0.98$ & 0.779 & 0.010 & 6.49 & 0 & 0.779 & 0.732 & 0.778 & 6.52 & 12.4 & 0 \\
 & P3, $\rho{=}0.95$ & 0.818 & 0.010 & 7.06 & 0 & 0.818 & 0.796 & 0.818 & 7.10 & 13.8 & 0 \\
\addlinespace
FAR(3) & P1 & 0.749 & 0.013 & 6.29 & 0 & 0.749 & 0.670 & 0.747 & 6.30 & 11.8 & 0 \\
 & P2, $\gamma{=}0.005$ & 0.753 & 0.007 & 6.34 & 0 & 0.755 & 0.728 & 0.793 & 6.38 & 13.0 & 0 \\
 & P2, $\gamma{=}0.02$ & 0.751 & 0.004 & 6.32 & 0 & 0.752 & 0.750 & 0.772 & 6.34 & 13.1 & ${<}0.1$ \\
 & P2, $\gamma{=}0.1$ & 0.750 & 0.002 & 6.27 & 0.3 & 0.751 & 0.750 & 0.751 & 6.29 & 12.6 & 1.1 \\
 & P3, $\rho{=}0.99$ & 0.768 & 0.012 & 6.50 & 0 & 0.768 & 0.706 & 0.767 & 6.52 & 12.4 & 0 \\
 & P3, $\rho{=}0.98$ & 0.779 & 0.011 & 6.63 & 0 & 0.779 & 0.731 & 0.778 & 6.66 & 12.7 & 0 \\
 & P3, $\rho{=}0.95$ & 0.817 & 0.011 & 7.19 & 0 & 0.817 & 0.795 & 0.818 & 7.23 & 14.1 & 0 \\
\bottomrule
\end{tabular}
}
\end{table}

\begin{table}[tbp]
\centering
\caption{Experiment~2: rolling window with $p=2$, $b=1$ ($K=48$); settings, methods and columns as in Table~\ref{tab:seq}. Infinite bands of P2 occur with $\gamma=0.1$, in $0.18\%$ to $0.57\%$ (S1) and $1.0\%$ to $1.6\%$ (S2) of the pairs, and with $\gamma=0.02$ in S2 only, in at most $0.0084\%$ of the pairs (${<}0.1$).}
\label{tab:seq_p2}
{\scriptsize\setlength{\tabcolsep}{4pt}\setlength{\defaultaddspace}{2pt}\renewcommand{\arraystretch}{0.9}
\begin{tabular}{llcccccccccc}
\toprule
 & & \multicolumn{4}{c}{S1 (stationary)} & \multicolumn{6}{c}{S2 (shift after $h = 250$)} \\
\cmidrule(lr){3-6} \cmidrule(lr){7-12}
 & & & & & & \multicolumn{3}{c}{Coverage, $h$} & \multicolumn{2}{c}{Size, $h$} & \\
\cmidrule(lr){7-9} \cmidrule(lr){10-11}
Model & Method & Cov. & SD & Size & Inf. & ${\le}250$ & ${>}250$ & ${>}350$ & ${\le}250$ & ${>}250$ & Inf. \\
\midrule
Oracle & P1 & 0.750 & 0.011 & 142 & 0 & 0.750 & 0.670 & 0.748 & 143 & 267 & 0 \\
 & P2, $\gamma{=}0.005$ & 0.753 & 0.006 & 143 & 0 & 0.756 & 0.727 & 0.792 & 144 & 294 & 0 \\
 & P2, $\gamma{=}0.02$ & 0.751 & 0.004 & 143 & 0 & 0.752 & 0.749 & 0.771 & 143 & 297 & ${<}0.1$ \\
 & P2, $\gamma{=}0.1$ & 0.750 & 0.002 & 144 & 0.2 & 0.750 & 0.750 & 0.751 & 144 & 290 & 1.0 \\
 & P3, $\rho{=}0.99$ & 0.769 & 0.010 & 148 & 0 & 0.770 & 0.707 & 0.767 & 148 & 280 & 0 \\
 & P3, $\rho{=}0.98$ & 0.780 & 0.010 & 151 & 0 & 0.780 & 0.732 & 0.778 & 151 & 289 & 0 \\
 & P3, $\rho{=}0.95$ & 0.817 & 0.010 & 165 & 0 & 0.817 & 0.793 & 0.815 & 165 & 321 & 0 \\
\addlinespace
VAR(1) & P1 & 0.747 & 0.013 & 161 & 0 & 0.748 & 0.660 & 0.745 & 162 & 303 & 0 \\
 & P2, $\gamma{=}0.005$ & 0.752 & 0.006 & 163 & 0 & 0.755 & 0.725 & 0.795 & 163 & 334 & 0 \\
 & P2, $\gamma{=}0.02$ & 0.750 & 0.004 & 162 & 0 & 0.752 & 0.748 & 0.772 & 163 & 337 & ${<}0.1$ \\
 & P2, $\gamma{=}0.1$ & 0.750 & 0.002 & 162 & 0.3 & 0.750 & 0.750 & 0.751 & 162 & 325 & 1.3 \\
 & P3, $\rho{=}0.99$ & 0.767 & 0.012 & 166 & 0 & 0.768 & 0.699 & 0.765 & 167 & 317 & 0 \\
 & P3, $\rho{=}0.98$ & 0.778 & 0.011 & 170 & 0 & 0.779 & 0.725 & 0.776 & 170 & 326 & 0 \\
 & P3, $\rho{=}0.95$ & 0.817 & 0.011 & 184 & 0 & 0.818 & 0.793 & 0.816 & 184 & 360 & 0 \\
\addlinespace
VAR(2) & P1 & 0.748 & 0.015 & 171 & 0 & 0.749 & 0.659 & 0.745 & 172 & 323 & 0 \\
 & P2, $\gamma{=}0.005$ & 0.752 & 0.007 & 173 & 0 & 0.755 & 0.725 & 0.795 & 173 & 356 & 0 \\
 & P2, $\gamma{=}0.02$ & 0.751 & 0.005 & 172 & 0 & 0.752 & 0.748 & 0.771 & 173 & 358 & ${<}0.1$ \\
 & P2, $\gamma{=}0.1$ & 0.750 & 0.002 & 171 & 0.4 & 0.750 & 0.750 & 0.751 & 171 & 346 & 1.4 \\
 & P3, $\rho{=}0.99$ & 0.767 & 0.014 & 177 & 0 & 0.768 & 0.699 & 0.765 & 177 & 337 & 0 \\
 & P3, $\rho{=}0.98$ & 0.778 & 0.013 & 180 & 0 & 0.778 & 0.727 & 0.777 & 180 & 347 & 0 \\
 & P3, $\rho{=}0.95$ & 0.817 & 0.012 & 194 & 0 & 0.817 & 0.793 & 0.816 & 195 & 382 & 0 \\
\addlinespace
VAR(3) & P1 & 0.749 & 0.016 & 193 & 0 & 0.751 & 0.658 & 0.745 & 193 & 364 & 0 \\
 & P2, $\gamma{=}0.005$ & 0.753 & 0.008 & 194 & 0 & 0.756 & 0.723 & 0.796 & 195 & 400 & 0 \\
 & P2, $\gamma{=}0.02$ & 0.751 & 0.005 & 194 & 0 & 0.753 & 0.748 & 0.772 & 194 & 403 & ${<}0.1$ \\
 & P2, $\gamma{=}0.1$ & 0.750 & 0.002 & 192 & 0.6 & 0.751 & 0.750 & 0.751 & 192 & 388 & 1.6 \\
 & P3, $\rho{=}0.99$ & 0.768 & 0.016 & 199 & 0 & 0.770 & 0.697 & 0.765 & 199 & 380 & 0 \\
 & P3, $\rho{=}0.98$ & 0.779 & 0.015 & 202 & 0 & 0.780 & 0.723 & 0.776 & 203 & 391 & 0 \\
 & P3, $\rho{=}0.95$ & 0.818 & 0.014 & 217 & 0 & 0.819 & 0.792 & 0.816 & 218 & 428 & 0 \\
\addlinespace
FAR(1) & P1 & 0.746 & 0.012 & 159 & 0 & 0.748 & 0.657 & 0.743 & 159 & 297 & 0 \\
 & P2, $\gamma{=}0.005$ & 0.752 & 0.006 & 160 & 0 & 0.755 & 0.724 & 0.795 & 161 & 328 & 0 \\
 & P2, $\gamma{=}0.02$ & 0.750 & 0.005 & 160 & 0 & 0.752 & 0.749 & 0.772 & 160 & 330 & ${<}0.1$ \\
 & P2, $\gamma{=}0.1$ & 0.750 & 0.002 & 159 & 0.3 & 0.750 & 0.750 & 0.752 & 159 & 319 & 1.3 \\
 & P3, $\rho{=}0.99$ & 0.766 & 0.011 & 164 & 0 & 0.768 & 0.697 & 0.764 & 164 & 310 & 0 \\
 & P3, $\rho{=}0.98$ & 0.777 & 0.011 & 167 & 0 & 0.779 & 0.724 & 0.775 & 167 & 319 & 0 \\
 & P3, $\rho{=}0.95$ & 0.816 & 0.010 & 180 & 0 & 0.818 & 0.791 & 0.815 & 181 & 352 & 0 \\
\addlinespace
FAR(2) & P1 & 0.748 & 0.012 & 154 & 0 & 0.750 & 0.660 & 0.745 & 154 & 289 & 0 \\
 & P2, $\gamma{=}0.005$ & 0.752 & 0.006 & 155 & 0 & 0.756 & 0.724 & 0.794 & 156 & 318 & 0 \\
 & P2, $\gamma{=}0.02$ & 0.751 & 0.004 & 154 & 0 & 0.752 & 0.749 & 0.772 & 155 & 320 & ${<}0.1$ \\
 & P2, $\gamma{=}0.1$ & 0.750 & 0.002 & 154 & 0.2 & 0.750 & 0.750 & 0.751 & 154 & 310 & 1.2 \\
 & P3, $\rho{=}0.99$ & 0.767 & 0.012 & 159 & 0 & 0.769 & 0.700 & 0.765 & 159 & 302 & 0 \\
 & P3, $\rho{=}0.98$ & 0.778 & 0.011 & 162 & 0 & 0.780 & 0.726 & 0.776 & 162 & 311 & 0 \\
 & P3, $\rho{=}0.95$ & 0.817 & 0.010 & 175 & 0 & 0.817 & 0.792 & 0.816 & 176 & 343 & 0 \\
\addlinespace
FAR(3) & P1 & 0.748 & 0.013 & 157 & 0 & 0.750 & 0.660 & 0.746 & 157 & 296 & 0 \\
 & P2, $\gamma{=}0.005$ & 0.753 & 0.007 & 158 & 0 & 0.756 & 0.724 & 0.795 & 159 & 326 & 0 \\
 & P2, $\gamma{=}0.02$ & 0.751 & 0.004 & 158 & 0 & 0.752 & 0.748 & 0.771 & 158 & 328 & ${<}0.1$ \\
 & P2, $\gamma{=}0.1$ & 0.750 & 0.002 & 157 & 0.3 & 0.751 & 0.750 & 0.751 & 157 & 317 & 1.2 \\
 & P3, $\rho{=}0.99$ & 0.768 & 0.012 & 162 & 0 & 0.769 & 0.699 & 0.765 & 163 & 309 & 0 \\
 & P3, $\rho{=}0.98$ & 0.779 & 0.012 & 165 & 0 & 0.780 & 0.725 & 0.776 & 166 & 318 & 0 \\
 & P3, $\rho{=}0.95$ & 0.816 & 0.011 & 178 & 0 & 0.817 & 0.792 & 0.816 & 179 & 350 & 0 \\
\bottomrule
\end{tabular}
}
\end{table}

\begin{table}[tbp]
\centering
\caption{Experiment~2: rolling window with $p=1$, $b=3$ ($K=16$), Procedure~1 only (Procedures~2 and~3 use $b=1$); settings and columns as in Table~\ref{tab:seq}. No band is infinite.}
\label{tab:seq_b3}
{\scriptsize\setlength{\tabcolsep}{4pt}\setlength{\defaultaddspace}{2pt}
\begin{tabular}{llcccccccccc}
\toprule
 & & \multicolumn{4}{c}{S1 (stationary)} & \multicolumn{6}{c}{S2 (shift after $h = 250$)} \\
\cmidrule(lr){3-6} \cmidrule(lr){7-12}
 & & & & & & \multicolumn{3}{c}{Coverage, $h$} & \multicolumn{2}{c}{Size, $h$} & \\
\cmidrule(lr){7-9} \cmidrule(lr){10-11}
Model & Method & Cov. & SD & Size & Inf. & ${\le}250$ & ${>}250$ & ${>}350$ & ${\le}250$ & ${>}250$ & Inf. \\
\midrule
Oracle & P1 & 0.751 & 0.013 & 5.78 & 0 & 0.751 & 0.683 & 0.751 & 5.80 & 10.6 & 0 \\
VAR(1) & P1 & 0.747 & 0.013 & 6.44 & 0 & 0.748 & 0.673 & 0.746 & 6.46 & 11.9 & 0 \\
VAR(2) & P1 & 0.750 & 0.014 & 6.32 & 0 & 0.750 & 0.677 & 0.749 & 6.33 & 11.7 & 0 \\
VAR(3) & P1 & 0.750 & 0.016 & 6.65 & 0 & 0.749 & 0.676 & 0.749 & 6.67 & 12.3 & 0 \\
FAR(1) & P1 & 0.747 & 0.014 & 6.65 & 0 & 0.748 & 0.670 & 0.745 & 6.68 & 12.3 & 0 \\
FAR(2) & P1 & 0.749 & 0.014 & 6.24 & 0 & 0.750 & 0.674 & 0.747 & 6.27 & 11.6 & 0 \\
FAR(3) & P1 & 0.749 & 0.013 & 6.39 & 0 & 0.749 & 0.673 & 0.748 & 6.41 & 11.9 & 0 \\
\bottomrule
\end{tabular}
}
\end{table}

\paragraph{Additional results} In S1, the long-run coverage of Procedure~1 takes values in $[0.694,0.802]$, and lies within $0.75 \pm 0.017$, the bound for $\gamma=0.1$, in $70.0\%$ to $89.6\%$ of the trajectories; that of Procedure~2 takes values in $[0.728,0.778]$, $[0.734,0.766]$ and $[0.744,0.758]$ for $\gamma=0.005$, $0.02$ and $0.1$ (Figure~\ref{fig:sim_seq_traj}). Over the trajectories of Procedure~2 in S1 and S2, the largest deviation of the long-run miscoverage from $\alpha$ is $0.034$, $0.018$ and $0.008$ for $\gamma=0.005$, $0.02$ and $0.1$. In S2, for $p=1$ and $b=1$, the minimum local coverage of Procedure~1 is between $0.337$ and $0.381$ (Figure~\ref{fig:sim_seq}), and its median band size after the shift is $1.86$ to $1.87$ times that before. The coverage of Procedure~1 is $0.548$--$0.574$ in the transition and $0.707$--$0.715$ over the whole trajectory, and its long-run coverage lies within $0.75 \pm 0.017$ in $2.2\%$ to $9.0\%$ of the trajectories. The minimum local coverage of Procedure~2 is $0.380$--$0.420$, $0.483$--$0.508$ and $0.674$--$0.680$ for $\gamma=0.005$, $0.02$ and $0.1$, and its coverage in the transition is $0.627$--$0.641$, $0.715$--$0.718$ and $0.748$. The local coverage of Procedure~2 returns to $0.74$ at $h=327$--$331$, $297$--$298$ and $280$, against $358$--$360$ for Procedure~1, and then peaks at $0.808$--$0.816$ and $0.817$--$0.824$ for $\gamma=0.005$ and $0.02$. After the shift ($h>250$), Procedure~2 covers more than Procedure~1 ($0.727$--$0.731$, $0.749$--$0.750$ and $0.750$, against $0.666$--$0.679$), and, for $\gamma=0.005$ and $0.02$, the median size of its bands is larger by $9.7\%$ to $10.2\%$ and $10.4\%$ to $11.4\%$. For $h>350$, Procedure~2 overcovers with these two step sizes ($0.791$--$0.794$ and $0.771$--$0.773$, against $0.745$--$0.750$), with median sizes larger by $8.4\%$ to $9.0\%$ and $4.1\%$ to $4.8\%$. At different coverage, these sizes are not comparable. With $\gamma=0.005$, the compensation of the misses of the transition is incomplete at $h=500$, and the coverage over the whole trajectory is $0.741$--$0.743$. With $\gamma=0.1$, at least one infinite band occurs in $93.6\%$ to $98.6\%$ of the trajectories of S2. For Procedure~1, $b=3$ gives a similar coverage and, in S1, a slightly larger SD ($0.013$ to $0.016$, against $0.010$ to $0.015$ with $b=1$) and median band sizes $1.5\%$ to $1.7\%$ larger. For $p=2$, the coverage in the transition is $0.527$--$0.553$ for Procedure~1, and $0.614$--$0.630$, $0.712$--$0.716$ and $0.748$--$0.749$ for Procedure~2 with $\gamma=0.005$, $0.02$ and $0.1$.

\paragraph{Procedure~3} The test point keeps the normalised weight $1/(W+1)$, whose mean over the windows is $0.032$--$0.033$, $0.047$--$0.048$ and $0.100$--$0.102$ for $\rho=0.99$, $0.98$ and $0.95$; since $W$ is at least $25.54$, $14.89$ and $3.83$, no band is the whole space. The effective number of calibration scores $(\sum_d w_d)^2/\sum_d w_d^2$, mean over the windows, is $43.5$--$43.6$, $35.9$--$36.2$ and $18.5$--$18.8$, against $K-1=47$ for Procedures~1 and~2, and its version with the test point, $(W+1)^2/(\sum_d w_d^2+1)$, is $44.3$--$44.4$, $36.3$--$36.7$ and $18.5$--$18.9$, for both values of $p$. For the Oracle, whose scores are exchangeable and almost surely distinct, the coverage of Procedure~3 in S1 lies, for every $(p,\rho)$, between $1-\alpha$ and $1-\alpha$ plus the mean of $1/(W+1)$, the upper bound of Proposition~\ref{prop:weighted} averaged over the splits. With $p=1$, it is $0.769$, $0.780$ and $0.817$, against upper bounds $0.783$, $0.797$ and $0.851$; for both values of $p$, the estimated point predictors have similar coverage ($0.766$--$0.770$, $0.777$--$0.780$ and $0.816$--$0.819$). In S2, with $p=1$, the coverage of Procedure~3 in the transition is $0.611$--$0.631$, $0.657$--$0.672$ and $0.759$--$0.765$, and its minimum local coverage is $0.399$--$0.443$, $0.459$--$0.494$ and $0.625$--$0.645$; both are to be compared with its coverage in S1. After the shift ($h>250$), its coverage is $0.703$--$0.714$, $0.729$--$0.736$ and $0.794$--$0.797$, against $0.666$--$0.679$ for Procedure~1, and its median band size is larger than that of Procedure~1 by $4.6\%$ to $5.1\%$, $7.7\%$ to $8.6\%$ and $19.1\%$ to $21.5\%$. For $h>350$, this excess is $3.4\%$ to $3.9\%$, $5.5\%$ to $6.3\%$ and $14.3\%$ to $16.7\%$, as in S1 ($3.3\%$ to $3.9\%$, $5.4\%$ to $6.2\%$ and $14.3\%$ to $16.7\%$). At different coverage, these sizes are not comparable. Over the whole trajectory, the coverage of Procedure~3 is $0.735$--$0.742$, $0.754$--$0.758$ and $0.806$--$0.808$. For $p=2$, the coverage of Procedure~3 in the transition is $0.596$--$0.617$, $0.645$--$0.663$ and $0.754$--$0.761$, i.e.\ $0.152$--$0.172$, $0.116$--$0.134$ and $0.056$--$0.062$ below that in S1. The corresponding loss is $0.197$--$0.220$ for Procedure~1, and $0.123$--$0.138$, $0.035$--$0.039$ and $0.001$--$0.003$ for Procedure~2 with $\gamma=0.005$, $0.02$ and $0.1$.

\paragraph{Comparison of the point predictors} We compare the point predictors within each procedure (Procedure~1 with $b=1$ and, for $p=1$, $b=3$; Procedure~2 for each $\gamma$; Procedure~3 for each $\rho$), with the correctly specified VAR(2) as reference. The point predictors of a trajectory use the same random numbers, and the Monte Carlo SE of a difference in coverage is the SD of the per-trajectory differences divided by $M^{1/2}$. Over all pairs, the coverage of the other estimated point predictors differs from that of VAR(2) by $-0.0026$ to $0.0011$ in S1 and by $-0.0047$ to $0.0013$ in S2 (SE at most $0.0008$); the largest differences in absolute value are those of FAR(1) with Procedure~1 and $p=1$. For $p=1$, the loss of FAR(1) exceeds that of VAR(2) by $0.0100$ and $0.0101$ with Procedure~1 and $b=1$ and~$3$, by $0.0040$, $0.0015$ and $0.0005$ with Procedure~2 and $\gamma=0.005$, $0.02$ and $0.1$, and by $0.0072$, $0.0065$ and $0.0045$ with Procedure~3 and $\rho=0.99$, $0.98$ and $0.95$; with Procedures~1 and~3, the loss of FAR(2) and FAR(3) exceeds that of VAR(2) by $0.0019$ to $0.0080$, and with Procedure~2 the difference is $-0.0009$ to $0.0014$; the loss of VAR(1) exceeds that of VAR(2) by $0.0002$ to $0.0031$ (SE of all these differences at most $0.0018$). For $p=2$, the loss of the misspecified point predictors, VAR(1) and the FAR models, differs from that of VAR(2) by $-0.0020$ to $0.0038$ (SE at most $0.0020$). With Procedure~1, the loss of the Oracle is smaller than that of every estimated point predictor, by $0.0105$ to $0.0231$. In S1, the SD of the long-run coverage of the misspecified point predictors exceeds that of VAR(2) by at most $0.0006$ for $p=1$ and is smaller by $0.0001$ to $0.0027$ for $p=2$; with Procedures~1 and~3, VAR(3) has the largest SD among the estimated point predictors. In S1, the median size of the misspecified point predictors is $0.986$ to $1.057$ ($p=1$) and $0.897$ to $0.946$ ($p=2$) times that of VAR(2), against $0.991$ to $1.055$ and $0.898$ to $0.944$ in Experiment~1 with $T=100$ and $b=1$, where their coverage differs from that of VAR(2) by $-0.006$ to $0.017$ (SE of each coverage $0.006$); for each estimated point predictor, the ratio of Experiment~2 varies across procedures by at most $0.013$. Relative to the Oracle, whose coverage in S1 differs from those of VAR(2) and FAR(1) by at most $0.004$, the median size in S1 is larger by $8.0\%$ to $9.2\%$ ($p=1$) and $18.0\%$ to $20.3\%$ ($p=2$) for VAR(2), and by $12.8\%$ to $15.5\%$ and $9.6\%$ to $11.6\%$ for FAR(1). In every procedure, the ranking of the seven point predictors by median size in S1 is that of Experiment~1 with $T=100$ and $b=1$, except that FAR(1) and VAR(3), whose median sizes differ by $0.2\%$, are interchanged for $p=1$ and $\rho=0.95$.

\paragraph{Long-run coverage under the shift} Summing \eqref{eq:aci_update} over $t$ shows that the long-run coverage of a trajectory of Procedure~2, $1-\alpha+(\alpha_{T+H+1}-\alpha)/(\gamma H)$, depends only on the final level. Once the whole window follows the shift, the observations of S2 are, up to the geometrically decaying effect of earlier values, twice those of S1, and, by the equivariance of the point predictors and the scale invariance of the score \eqref{eq:score}, with common splits, the scores of the two scenarios nearly coincide (for $h>360$, the coverage indicators of Procedure~1 agree in $99.99\%$ of the cases), and so do the levels $\alpha_t$ once they meet. Hence, the long-run coverage of Procedure~2 is the same in S1 and S2 in all trajectories with $\gamma=0.1$ and in $81.2\%$ to $86.6\%$ of them with $\gamma=0.02$ (coverage over all pairs at most $0.0004$ lower in S2), and that of Procedure~1 in none (Figure~\ref{fig:sim_seq_traj}). With $\gamma=0.005$, the levels of the two scenarios rarely meet before $h=500$, and the long-run coverage is the same in at most $0.4\%$ of the trajectories (coverage over all pairs $0.011$ to $0.013$ lower in S2).

\paragraph{Discreteness of the level in Procedure~2} The level $\alpha_t$ acts on the band only through $\lceil K(1-\alpha_t)\rceil$, i.e.\ through cells of width $1/K$. For exchangeable scores, the probability of a miss would be $\lfloor K\alpha_t\rfloor/K$, which, for $K=48$, equals $\alpha$ for $\alpha_t \in [0.25,0.2708)$ and is smaller (larger) below (above) this interval: $\alpha_t$ has no drift in it and is pushed back into it from both sides. Accordingly, in scenario S1, the mean of $\alpha_{T+H+1}$ over the trajectories is $0.2546$ to $0.2598$, $0.2537$ to $0.2618$ and $0.2556$ to $0.2724$ for $\gamma=0.005$, $0.02$ and $0.1$, and, since the long-run coverage is $1-\alpha+(\alpha_{T+H+1}-\alpha)/(\gamma H)$, the coverage over all pairs exceeds $0.75$ by $0.0019$ to $0.0039$, $0.0004$ to $0.0012$ and $0.0001$ to $0.0004$.

\section{Details and additional results of the application}
\label{app:app_extra}
\setcounter{figure}{0}\setcounter{table}{0}

This appendix gives details and additional results of the application of Section~\ref{sec:application}: the domain and the monotone point predictions (Sections~\ref{sec:app_data} and~\ref{sec:app_setup}), the results of Procedure~1 for other window lengths (Section~\ref{sec:app_bands}) and without the offers and bids of Snam, the projections of the region for the clearing point of Section~\ref{sec:app_region}, the region of the what-if analysis with a coverage guarantee (Section~\ref{sec:app_whatif}), and further results of the three procedures (Section~\ref{sec:app_procedures}).

\paragraph{Domain} A domain extending beyond the total quantity offered or demanded on some day requires arbitrary extensions of the curves beyond that quantity, which affect the scores; a shorter one ignores the curves beyond its upper end, and the clearing point cannot lie in the region $\mathcal{R}_t$ of Section~\ref{sec:app_region} when $Q_t$ exceeds it.

\paragraph{Monotone point predictions} For $q \in \mathcal{G}$, the point prediction of the offer curve is obtained from the prediction $\tilde{y}_t^{(1)}$ of model~\eqref{eq:gas_model} in Section~\ref{sec:app_setup} as
\begin{equation}
\hat{y}_t^{(1)}(q)=\begin{cases}
\tilde{y}_t^{(1)}(q) & \text{if } \tilde{y}_t^{(1)}(q)=\max_{v \in \mathcal{G},\, v \leq q} \tilde{y}_t^{(1)}(v),\\[4pt]
\dfrac{q''-q}{q''-q'}\,\tilde{y}_t^{(1)}(q')+\dfrac{q-q'}{q''-q'}\,\tilde{y}_t^{(1)}(q'') & \text{otherwise},
\end{cases}
\label{eq:gas_monotone}
\end{equation}
where $q':=\min\{\operatorname{argmax}_{v \in \mathcal{G},\, v \leq q} \tilde{y}_t^{(1)}(v)\}$ and $q'':=\min\{v \in \mathcal{G}:\allowbreak v \geq q,\allowbreak\ \tilde{y}_t^{(1)}(v) \geq \tilde{y}_t^{(1)}(q')\}$, with $\hat{y}_t^{(1)}(q):=\tilde{y}_t^{(1)}(q')$ if the latter set is empty. The resulting $\hat{y}_t^{(1)}$ is non-decreasing; $\hat{y}_t^{(2)}$ is obtained with the minimum in place of the maximum and the inequality reversed. Rule~\eqref{eq:gas_monotone} and its counterpart for the demand curve are also applied to the predictions of the training and calibration days. The point predictions are not step functions, but neither is their target, the conditional expectation of the curve, which is continuous in $q$ (under integrability conditions) when the quantities at which the steps occur have absolutely continuous distributions \citep[cf.][]{pelagatti2013supply}.

\paragraph{Smaller miscoverage levels} For a given $T$, the larger $K$ required by smaller values of $\alpha$ (Section~\ref{sec:procedure}) implies a shorter training set: with $K=40$, $\alpha=0.05$ would make $k_t$ the second largest calibration score.

\paragraph{Window length} Table~\ref{tab:gas_T} reports the frequency of coverage and the width of the bands of Procedure~1 for the window lengths $T \in \{45,60,90,180,365\}$ compared in Section~\ref{sec:app_bands}. Across $T$, the median widths differ by at most 0.14 ($\alpha=0.5$) and 0.5 ($\alpha=0.25$) Euro/MWh, and their IQR is smaller for $T \geq 180$. In the choice of $T$, a longer window increases $n_1$ and $K$, which control the bound~\eqref{eq:bound}, whereas a shorter one contains only recent days, which may be more representative when the market changes. Under stationarity, and for a window length fixed in advance, the frequencies of coverage would be unbiased estimates of the marginal coverage, but their precision cannot be assessed as in Section~\ref{sec:simulation}, since they come from a single trajectory with overlapping windows; here, moreover, $T$ was selected on the same days.

\begin{table}[tbp]
\centering
\caption{Procedure~1 with window length $T$ ($t_0=8$, $n_1=T-n_2-t_0$, $K=n_2+1$, random split, $b=1$): percentage of the 184 days on which the observed curves lie in the band, and median and interquartile range (IQR) over the days of the average width of the band (Euro/MWh; difference between the monotone bounds, averaged over the grid and the two curves).}
\label{tab:gas_T}
{\small\setlength{\tabcolsep}{4pt}
\begin{tabular}{rrrrrrrrrr}
\toprule
 & & & & \multicolumn{2}{c}{Curves in band (\%)} & \multicolumn{2}{c}{Median width} & \multicolumn{2}{c}{IQR of width} \\
\cmidrule(lr){5-6}\cmidrule(lr){7-8}\cmidrule(lr){9-10}
$T$ & $n_1$ & $n_2$ & $K$ & $\alpha=0.25$ & $\alpha=0.5$ & $\alpha=0.25$ & $\alpha=0.5$ & $\alpha=0.25$ & $\alpha=0.5$ \\
\midrule
45 & 18 & 19 & 20 & 70.7 & 45.7 & 4.33 & 3.20 & 1.77 & 1.15 \\
60 & 33 & 19 & 20 & 73.9 & 50.5 & 4.42 & 3.19 & 1.31 & 0.78 \\
90 & 43 & 39 & 40 & 75.5 & 52.7 & 4.51 & 3.16 & 1.36 & 0.65 \\
180 & 93 & 79 & 80 & 78.8 & 53.3 & 4.53 & 3.22 & 0.63 & 0.34 \\
365 & 178 & 179 & 180 & 88.0 & 66.3 & 4.83 & 3.30 & 0.78 & 0.25 \\
\bottomrule
\end{tabular}
}
\end{table}

\paragraph{Bands without the offers and bids of Snam} When the curves are built without Snam's offers and bids, ceteris paribus (Section~\ref{sec:app_whatif}), and the bands are recomputed, they are no longer wider on $[0,20{,}000]$ MWh than on the rest of the domain (cf.\ Figure~\ref{fig:gaspredset}), and the average width of the band of each curve on this interval decreases, in median over the 184 days, by 47.4\% (offer curve) and 73.1\% (demand curve) for $\alpha=0.5$, and by 44.2\% and 71.8\% for $\alpha=0.25$. Part of this reduction is due to a smaller $k_t$, which also narrows the bands on the rest of the domain (by 34.2\% and 43.6\% for $\alpha=0.5$), and the curves without Snam lie in their bands less often (on 44.6\% and 65.2\% of the days, against 52.7\% and 75.5\%).

\paragraph{Projections of the region for the clearing point} The projections of $\mathcal{R}_t$ (Section~\ref{sec:app_region}) on the axes give sets for $Q_t$ and $P_t$ with the same guarantee; in this market, however, $\mathcal{R}_t$ contains points at all quantities in $[0,2\cdot 10^5)$ on 182 and 146 of the 184 days ($\alpha=0.25$ and $0.5$), so that only the set for $P_t$, of median length 4.28 and 3.06 Euro/MWh, is informative.

\paragraph{What-if analysis with a coverage guarantee} Inserting a demand bid of quantity $\Delta$ at price $c$ (Section~\ref{sec:app_whatif}) preserves the pointwise order between non-increasing curves (on a uniform grid, for $\Delta$ a multiple of its spacing). Hence, under the ceteris paribus assumption, the modified demand curve lies between the bounds $\ell_t^{(2)}$ and $u_t^{(2)}$ modified by the same insertion whenever the demand curve lies between $\ell_t^{(2)}$ and $u_t^{(2)}$, and the region computed from the modified bounds has, under the intersection condition of Section~\ref{sec:app_region}, coverage at least that of the band. The same holds for offers inserted in the offer curve and in $\ell_t^{(1)}$ and $u_t^{(1)}$.

\paragraph{Misses per month} Table~\ref{tab:gas_monthly} reports the misses of the three procedures in each month. For Procedure~1 with $\alpha=0.25$, they are 14 and 11 in September and October 2019, against 2 to 8 in each of the other months. With $\gamma=0.1$ and $\alpha=0.5$, the 9 empty bands of Procedure~2 (26 August to 3 September 2019, 6 of them in August) are misses. Table~\ref{tab:gas_procedures_full} adds Procedure~3 with $\rho=0.98$ to Table~\ref{tab:gas_procedures}. Its results in this table and in Table~\ref{tab:gas_monthly} lie between those for $\rho=0.99$ and $0.95$.

\begin{table}[tbp]
\centering
\caption{Number of days with a miss (observed pair of curves not in the band) in each month of the evaluation period, for the procedures of Table~\ref{tab:gas_procedures_full} ($T=90$, $K=40$; P1, P2, P3: Procedures~1, 2 and~3; Procedure~2 with $\tau=7$). The months have 31, 30, 31, 30, 31 and 31 days.}
\label{tab:gas_monthly}
{\small\setlength{\tabcolsep}{4pt}
\begin{tabular}{lrrrrrrr}
\toprule
 & Aug.\ 2019 & Sep. & Oct. & Nov. & Dec. & Jan.\ 2020 & Total \\
\midrule
\multicolumn{8}{l}{$\alpha=0.25$ (nominal level 0.75)} \\
P1 & 2 & 14 & 11 & 5 & 8 & 5 & 45 \\
P2, $\gamma{=}0.005$ & 2 & 14 & 11 & 1 & 8 & 5 & 41 \\
P2, $\gamma{=}0.02$ & 2 & 15 & 11 & 0 & 8 & 8 & 44 \\
P2, $\gamma{=}0.1$ & 2 & 16 & 9 & 1 & 8 & 11 & 47 \\
P3, $\rho{=}0.99$ & 1 & 14 & 11 & 0 & 8 & 5 & 39 \\
P3, $\rho{=}0.98$ & 1 & 13 & 9 & 0 & 8 & 5 & 36 \\
P3, $\rho{=}0.95$ & 0 & 11 & 7 & 0 & 7 & 4 & 29 \\
\addlinespace
\multicolumn{8}{l}{$\alpha=0.5$ (nominal level 0.50)} \\
P1 & 3 & 21 & 22 & 17 & 10 & 14 & 87 \\
P2, $\gamma{=}0.005$ & 3 & 24 & 22 & 15 & 10 & 14 & 88 \\
P2, $\gamma{=}0.02$ & 3 & 25 & 23 & 10 & 11 & 16 & 88 \\
P2, $\gamma{=}0.1$ & 13 & 18 & 18 & 11 & 17 & 13 & 90 \\
P3, $\rho{=}0.99$ & 3 & 23 & 22 & 15 & 10 & 15 & 88 \\
P3, $\rho{=}0.98$ & 3 & 23 & 20 & 12 & 11 & 14 & 83 \\
P3, $\rho{=}0.95$ & 2 & 22 & 16 & 11 & 12 & 13 & 76 \\
\bottomrule
\end{tabular}
}
\end{table}

\begin{table}[tbp]
\centering
\caption{Procedures~1, 2 ($\gamma \in \{0.005,0.02,0.1\}$) and~3 ($\rho \in \{0.99,0.98,0.95\}$) on the 184 days of the evaluation period ($T=90$, $n_1=43$, $n_2=39$, random split, $b=1$, $K=40$; Procedure~2 with delay $\tau=7$; Procedure~3 with weights $\rho^{\delta}$, $\delta$ the number of days between the calibration day and the predicted day, $8\le\delta\le 89$, and effective number of calibration scores $(\sum_d w_d)^2/\sum_d w_d^2$ between 36.4 and 37.5 for $\rho=0.99$, between 30.0 and 33.8 for $\rho=0.98$ and between 14.6 and 21.8 for $\rho=0.95$). Frequency: percentage of days on which the observed pair of curves lies in the band (column Curves) and on which the clearing point lies in the region $\mathcal{R}_t$ (column $(Q_t,P_t)$); median, interquartile range (IQR) and mean of the average width of the finite, non-empty bands (Euro/MWh; mean over the grid of the difference between the monotone bounds, averaged over the two curves); Bands: number of bands equal to the whole space (Inf.) and of empty bands (Empty); Dev.: $|H^{-1}\sum_t \mathrm{err}_t-\alpha|$; Bound: right-hand side of~\eqref{eq:aci_bound} with $H=184$ (--: not applicable); P1, P2, P3: Procedures~1, 2 and~3.}
\label{tab:gas_procedures_full}
{\footnotesize\setlength{\tabcolsep}{4pt}
\begin{tabular}{lrrrrrrrrr}
\toprule
 & \multicolumn{2}{c}{Frequency (\%)} & \multicolumn{3}{c}{Width} & \multicolumn{2}{c}{Bands} & & \\
\cmidrule(lr){2-3}\cmidrule(lr){4-6}\cmidrule(lr){7-8}
 & Curves & $(Q_t,P_t)$ & Median & IQR & Mean & Inf. & Empty & Dev. & Bound \\
\midrule
\multicolumn{10}{l}{$\alpha=0.25$ (nominal level 0.75)} \\
P1 & 75.5 & 97.8 & 4.51 & 1.36 & 4.60 & 0 & 0 & 0.005 & -- \\
P2, $\gamma{=}0.005$ & 77.7 & 97.3 & 4.55 & 1.49 & 4.67 & 0 & 0 & 0.027 & 0.887 \\
P2, $\gamma{=}0.02$ & 76.1 & 97.3 & 4.50 & 1.92 & 4.74 & 0 & 0 & 0.011 & 0.276 \\
P2, $\gamma{=}0.1$ & 74.5 & 94.6 & 3.71 & 1.50 & 4.15 & 35 & 0 & 0.005 & 0.113 \\
P3, $\rho{=}0.99$ & 78.8 & 97.8 & 4.68 & 1.44 & 4.80 & 0 & 0 & 0.038 & -- \\
P3, $\rho{=}0.98$ & 80.4 & 97.3 & 4.87 & 1.83 & 4.99 & 0 & 0 & 0.054 & -- \\
P3, $\rho{=}0.95$ & 84.2 & 97.3 & 5.82 & 2.57 & 5.90 & 0 & 0 & 0.092 & -- \\
\addlinespace
\multicolumn{10}{l}{$\alpha=0.5$ (nominal level 0.50)} \\
P1 & 52.7 & 92.4 & 3.16 & 0.65 & 3.20 & 0 & 0 & 0.027 & -- \\
P2, $\gamma{=}0.005$ & 52.2 & 91.3 & 3.14 & 0.72 & 3.21 & 0 & 0 & 0.022 & 0.606 \\
P2, $\gamma{=}0.02$ & 52.2 & 89.7 & 3.04 & 0.96 & 3.18 & 0 & 0 & 0.022 & 0.198 \\
P2, $\gamma{=}0.1$ & 51.1 & 85.9 & 3.07 & 1.10 & 3.61 & 4 & 9 & 0.011 & 0.090 \\
P3, $\rho{=}0.99$ & 52.2 & 92.4 & 3.20 & 0.75 & 3.27 & 0 & 0 & 0.022 & -- \\
P3, $\rho{=}0.98$ & 54.9 & 91.8 & 3.22 & 0.80 & 3.29 & 0 & 0 & 0.049 & -- \\
P3, $\rho{=}0.95$ & 58.7 & 91.8 & 3.34 & 1.26 & 3.53 & 0 & 0 & 0.087 & -- \\
\bottomrule
\end{tabular}
}
\end{table}

\paragraph{Procedure~2} For $H=184$, the bound \eqref{eq:aci_bound} allows, for $\alpha=0.25$, frequencies of misses in $[0.137,0.363]$ ($\gamma=0.1$) and $[0,0.526]$ ($\gamma=0.02$), and, for $\alpha=0.5$, in $[0.410,0.590]$ and $[0.302,0.698]$; for $\gamma=0.005$, it allows any frequency. The working miscoverage level $\alpha_t$ ranges over $[0.23,0.28125]$, $[0.15,0.37]$ and $[-0.375,0.825]$ for $\gamma=0.005$, $0.02$ and $0.1$ ($\alpha=0.25$), and over $[0.4675,0.5625]$, $[0.34,0.75]$ and $[0,1.1]$ ($\alpha=0.5$). In all six cases, after few misses in the first half of August, $\alpha_t$ first reaches its maximum on a day between 28 August and 10 September 2019, and the misses that follow lower it only eight days later. With $\gamma=0.1$ and $\alpha=0.25$, the band is the whole space ($\alpha_t<1/K$) on 35 days, and $\alpha_t=1/K=0.025$ on 10 October and 19 November 2019, days on which $k_t$ is the largest of the 39 calibration scores; with $\gamma=0.1$ and $\alpha=0.5$, the band is the whole space on 4 days and empty ($\alpha_t \geq 1$) on the 9 days from 26 August to 3 September 2019.

\paragraph{Procedure~3} In the notation of the window, the predicted day is the time point $T+1+\tau=98$ (Section~\ref{sec:weighted}), and the distance $\delta=98-\omega_d$ of a calibration day is at least 8, since the 7 days that precede the predicted day are not in the window; on each day, the smallest distance is between 8 and 15 and the largest between 82 and 89. Over the 184 days, $W$ lies in $[22.80,26.24]$, $[13.98,18.59]$ and $[4.26,8.42]$ for $\rho=0.99$, $0.98$ and $0.95$, so that no band is the whole space, which would require $W<3$ ($\alpha=0.25$) or $W<1$ ($\alpha=0.5$). The mass $1/(W+1)$ of the test point lies in $[0.037,0.042]$, $[0.051,0.067]$ and $[0.106,0.190]$, and the effective number of calibration scores $(\sum_d w_d)^2/\sum_d w_d^2$ in $[36.4,37.5]$, $[30.0,33.8]$ and $[14.6,21.8]$ (the version that includes the test point, in $[37.1,38.3]$, $[30.1,33.8]$ and $[12.6,20.4]$). These quantities depend only on the split. The band of day $t$ is that of Procedure~1 at the miscoverage level $1-m_t/K$, where $m_t$ is the position of $k_t$ among the 39 sorted calibration scores (30 and 20 for Procedure~1 with $\alpha=0.25$ and $0.5$). For $\alpha=0.25$, $m_t$ ranges from 29 to 32, from 27 to 34 and from 23 to 39 (the largest score on 3 days with $\rho=0.95$), and the mean of $1-m_t/K$ over the days is 0.231, 0.220 and 0.169; for $\alpha=0.5$, $m_t$ ranges from 18 to 23, from 16 to 26 and from 10 to 32, and the mean of $1-m_t/K$ is 0.488, 0.485 and 0.451. With $\rho=0.95$, $k_t$ is larger than that of Procedure~1 on 150 days, equal on 6 and smaller on 28 ($\alpha=0.25$), and larger on 109, equal on 18 and smaller on 57 ($\alpha=0.5$); with $\rho=0.98$, on 123, 28 and 33 days and on 91, 30 and 63 days; with $\rho=0.99$, on 115, 56 and 13 days and on 85, 59 and 40 days. With $\rho=0.95$, Procedure~3 misses on 1 ($\alpha=0.25$) and 6 ($\alpha=0.5$) days on which Procedure~1 covers, and covers on 17 days (at both levels) on which Procedure~1 misses. Since the weight of the test point is fixed to $1$, measuring the distance from the day after the end of the window, i.e.\ using the weights $\rho^{\delta-7}$, with distances from 1 to 82, gives a different procedure (\ref{app:remarks}); on these data, it has 40, 37 and 34 misses for $\alpha=0.25$ ($\rho=0.99$, $0.98$ and $0.95$) and 89, 83 and 80 for $\alpha=0.5$, against 39, 36 and 29, and 88, 83 and 76. In the code, Procedure~3 with $\rho=1$ gives the same $k_t$, misses and widths as Procedure~1 on all days.

\end{document}